\documentclass[aps,prb,twocolumn,superscriptaddress]{revtex4-2}
\usepackage{graphicx}
\usepackage{latexsym}
\usepackage{amssymb}
\usepackage{amsmath}
\usepackage{amsfonts}
\usepackage[vcentermath]{youngtab}
\usepackage{upgreek}
\usepackage{bm,dsfont}
\usepackage{multirow}
\usepackage{enumitem}
\usepackage{color}
\usepackage{hyperref}
\hypersetup{
colorlinks = true,
linkcolor = [rgb]{0.70,0.13,0.13},
citecolor = [rgb]{0.13,0.55,0.13},
urlcolor = [rgb]{0.25, 0.41, 0.88}}
\usepackage{makecell}

\newcommand{\bra}[1]{\langle #1|}
\newcommand{\ket}[1]{|#1 \rangle}

\newcommand{\dd}{\mathrm{d}}
\newcommand{\DD}{\mathcal{D}}
\newcommand{\ii}{\mathrm{i}}
\newcommand{\e}{\mathrm{e}}
\newcommand{\rep}{\mathbf{r}}
\newcommand{\ybox}{\tiny\yng}

\newcommand{\U}{\mathrm{U}}
\newcommand{\SU}{\mathrm{SU}}

\newcommand{\SO}{\mathrm{SO}}
\newcommand{\Sp}{\mathrm{Sp}}
\newcommand{\Spin}{\mathrm{Spin}}
\renewcommand{\u}{\mathfrak{u}}
\newcommand{\su}{\mathfrak{su}}

\newcommand{\so}{\mathfrak{so}}

\newcommand{\dsZ}{\mathbb{Z}}
\newcommand{\scA}{\mathcal{A}}

\newcommand{\scL}{\mathcal{L}}

\newcommand{\scN}{\mathcal{N}}

\newcommand{\scS}{\mathcal{S}}
\newcommand{\scT}{\mathcal{T}}

\newcommand{\vect}[1]{{\bm{#1}}}

\newcommand{\eq}[1]{\begin{equation}#1\end{equation}}
\newcommand{\eqs}[1]{\begin{equation}\begin{split}#1\end{split}\end{equation}}
\newcommand{\eqnref}[1]{Eq.\,\eqref{#1}}
\newcommand{\figref}[1]{Fig.\,\ref{#1}}
\newcommand{\tabref}[1]{Tab.\,\ref{#1}}
\newcommand{\secref}[1]{Sec.\,\ref{#1}}
\newcommand{\refcite}[1]{Ref.\,\onlinecite{#1}}

\def \Z{\mathbb{Z}}
\newcommand{\bea}{\begin{eqnarray}}
\newcommand{\eea}{\end{eqnarray}}
\def\be{\begin{equation}}
\def\ee{\end{equation}}
\newcommand{\beq}{\begin{equation}}
\newcommand{\eeq}{\end{equation}}
\newcommand{\beqn}{\begin{eqnarray}}
\newcommand{\eeqn}{\end{eqnarray}}

\newcommand{\CW}{\mathcal{W}}
\newcommand{\SM}{\mathrm{SM}}
\newcommand{\LR}{\mathrm{LR}}
\newcommand{\CN}{\mathcal{N}}
\newcommand{\TP}{\mathrm{TP}}

\newcommand{\PD}{\mathrm{PD}}

\usepackage{hhline}
\usepackage{array}

\begin{document}

\title{Symmetric Mass Generation}

\author{Juven Wang}
\affiliation{Center of Mathematical Sciences and Applications, Harvard University, MA 02138, USA}
\email[]{jw@cmsa.fas.harvard.edu}

\author{Yi-Zhuang You}
\affiliation{Department of Physics, University of California, San Diego, CA 92093, USA}
\email[]{yzyou@ucsd.edu}



\begin{abstract}

The most well-known mechanism for fermions to acquire a mass is the Nambu-Goldstone-Anderson-Higgs mechanism, i.e.~after a spontaneous symmetry breaking, a bosonic field that couples to the fermion mass term condenses, which grants a mass gap for the fermionic excitation. In the last few years, it was gradually understood that there is a new mechanism of mass generation for fermions without involving any symmetry breaking within an anomaly-free symmetry group. 
This new mechanism is generally referred to as the ``Symmetric Mass Generation (SMG)." It is realized that the SMG has deep connections with interacting topological insulator/superconductors, symmetry-protected topological states, perturbative local and non-perturbative global anomaly cancellations,
and deconfined quantum criticality. It has strong implications for the lattice regularization of chiral gauge theories. This article defines the SMG, summarizes current numerical results, introduces a novel unifying theoretical framework (including the parton-Higgs and the s-confinement mechanisms, as well as the symmetry-extension construction), and overviews various features and applications of SMG.

\end{abstract}
\maketitle
\tableofcontents

\section{Introduction}

Global symmetry is a central concept in quantum field theories (QFTs). One of the most immediate implications of symmetry is that it restricts the terms that can appear in a field theory action in the path integral formulation, as the 
{partition function ${Z}$} must remain invariant under the symmetry transformation --- the 
{partition function} invariance however can be up to an invertible {complex phase factor ${Z}\to\e^{\ii\alpha}{Z}$} known as a quantum anomaly \cite{Fujikawa1980Path,Fujikawa2004Path}. The quantum anomaly associated with the global symmetry is also known as the 't Hooft anomaly \cite{tHooft1980}, which has profound consequences. The invertible phase
$\e^{\ii\alpha}$ that detects the anomaly also specifies  
a cobordism class of the partition function of
one-higher-dimensional invertible topological quantum field theory (invertible TQFT) \cite{Freed2016Reflection} 
via the anomaly inflow \cite{1984saCallanHarvey, Witten2019bou1909.08775}.

The anomaly provides a concise and powerful organization principle to classify quantum field theories, and to dictate the influence of the ultraviolet (UV) kinematics on the infrared (IR) dynamics of a field theory, because the anomaly does not change under symmetric deformations of the theory, including the renormalization group (RG) flow. 
The preservation of the anomaly index from UV to IR is known as 
the anomaly matching.
If a quantum field theory with a global symmetry $G$ has a non-vanishing 't Hooft anomaly in $G$, its IR dynamics cannot be trivially gapped while preserving $G$. As a consequence, either the symmetry $G$ is spontaneously broken in IR~\cite{Frishman1981The-axial,Wess1971Consequences,Witten1983Global,Hason2020Anomaly,Yonekura2020General}, or the anomaly must be saturated 
by anomalous \emph{gapless} quantum criticality (e.g., the IR limit of the system corresponds to massless free theories or interacting conformal field theories [CFT]) or \emph{gapped} topological order
(e.g., the IR limit of the system is described by a topological quantum field theory [TQFT]) \cite{Vishwanath2013Physics,Bonderson2013TimeReversal,Wang2013Gapped,Fidkowski2013Non-Abelian,Wang2014Interacting,Metlitski2014Interaction,Burnell2014Exactly,Mross2015Composite,Metlitski2015Symmetry-respecting,Wang2016Bulk-Boundary,
SeibergWitten1602.04251, Witten:2016yb, Wang2018Symmetric}.

Many examples of anomalies involve massless fermions, such as the chiral fermion Adler-Bell-Jackiw anomaly \cite{Adler1969Axial-Vector,Bell1969A-PCAC}. The anomaly causes an obstruction to open a gap in the fermion spectrum (i.e.~an obstruction to make the fermion correlation length finite) without breaking the symmetry.
The symmetry-breaking mechanism to generate a mass gap is known as Nambu-Goldstone-Anderson-Higgs mechanism 
\cite{Nambu1960PRQuasiparticles, Nambu1961PRJonaLasinio, Goldstone1961NuovoCim, GoldstoneSalamWeinberg1962,
Anderson1963pcPRPlasmons, EnglertBrout1964PRL, Higgs1964PRL}. 
Here Nambu-Goldstone refers to the spontaneous symmetry breaking, 
while Anderson-Higgs refers to giving a mass by the elementary or composite boson condensation.
However, there are instances where the anomaly vanishes for a collection of massless fermions, yet the symmetry is still restrictive enough to forbid any fermion bilinear mass term. In this case, although there is no obstruction towards gapping the fermions symmetrically, the mechanism to achieve the symmetric gapped state must go beyond the free-fermion (perturbatively free or weak-coupling) approach, which potentially leads to a  non-perturbative strong-coupling 
approach in order to generate a finite excitation gap in the fermion many-body spectrum by non-trivial interaction effects. The \emph{strong-coupling} \cite{Tong2021Comments}
here refers to the coupling in the continuum field theory being non-perturbative, or the interaction energy being of the same order as the kinetic energy on the lattice scale (which may also be called the \emph{intermediate}-strength interaction on a lattice).
This phenomenon of gapping out massless fermions by interactions in an anomaly-free system without breaking the anomaly-free symmetry is now called the \emph{symmetric mass generation} (SMG) \cite{You2018Symmetric,Tong2021Comments}.

The idea of gapping out massless fermions by interaction dates back to Eichten and Preskill \cite{Eichten1986Chiral} in an attempt to regularize chiral fermions on the lattice. The understanding of SMG is significantly deepened over the past few years, following the development in condensed matter theory regarding symmetry-protected topological (SPT) states \cite{Gu:2009ai,Pollmann:2012jw,Chen:2011vx}. SPT states are short-range-entangled quantum many-body states respecting certain global symmetry $G$. The bulk of a SPT state is featureless (i.e.~gapped, symmetric and non-degenerated). A non-trivial SPT state is most explicitly characterized by its non-trivial boundary features, which are endowed by the non-vanishing 't Hooft anomaly of $G$ (or mixed $G$ symmetry gauge-gravity anomaly in broader cases) in the boundary effective theory of the SPT state. The one-to-one correspondence between the bulk of a SPT state and its boundary anomaly \cite{Ryu2012Electromagnetic,Wenanomalies1303.1803,Kapustin2014Anomalies,Wang2015Field-Theory, Witten1508.04715} provides the basis to classify distinct SPT states by their distinct boundary anomalies. 

The connection between SPT state and anomaly relates SMG to another topic: the interaction-reduced classification of fermionic SPT states. 
Fermionic SPT states \cite{Turner2011Topological,Gu2012Symmetry-protected,Cheng2015Classification,Kapustin2015Fermionic,Freed2016Reflection,Gaiotto2016Spin,Wang2017Towards,Kapustin2017Fermionic,1711.11587GPW, Wang2018Tunneling,Wang2018Construction,Gaiotto2019Symmetry,Lan2019Fermion,Guo2020Fermionic,Ouyang2020Computing} are SPT states of fermionic systems with (at least) fermion parity symmetry $\mathbb{Z}_2^F$ (or more generally, invertible topological phases of fermions \cite{Freed2016Reflection,Guo2020Fermionic}). In the free-fermion limit \cite{Schnyder:2008os,Kitaev2009Periodic,Ryu:2010fe,Wen2012fSPT,Ludwig:2016pt}, non-trivial fermionic SPT states are characterized by the symmetry-protected gapless fermion boundary modes, if the symmetry forbids any fermion bilinear mass on the boundary. However, the free-fermion analysis does not rule out the possibility of gapping out the boundary fermions by interaction though the SMG mechanism. Such situation can indeed happen when the boundary fermions are actually anomaly free, such that the bulk state should be classified as a trivial gapped state under symmetric interactions, even though it looks like a non-trivial SPT state in the non-interacting limit. This leads to the interaction-reduced classification of fermionic SPT states in the bulk, which is closely related to the SMG for gapless fermions on the boundary.

The first example of interaction-reduced classification was provided by Fidkowski and Kitaev \cite{Fidkowski:2010bf,Fidkowski:2011dd} in (1+1)D fermion systems, where an explicit interaction was proposed to drive the SMG among fermion zero modes on the (0+1)D boundary. This motivated a sequence of works generalizing the discussion to (2+1)D \cite{Ryu:2012ph,Qi:2013qe,Yao:2013yg,Gu:2014tw}, (3+1)D \cite{Fidkowski2013Non-Abelian,Wang2014Interacting,Metlitski2014Interaction,Yoshida:2015aj,Gu:2015cy}, and higher dimensions \cite{You:2014ho,Song:2016ut,Queiroz:2016se}. Studies along this direction reveals families of interactions that could potentially drive the SMG in different dimensions, paving ways for numerical verifications in concrete lattice models. It is also realized that the interaction must be carefully designed to drive the SMG: some symmetric interactions are helpful towards this goal, while other symmetric interactions are not~\cite{Wang2013Non-Perturbative}. 
%
%

Currently most known examples of interaction-reduced classifications all correspond to cancellations of \emph{non-perturbative global anomalies}. However, there are also cases of SMG that involve cancellations of \emph{perturbative local anomalies}. Important examples of such are the chiral fermions \cite{Kaplan1992A-method, Luscher2001Chiral,Kaplan2009Chiral,Poppitz2010Chiral} in even dimensional spacetime. Regularizing chiral fermions on a lattice is a long-standing problem in the lattice gauge theory. The Nielsen-Ninomiya no-go theorem \cite{Nielsen:1981we,Nielsen1981Absence,Nielsen:1981bc} states that it is not possible to gap the fermion doubler in a non-interacting local lattice model without breaking the chiral symmetry. One possibility to circumvent the no-go theorem is to consider interaction effects \cite{Eichten1986Chiral}, which introduces fermion interaction to gap out the fermion doubler (the mirror fermion) in local lattice fermion models, leaving the normal fermion (the light fermion) untouched. However, the early attempts \cite{Bock1990Unquenched,Lee1990Study,Hasenfratz1991The-equivalence,Banks1992Decoupling,Golterman1993Absence,Lin1994Nondecoupling,Bock1994Staggered,Golterman1995Domain,Poppitz2009Lattice,Poppitz2010Chiral,Chen2013Model345} were not successful, either because certain anomalies were not carefully cancelled or because the appropriate gapping interaction has not been found. With a deeper understanding of the SMG mechanism, the problem was revisited by Wen \cite{Wen:2013kr} for $\Spin(10)$ chiral fermions in (3+1)D. The idea is further developed by subsequent works in the same dimension \cite{You:2014ow,You:2015lj,BenTov:2016co,Wang2018A-Non-Perturbative,Razamat2021Gapped} as well as in lower-dimensional analogs \cite{Wang2013Non-Perturbative,BenTov:2015lh,DeMarco2017A-Novel,Wang2019Solution}. More recent numerical works have successfully shown that the SMG indeed provides a feasible solution to regularize chiral fermions \cite{Kikukawa2017On-the-gauge,Kikukawa2019Why-is-the-mission,Catterall2021Chiral,Zeng2022Symmetric}.


The article is organized as follows. We will start by introducing some selective representative models of SMG (one in each spacetime dimension, in (0+1)D, (1+1)D, (2+1)D and (3+1)D respectively) in \secref{sec:examples}. We conclude with \secref{sec:generalD} by providing a general definition of SMG in all dimensions. We then review the numerical efforts in \secref{sec:numerics}, which is mainly focused on two tasks: (i) to establish the existence of SMG phases in \secref{sec:phase} and 
(ii) to investigate the nature of SMG transitions in \secref{sec:transition}. 
Based on these backgrounds, we then summarize the recent theoretical progress in \secref{sec:theory}, which aims to (i) understand the SMG phase by the fluctuating bilinear mass picture in \secref{sec:FBM} and (ii) describe the SMG transition by fermion fractionalization field theory in \secref{sec:fermion_frac}. In particular, we unify two currently existing SMG mechanisms (namely the parton-Higgs mechanism and the s-confinement mechanism) under the same theoretical framework of fermion fractionalization. 
We also make connection to 
further understanding of SMG based on the {symmetry extension construction} \cite{Witten:2016yb, Wang2018Symmetric} in \secref{sec:Symmetry-Extension}. In \secref{sec:other}, we discuss other aspects of SMG including Green's function zeros \secref{sec:zeros}, the deconfined quantum criticality \secref{sec:DQCP}, and the Standard Model regularization \secref{sec:SM}. Finally, we summarize our review in \secref{sec:summary}.

\section{Example Models}
\label{sec:examples}

\subsection{(0+1)D SMG: Fidkowski-Kitaev Majorana Fermion Model}
\label{sec:FK}

The simplest example of SMG happens in (0+1)D spacetime among a collection of Majorana fermion zero modes, as first shown by Fidkowski and Kitaev \cite{Fidkowski:2010bf,Fidkowski:2011dd}. The model concerns a system of eight Majorana fermion modes, described by the Majorana fermion operators $\chi_a$ ($a=1,2,\cdots,8$) satisfying $\{\chi_a,\chi_b\}=2\delta_{ab}$. Consider an anti-unitary (time-reversal) symmetry $\dsZ_2^{T}:\chi_a\to\chi_a, \ii\to-\ii$ 
and the fermion parity symmetry $\dsZ_2^{F}:\chi_a\to-\chi_a$. Without involving interactions, the Hamiltonian must take a fermion bilinear form to preserve the fermion parity symmetry $\dsZ_2^{F}$. But any Majorana fermion bilinear term $\ii\chi_{a}\chi_{b}$ (one needs $\ii$ to keep the operator Hermitian) will break the time-reversal symmetry $\dsZ_2^{T}$, so the free-fermion Hamiltonian has to vanish, e.g. $H=0$, under both symmetry requirements, and these Majorana modes cannot be gapped on the free-fermion level. Eight Majorana fermion modes form four qubits, hence the system has a $2^4 = 16$ fold degeneracy, and this degeneracy is protected by the $\dsZ_2^T\times\dsZ_2^F$ symmetry. 

However, it is possible to create a many-body excitation gap by fermion interaction, leaving a unique ground state of this (0+1)D system. As introduced by Fidkowski and Kitaev \cite{Fidkowski:2010bf}, the following four-fermion interaction suffices to gap out all eight Majorana fermion modes without breaking the time-reversal symmetry $\dsZ_2^{T}$
\eq{\label{eq:H_FK}
H_\text{FK}=-\sum_{a<b<c<d}V_{abcd}\chi_a\chi_b\chi_c\chi_d.}
The coefficient is specified by $V_{abcd}=\bra{\mathsf{4e}}\chi_a\chi_b\chi_c\chi_d\ket{\mathsf{4e}}$, where $\ket{\mathsf{4e}}=(\ket{0000}+\ket{1111})/\sqrt{2}$ is a many-body reference state written in the Fock state basis $\ket{n_1n_2n_3n_4}$, labeled by the fermion occupation numbers $n_i=(1+\ii\chi_{2i-1}\chi_{2i})/2$ (for $i=1,2,3,4$). Here the fermion number operator $n_i=c_i^\dagger c_i$ can also be expressed in terms of the complex fermion annihilation operator $c_i = (\chi_{2i-1} + \ii \chi_{2i})/2$, which are constructed by pairing up the Majorana operators. The reference state $\ket{\mathsf{4e}}$ describes the quartet condensation of the complex fermions $c_i$ (where four fermions are created or annihilated together in the quantum superposition), which is also known as the charge-4e superconducting state \cite{Kivelson:1990wy,Talukdar2007Quartet,Berg:2009uq,Radzihovsky:2009bk,Berg:2009nm,Moon:2012ng,Jiang:2016km} in condensed matter physics. It turns out that the state $\ket{\mathsf{4e}}$ respects the $\dsZ_2^{T}$ symmetry, and is the unique ground state of $H_\text{FK}$ with a finite excitation gap of $14$ (energy units). 

The fact that the system has \emph{eight} Majorana zero modes is crucial for the SMG to occur. In contrast, if there are only four Majorana zero modes, the only Hamiltonian allowed by symmetry is 
\eq{H = - g \chi_1 \chi_2 \chi_3 \chi_4=-g P_1^FP_2^F,} 
where each pair of Majorana fermion operators defines a fermion parity operator $P_i^F= \ii \chi_{2i-1} \chi_{i} = 2 n_i - 1$ (associated with the $i$th complex fermion mode). This Hamiltonian always has a two-fold ground state degeneracy, regardless of the sign of $g$. When $g < 0$, the ground states $\ket{P_1^F=P_2^F=\pm1}$ are bosonic (as the total fermion parity $P_1^FP_2^F=+1$ is even). When $g>0$, the ground states $\ket{P_1^F=-P_2^F=\pm1}$ are fermionic (as $P_1^FP_2^F=-1$). In either cases, the two-fold degenerated ground states transform into each other under $\dsZ_2^T$, and form a \emph{Kramers doublet} \cite{Kramers1930Theorie}. The ground state degeneracy implies that the $\dsZ_2^T$ symmetry is spontaneously broken. Hence even with interactions, \emph{four} Majorana zero modes still \emph{cannot} be symmetrically gapped. However, by doubling to eight Majorana fermions, it is then possible to couple the two Kramers doublets (originated from the first- and the last-four Majorana fermions under interaction) together via a Heisenberg-type spin-spin interaction as \cite{You:2014ho,Prakash2021Unwinding}
\eq{\label{eq:H_SS}H=-P_1^F P_2^F-P_3^FP_4^F+\vect{S}_\text{I}\cdot\vect{S}_\text{II},}
where the first two terms stabilizes two Kramers doublets, and $\vect{S}_\text{I/II}$ stands for the effective spin-operator for each Kramers doublet. This will end up with a unique spin-singlet bosonic ground state with a finite gap to all excitations, which successfully gaps out all fermions without breaking the symmetry. \eqnref{eq:H_SS} can be rewritten as $H_\text{FK}$ in \eqnref{eq:H_FK} under some appropriate basis choice.

From the perspective of quantum anomaly, the (0+1)D Majorana fermions with a $\dsZ_2^{T}\times\dsZ_2^{F}$ \footnote{$\dsZ_2^{F}:\chi_i\to-\chi_i$ denotes the fermion parity symmetry, which is always assumed implicitly for local fermion systems.} internal symmetry (or the $\text{Pin}^{-}$ spacetime-internal symmetry \cite{Kapustin2015Fermionic,Prakash2021Unwinding}, or the BDI symmetry class \cite{Altland1997Nonstandard,Zirnbauer2010Symmetry}) has a $\dsZ_8$ class of non-perturbative global anomaly, whose anomaly index corresponds to the number of Majorana modes. With eight Majorana modes, the anomaly vanishes, meaning that the system can be trivially gapped without breaking symmetry. However, the $\dsZ_2^{T}$ symmetry is restrictive enough to rule out any fermion bilinear mass, making interaction a necessary ingredient in the fermion mass generation, which corresponds to the SMG mechanism.

The Fidkowski-Kitaev (FK) interaction $H_\text{FK}$ in \eqnref{eq:H_FK} has a (unnecessarily high) flavor symmetry of $\Spin(7)$, which rotates seven fermion bilinear operators $\Phi^\alpha:=\chi_a\Gamma_{ab}^\alpha\chi_b$ (for $\alpha=1,\cdots,7$) as a vector representation of $\SO(7)$, with $\Gamma^{\alpha}=(\sigma^{123},\sigma^{203},\sigma^{323},\sigma^{211},\sigma^{021},\sigma^{231},\sigma^{002})$, where $\sigma^{\mu\nu\cdots}=\sigma^{\mu}\otimes\sigma^{\nu}\otimes\cdots$ denotes the direct product of Pauli matrices. It can be shown that \eqnref{eq:H_FK} can be equivalently written as
\eqs{\label{eq:H_FK_YH}
H_\text{FK}=-\frac{1}{4!}\sum_{\alpha=1}^{7}(\Phi^\alpha\Phi^\alpha-16),
} 
which exhibits the $\Spin(7)$ symmetry explicitly. However, it is possible to lower the symmetry to $\Spin(6)\cong\SU(4)$ for example, without affecting the SMG physics \cite{You:2015lj},
\eqs{\label{eq:H_SU4}
&H_{\SU(4)}=-\frac{1}{192}\sum_{\alpha=1}^{6}\Phi^\alpha\Phi^\alpha\\
&=-(c_1c_2c_3c_4+\text{h.c.})-\frac{1}{3}\sum_{i<j}(n_i-\tfrac{1}{2})(n_j-\tfrac{1}{2}),}
where $c_i$ (for $i=1,2,3,4$) are complex fermions in $\SU(4)$ fundamental representation and $n_i=c_i^\dagger c_i$ are their number operators. $H_{\SU(4)}$ stabilizes the \emph{same} symmetric and non-degenerated ground state $\ket{\mathsf{4e}}$ as $H_\text{FK}$ with a finite gap to all excitations. This example illustrates that the interaction that drives SMG is not unique, since the SMG is a generic phenomenon in anomaly free fermion systems.

The decomposition of the four-fermion interaction as a product of two fermion-bilinear operators $\Phi^\alpha\Phi^\alpha$ in both \eqnref{eq:H_FK_YH} and \eqnref{eq:H_SU4} suggests a common physical picture to understand the SMG as fluctuating bilinear masses. Under the Hubbard-Stratonovich transformation, the SMG interactions in the above examples take the general form of Yukawa-Higgs interaction 
\eq{\label{eq:H_YH}H_\text{YH}=-\phi_\alpha\chi_a\Gamma^{\alpha}_{ab}\chi_b+\frac{1}{2g}\phi_\alpha\phi_\alpha,}
where $\phi_\alpha$ is a bosonic Yukawa field that couples to the fermion bilinear mass, and by integrating out $\phi^\alpha$, the desired four fermion interaction is generated. If the Yukawa field condenses, i.e.~$\langle\phi_\alpha\rangle\neq0$, the fermions will be gapped (with a gap size proportional to the amplitude of $\langle\phi_\alpha\rangle$) and the symmetry is also broken spontaneously. However, if it is possible to fluctuate the orientation of the Yukawa field smoothly in the spacetime (which is only the time here for the (0+1)D system) without bringing its local amplitude to zero, the fermion could potentially retain the excitation gap while restoring the required symmetry. This intuitive picture leads to fruitful understandings of SMG in higher dimensions \cite{Wang2014Interacting,Catterall:2016sw,DeMarco2017A-Novel}, which will be further discussed in \secref{sec:FBM}.

The SMG in (0+1)D is intimately related to the interaction-reduced classification of fermionic SPT states in (1+1)D. In the non-interacting limit, the (0+1)D Majorana zero modes can be viewed as the topological edge modes of a (1+1)D fermionic SPT state protected by the same $\dsZ_2^{T}\times\dsZ_2^{F}$ internal symmetry. A specific lattice model for such SPT root state is Kitaev's Majorana fermionic chain \cite{Kitaev2001Unpaired}, 
which supports a dangling Majorana zero mode on each open end. So eight Majorana zero modes can be viewed as the boundary state of eight copies of Majorana chains. Without interaction, the $\dsZ_2^{T}\times\dsZ_2^{F}$ symmetric (BDI symmetry class) \emph{free} fermionic SPT states are $\dsZ$ classified \cite{Schnyder:2008os,Kitaev2009Periodic,Ryu:2010fe}, where the index corresponds to the number of Majorana edge modes \cite{Kaplan2021Index}. The fact that eight Majorana zero modes can be trivially gapped out by interaction without breaking the symmetry implies that eight copies of Majorana chain actually belong to the trivial SPT phase, as their interface with the vacuum state can be made featureless (i.e.~gapped and non-degenerated) by the SMG. This indicates that the $\dsZ_2^{T}\times\dsZ_2^{F}$ (or $\text{Pin}^{-}$) symmetric \emph{interacting} fermionic SPT states are $\dsZ_8$ classified \cite{Fidkowski:2010bf,Fidkowski:2011dd,Turner2011Topological}, which is consistent with the formal result $\Omega_2^{\text{Pin}^-}(\mathsf{pt})=\dsZ_8$ by the cobordism classification \cite{KT1990, Kapustin2015Fermionic}. The phenomenon that the fermion SPT classification is reduced from $\dsZ$ in the non-interacting limit to $\dsZ_8$ under interaction is called the \emph{interaction-reduced classification}.

\begin{figure}[htbp]
\begin{center}
\includegraphics[width=0.68\columnwidth]{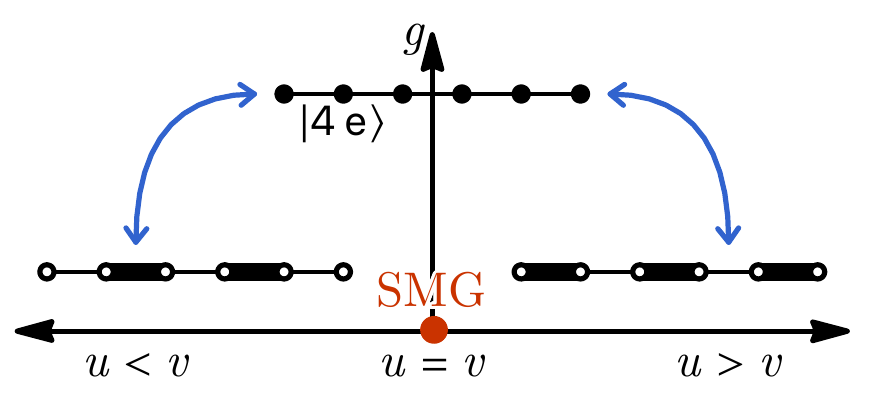}
\caption{Phase diagram of the interacting Majorana chain model described by \eqnref{eq:H_maj_chain}}
\label{fig:chain}
\end{center}
\end{figure}

The $\dsZ\to\dsZ_8$ interaction-reduced classification implies that eight copies of the Majorana chain can be smoothly tuned (without closing the bulk gap) to the trivial phase under interaction. The bulk model is defined on a one-dimensional lattice
\eqs{\label{eq:H_maj_chain}H&=H_0+H_\text{int},\\
H_0&=\frac{\ii}{2}\sum_{i}\sum_{a=1}^{8}(u\,\chi_{2i-1,a}\chi_{2i,a}+v\,\chi_{2i,a}\chi_{2i-1,a}),\\
H_\text{int}&=-g\sum_{i}\sum_{a<b<c<d}V_{abcd}\chi_{i,a}\chi_{i,b}\chi_{i,c}\chi_{i,d},}
where each site $i$ hosts eight Majorana modes $\chi_{i,a}$ ($a=1,\cdots,8$). In the free-fermion limit ($g=0$), the $u<v$ and $u>v$ phases are separated by a quantum phase transition at which the single-particle band gap closes in the bulk. \refcite{Fidkowski:2010bf} shows that the quantum critical point can be circumvented by applying the Fidkowski-Kitaev interaction on every site with strength $g$ (where $V_{abcd}$ in $H_\text{int}$ follows the same definition as that in \eqnref{eq:H_FK}). This can be argued by inspecting the limiting cases $(u=0,v=1)$ or $(u=1,v=0)$ where Majorana fermions are fully dimerized along the lattice over even or odd bonds, see \figref{fig:chain}. The two dimerized states can both be smoothly tuned to the same strong coupling state $\bigotimes_i\ket{\mathsf{4e}}_i$ as $g\to\infty$ without closing the many-body gap, which can be explicitly verified by diagonalizing the local Hamiltonian across a bond. This establishes a smooth deformation path between the $u<v$ and $u>v$ phases by going through the strong coupling regime as shown in \figref{fig:chain}, demonstrating that eight copies of the Majorana chain is indeed in the trivial phase. 

Moreover, the argument also implies that the gapless Majorana fermions (in the (1+1)D bulk) at the free-fermion critical point ($u=v$) can be gapped out by turning on the interaction $g$ (see \figref{fig:chain}), leading to another example of SMG \footnote{The anomalous chiral $\dsZ_2$ symmetry should be included to protect the criticality.}. The bulk SMG of the (1+1)D Majorana chain (at the critical point) is dynamically equivalent to the SMG for the boundary mode of a (2+1)D fermionic topological superconductor (TSC) made of eight copies of the $p_x\pm\ii p_y$ superconductor \cite{Volovik1988,Read2000Paired,Fu2008Superconducting}. In general, each interaction-reduced classification of the fermion SPT state implies the existence of SMG mechanisms for the gapless fermions both on the \emph{boundary} and at the \emph{bulk} critical point between the SPT and the trivial state, where the same local interaction that gaps out the boundary mode can be used to gap out the bulk critical point as well. Using the connection of SMG phenomena between the boundary and the bulk, \refcite{You:2014ho} obtains a series of interaction-reduced classification of fermionic SPT states in all spacetime dimensions systematically, hence extending the phenomenon of SMG to higher dimensions.


\subsection{(1+1)D SMG: 3-4-5-0 Chiral Fermion Model}

The 3-4-5-0 $\U(1)$ chiral fermion model by Wang-Wen \cite{Wang2013Non-Perturbative,Wang2019Solution} 
provides an instructive example of SMG in (1+1)D, 
which involves the cancellation of a perturbative anomaly and is not related to interaction-reduced fermionic SPT classifications. The (1+1)D chiral fermion theory is described by the Lagrangian
\eq{
\scL=\sum_{a=1}^{4}\psi_a^{\dagger}\ii (\partial_t- v_a\partial_x)\psi_a,
}
where $\psi_a$ are complex fermion fields (for $a=1,2,3,4$) with different velocities $v=(1,1,-1,-1)$. The model can be assigned a $\U(1)$ symmetry, under which the fermions transform as $\U(1):\psi_a\to \e^{\ii \theta q_a}\psi_a$ with $q=(3,4,5,0)$. The two left-moving Weyl fermions $\psi_1$ and $\psi_2$ are assigned with charge 3 and 4, and the two right-moving Weyl fermions $\psi_3$ and $\psi_4$ are assigned with charge 5 and 0, hence the name of the 3-4-5-0 model.

The seemingly peculiar charge assignment of the chiral fermions is designed to cancel the $\U(1)$ 't Hooft anomaly. The (1+1)D fermion with $\U(1)$ internal symmetry (or $\Spin^c \equiv (\Spin \times \U(1))/\Z_2$ spacetime-internal symmetry \cite{Kapustin2015Fermionic,Prakash2021Unwinding}, or the A symmetry class \cite{Altland1997Nonstandard,Zirnbauer2010Symmetry}) has a $\dsZ$ perturbative anomaly, whose anomaly index is given by $\sum_{a}v_a q_a^2$, which vanishes for the charge assignment of the 3-4-5-0 model. The (1+1)D chiral fermions can be also viewed as the chiral edge modes of a (2+1)D integer quantum Hall state, also known as the fermionic SPT state in symmetry class A. The integer quantum Hall states are $\dsZ$ classified by the quantized Hall conductance, and the classification does not further reduce under interaction. The bulk Hall conductance (in unit of $e^2/h$) matches the boundary anomaly index.

The fact that the $\U(1)$ 't Hooft anomaly vanishes for the 3-4-5-0 model indicates that it should be possible to trivially gap out all fermions without breaking the $\U(1)$ symmetry. However, the $\U(1)$ symmetry is restrictive enough to prevent the gapping to happen on the free-fermion level, because any fermion bilinear term that produces a gap must take the form of $\psi_a^\dagger \psi_b$ or $\psi_a \psi_b$ (with $a\in\{1,2\}$ and $b\in\{3,4\}$) that mixes the left- and right-moving fermions. Since the four flavors of fermions all carry distinct $\U(1)$ charges that do not add or subtract to zero, any flavor mixing fermion bilinear term will necessarily break the $\U(1)$ symmetry, which makes it impossible to symmetrically gap out these chiral fermions on the free-fermion level.

Nevertheless, the gapping can be achieved with fermion interaction, hence an example of SMG. The interaction to achieve the SMG would involve at least six-fermion terms, which can be derived using the null-vector condition 
\cite{Haldane1995Stability, KapustinSaulina1008.0654KS, Wang2015Boundary, Levin1301.7355, LanWangWen1408.6514LWW} for quantum Hall edge states. By bosonization $\psi_a\sim \e^{\ii\varphi_a}$, the (1+1)D chiral fermion system can be effectively described by a Luttinger liquid theory
\eq{\label{eq:LL}\scL=\frac{1}{4\pi}(K_{ab}\partial_t\varphi_a\partial_x\varphi_b-V_{ab}\partial_x\varphi_a\partial_x\varphi_b),}
where $K=\text{diag}(1,1,-1,-1)$ and $V=\text{diag}(1,1,1,1)$ are diagonal matrices. In the bosonization language, backscattering fermion interactions can be introduced as
\eq{\label{eq:LLint}\scL_\text{int}=\sum_\alpha g_{\alpha}\cos(l_{\alpha,a}\varphi_a),}
each interaction labeled by a charge vector $l_\alpha$. It is possible to find two charge vectors $l_1=(1,-2,1,2)$ and $l_2=(2,1,-2,1)$ that satisfy the null-vector condition $l_\alpha^\intercal K^{-1}l_\beta=0$ (for $\alpha,\beta=1,2$). The null-vector condition ensures that the vertex operators $O_\alpha=\e^{\ii\l_\alpha^\intercal \varphi}$ are both self-boson and mutual-boson, which can be simultaneously condensed at large coupling $g_\alpha$. Once these vertex operators condense $\langle O_\alpha\rangle\neq0$, all fermion excitations will be gapped, since none of the fermion operator braid trivially with any of the vertex operator (as seen from $l_\alpha^\intercal K^{-1}\neq 0$). Furthermore, the interaction term $\scL_\text{int}$, as well as the condensate $\langle O_\alpha\rangle$, preserves the $\U(1)$ symmetry since $l_{\alpha}^\intercal q=0$ for $\alpha=1,2$. Therefore, $\scL_\text{int}$ provides a symmetric way to gap out all chiral fermions in the 3-4-5-0 model, realizing the SMG. In terms of the fermion field, the interaction can be translated from \eqnref{eq:LLint} to
\eqs{\label{eq:LLint_fermion}\scL_\text{int}&=g_1(\psi_1\psi_2^\dagger\partial_x\psi_2^\dagger\psi_3\psi_4\partial_x\psi_4+\text{h.c.})\\&+g_2(\psi_1\partial_x\psi_1\psi_2\psi_3^\dagger\partial_x\psi_3^\dagger\psi_3\psi_4+\text{h.c.}),}
where the $\partial_x$ operator (that generates infinitesimal translation) is inserted as a point-splitting regularization to avoid identical fermion operators appearing at the same spatial position.

The multi-fermion interaction in \eqnref{eq:LLint_fermion} can be mediated by two independent Yukawa fields $\phi_1,\phi_2$ via
\eqs{\label{eq:LLint_YH}\scL_\text{YH}=-&(\phi_1^2\psi_1\psi_3+\phi_1^\dagger\psi_2^\dagger \psi_4+\text{h.c.})+\frac{1}{\tilde{g}_1}\phi_1^\dagger\phi_1\\
-&(\phi_2^2\psi_2\psi_4+\phi_2^\dagger\psi_1\psi_3^\dagger+\text{h.c.})+\frac{1}{\tilde{g}_2}\phi_2^\dagger\phi_2.}
Integrating out the Yukawa fields \footnote{Whenever two fermion operators overlap at the same spacetime point, the replacement $\psi_a\psi_a\to\psi_a\partial_x\psi_a$ is assumed to split the fermion operators.} will generate the interaction in \eqnref{eq:LLint_fermion} to the leading order of $g_\alpha\sim \tilde{g}_\alpha^2$ (plus additional density-density interactions like $\psi_1^\dagger\psi_1\psi_3^\dagger\psi_3$ or $\psi_2^\dagger\psi_2\psi_4^\dagger\psi_4$ whose effect is only to renormalize the $V$ matrix in the Luttinger liquid theory and can be safely ignored).
The Yukawa fields $\phi_1$ and $\phi_2$ carry the $\U(1)$ charges $-4$ and $-2$ respectively. Directly condensing the Yukawa fields would provide Dirac/Majorana masses to all chiral fermions $\psi_a$ at the price of breaking the $\U(1)$ symmetry. Nevertheless, the SMG mechanism suggests an alternative scenario that these Yukawa fields are fluctuating in the disordered (a.k.a.~strong-coupling symmetric) phase, such that the $\U(1)$ symmetry remains unbroken but the chiral fermions could still acquire a spectral gap via the Yukawa interaction in suitable parameter regimes. Similar idea was numerically explored in \refcite{Chen2013Model345} without success, due to the incorrect design of the fermion interaction (which does not satisfy the null-vector condition). The correct design of fermion interaction in \eqnref{eq:LLint_fermion} or \eqnref{eq:LLint_YH} includes only a restricted subset of symmetry-allowed interactions \cite{Wang2013Non-Perturbative}, which are helpful for the SMG. More recent numerical study in \refcite{Zeng2022Symmetric} has confirmed that the correct interaction indeed leads to the SMG phase. 

The SMG transition happens when the interaction is beyond a finite critical strength, because the interaction is perturbatively irrelevant at the free-fermion fixed point, due to its high-order nature. 
However, strong enough interaction could lead to non-perturbative effects. Increasing the interaction strength generally tunes the Luttinger parameter (by renormalizing the $V$ matrix) and alters operator scaling dimensions in the Luttinger liquid theory \cite{Zeng2022Symmetric}. When the scaling dimension of the interaction term itself is tuned to marginal, the SMG transition will be triggered, which drives the system from the gapless phase to the featureless gapped phase \cite{Tong2021Comments,Zeng2022Symmetric}. In this case, the SMG transition will belong to the Berezinskii-Kosterlitz-Thouless (BKT) universality class.

\subsubsection{Proof on the equivalence between the anomaly-free and SMG gapping conditions}

Although the earlier discussions focus on the SMG of (1+1)D 3-4-5-0 $\U(1)$ chiral fermion model, 
there is no obstacle to generalize to show any (1+1)D anomaly-free chiral fermion model 
with multiple $\U(1)$ symmetries can allow SMG, following \cite{Wang2013Non-Perturbative}.
\refcite{Wang2013Non-Perturbative} proves that
the \emph{anomaly-free condition} of the (1+1)D multiple $\U(1)$ chiral fermion theory
is equivalent to the \emph{SMG gapping condition} of the same theory.
The proof can be achieved due to the exact bosonization-fermionization techniques in (1+1)D. 
Here we recall the proof to complete the discussion of (1+1)D SMG.

A generic (1+1)D anomaly-free multiple-$\U(1)$ chiral fermion theory have equal numbers of left and right moving 
Weyl fermions, $N_L=N_R=N \in \Z^+$, such that the total Weyl fermion number is $2N=N_L+N_R= 2 \Z$, an even integer.

\begin{enumerate}[leftmargin=2.mm]

\item
The \emph{SMG gapping condition} requires to add $N$ independent compatible gapping terms
\cite{Haldane1995Stability, KapustinSaulina1008.0654KS, Wang2015Boundary, Levin1301.7355}
to preserve internal chiral $\U(1)$ symmetries. To prove the SMG gapping holds,
we bosonize the fermionic theory
\bea \label{eq:multi-fermion}
\scL=\sum_{a=1}^{N_L}\psi_a^{\dagger}\ii (\partial_t- \partial_x)\psi_a
+\sum_{a=1}^{N_R}\psi_a^{\dagger}\ii (\partial_t+ \partial_x)\psi_a
\eea
to a multiplet chiral boson theory
\bea \label{eq:multi-boson}
\scL=\frac{1}{4\pi}(K_{ab}\partial_t\varphi_a\partial_x\varphi_b-V_{ab}\partial_x\varphi_a\partial_x\varphi_b),
\eea
where $K=K_f \equiv\bigl( {\begin{smallmatrix}
1 &0 \\
0 & -1
\end{smallmatrix}}  \bigl) \oplus \bigl( {\begin{smallmatrix}
1 &0 \\
0 & -1
\end{smallmatrix}}  \bigl) \oplus \dots$ and the appropriate rescaled $V=\mathbb{I}_{2N \times 2N}$ are diagonal rank-$2N$ matrices.
The $K_f$ is the unimodular symmetric bilinear canonical form for the fermionic system (with $|\det(K)|=1$).
The advantage of \eqnref{eq:multi-boson} is that choosing  
$K=K_b \equiv\bigl( {\begin{smallmatrix}
0 &1  \\
1 & 0 
\end{smallmatrix}}  \bigl) \oplus \bigl( {\begin{smallmatrix}
0 &1  \\
1 & 0 
\end{smallmatrix}}  \bigl) \oplus \dots$ works
also for the unimodular symmetric bilinear canonical form for the bosonic system (with $|\det(K)|=1$).
The SMG gapping condition requires: \\
$\bullet$ To find a set of $N$ linear-independent of integer-valued $2N$-component $l$ vectors
such that
\bea \label{eq:1+1d-gapping-free}
l_\alpha^\intercal K^{-1}l_\beta=\sum_{a,b}l_{\alpha,a} (K^{-1})_{ab}l_{\beta,b} =0,
\eea
for $\alpha,\beta \in \{1,2,\dots,N\}$ and each $l_\alpha$ vector contains $2N$ components ($l_{\alpha,a}$ with $a=1,\dots, 2N$).
Gapping \eqnref{eq:multi-boson} requires to add the sine-Gordon deformation $\scL_\text{int}=\sum_{\alpha=1}^N g_{\alpha}\cos(l_{\alpha,a}\varphi_a)$,
which can be fermionized to multi-fermion interactions.\\
$\bullet$ The massless Weyl fermion theory has at most a internal $\U(N_f) \times \U(N_R)$ symmetry, which contains
at most a chiral $\U(1)^{2N}$ symmetry.
But for SMG, one can preserve at most $N$-linear independent chiral $\U(1)^N$ symmetries, labeled by a set of charge vectors,
$q_\alpha$ with $\alpha \in \{1,2,\dots,N\}$, such that the fermions transform as $\psi_{a} \to \psi_{a} \e^{\ii {q_{\alpha,a} \theta}}$,
and bosonized fields $\varphi_{a} \to \varphi_{a} +  {q_{\alpha,a} \theta}$ with $a=1,\dots, 2N$ and $\theta \in [0, 2 \pi)$. The symmetric sine-Gordon interactions demand 
\bea \label{eq:1+1d-symm}
l_\alpha^\intercal q_{\beta} = \sum_a l_{\alpha,a} q_{\beta,a} =0
\eea
for any $\alpha,\beta \in \{1,2,\dots,N\}$.

\item 
Its \emph{anomaly-free condition}, on the other hand,  requires:\\
$\bullet$ Gravitational anomaly free (two-point one-loop Feynman diagram of grav$^2$ vertices vanish): 
The left and right chiral central charges ${\rm c}_L = {\rm c}_R$, which means $N_L=N_R$.\\
$\bullet$ Gauge anomaly free (two-point one-loop Feynman diagram of $\U(1)^2$ vertices vanish): 
For each $\U(1)$ symmetry, with left-handed and right-handed Weyl fermion charge vector $q_L$ and $q_R$ respectively, 
the anomaly-free requires the square sum of each component $\sum q_L^2 - q_R^2 =0$.  In terms of the symmetric bilinear form $K$ for both bosonic ($K_b$) and fermionic ($K_f$) 
systems, the anomaly-free condition demands that
\bea \label{eq:1+1d-anomaly-free}
q_\alpha^\intercal K^{-1} q_\beta=\sum_{a,b}q_{\alpha,a} (K^{-1})_{ab}q_{\beta,b} =0,
\eea
again for $\alpha,\beta \in \{1,2,\dots,N\}$ and $a,b \in \{1,2,\dots,2N\}$.

$\bullet$ The above two anomaly-free conditions are perturbative local anomalies. 
The (1+1)D nonperturbative global anomalies are classified by cobordism groups
(denoted $\TP_3$ or $\Omega_3$ with the special orthogonal SO or Spin group and some internal $\U(1)$ symmetries,
such $\SO \times \U(1)$, $\Spin \times \U(1)$ and $\Spin^c$) 
which turn out to always vanish \cite{WanWang1812.11967}.

\item To explain that the SMG gapping condition 
holds implies the anomaly-free condition also holds,
\refcite{Wang2013Non-Perturbative} shows that given the set of
$l_\alpha$ satisfying \eqref{eq:1+1d-gapping-free}, 
one can find the set of $q_{\alpha}$ simultaneously satisfying \eqref{eq:1+1d-symm} and \eqref{eq:1+1d-anomaly-free}.
This is true because that given $l_\alpha$, we can choose $q_{\alpha}=K l_\alpha =K^{-1} l_\alpha$ 
thanks to the property $K=K^{-1}$.

\item To explain that the anomaly-free condition 
holds implies the SMG gapping condition also holds,
\refcite{Wang2013Non-Perturbative} shows that given the set of
$q_\alpha$ satisfying \eqref{eq:1+1d-anomaly-free}, 
one can find the set of $l_{\alpha}$ simultaneously satisfying \eqref{eq:1+1d-symm} and \eqref{eq:1+1d-gapping-free}.
This is true because that given $q_\alpha$, we can choose $l_{\alpha}=K q_\alpha =K^{-1} q_\alpha$ 
thanks to $K=K^{-1}$.

\item The above two remarks prove that the \emph{if and only if} (sufficient and necessary) conditions to the equivalence of 
the anomaly-free condition and the SMG gapping condition. Once the set of $q_\alpha$ and $l_{\alpha}$ are found,
they form $2N$ linear-independent integer-valued vectors spanning completely 
the $2N$-dimensional vector space (known as the Narain lattice \cite{NarainSarmadiWittenNPB1986am}).

\item What remains to be explained is why the SMG gapping condition defines a gapped boundary 
without any topological boundary degeneracy \cite{Wang2015Boundary}. The idea is viewing 
the (1+1)D theory \eqref{eq:multi-boson} as the boundary theory of (2+1)D invertible TQFT
with a Chern-Simons action $S_{\text{bulk}}=\frac{K_{ab}}{4\pi}\int_{\mathcal{M}^3}  A_a \wedge \dd A_b$ on a 3-manifold $\mathcal{M}^3$.
The $A_a$ is a multiplet 1-form gauge field.
Hereafter all repeated indices are summed over.
A stable boundary condition 
requires the variation of $S_{\text{bulk}}$ on the boundary 2-manifold 
$(\partial \mathcal{M})^2$ vanished \cite{KapustinSaulina1008.0654KS}
under the boundary 1-form gauge field $A_{\partial} \to A_{\partial} + \delta A_{\partial}$ variation:
$
\delta_{\text{bdry}} (S_{\text{bulk}})
= 
\frac{K_{ab}}{4\pi}\int_{(\partial \mathcal{M})^2} 
A_{\partial, a} \wedge \delta A_{\partial, b}.$
The differential $\delta$ of this variation is a symplectic form
$
\omega_{\text{Sp}} = \frac{K_{ab}}{4\pi}\int_{(\partial \mathcal{M})^2} 
\delta A_{\partial, a} \wedge \delta A_{\partial, b}
$
on the space of boundary gauge fields. 
Consistent stable boundary conditions on $(\partial \mathcal{M})^2$
defines a Lagrangian submanifold with respect to the symplectic form $\omega_{\text{Sp}}$ in symplectic geometry.

$\bullet$ One consistent boundary condition sets one component of $A_{\partial}$ vanished,
such as
$
\big( K_{ab} A_{b, t}-V_{ab} A_{b, x}\big) \rvert_{\partial M}=0, 
$
which gives a gapless (1+1)D CFT \eqref{eq:multi-boson}.

$\bullet$ Another boundary condition sets the gauge degrees of freedom
$
l_{\alpha, a}  A_a \rvert_{\partial \mathcal{M}} =0
$
vanish \cite{Wang2018Tunneling}.
The boson modes $\varphi_{a}$, originally related by the gauge transformation $A_a \to A_a + \dd \lambda_a$ 
and $\varphi_{a} \to \varphi_{a}- \lambda_a$,
now may condense on the boundary with nonzero vacuum expectation values
$\langle \exp\big(\ii  (l_{\alpha,a} \varphi_{a})\big) \rangle \rvert_{\partial \mathcal{M}} \neq 0$, 
more precisely,  indeed 
$
\langle \exp\big(\ii  (\frac{l_{\alpha,a}}{|\gcd({l_{\alpha}}) |}  \varphi_{a})\big)  \rangle \rvert_{\partial \mathcal{M}} \neq 0,
$
where $\gcd({l_{\alpha}}) \equiv 
\gcd(l_{\alpha,1}, l_{\alpha,2}, \dots, l_{\alpha,2N})$ is the greatest common divisor (gcd) of the all components of ${l_{\alpha}}$.
This condensation of $\varphi$
can be triggered by the earlier sine-Gordon cosine term at a strong $g$ coupling.
The boundary vertex operator and bulk line operator are connected
$
\exp\big(\ii  (l_{\alpha,a} \varphi_{a}) \rvert_{\partial \mathcal{M}} +\ii \int l_{\alpha,a} A_{a} \rvert_{ \mathcal{M}}\big)
$.
The gapped bulk and gapped boundary demand that the partition function $Z$ evaluated on the 3-manifold $M$ with the 2-boundary
$\partial \mathcal{M}$ has a finite value 
(in fact $Z=1$ when the $Z$ corresponds to counting the dimension of the Hilbert space for the invertible TQFT).
This means that arbitrary link configuration of the bulk line operators should give a trivial braiding statistical phase
to $Z$, so there are no unwanted quantum fluctuations destabilizing the gapped vacuum 
--- namely, the mutual statistics 
$\exp(\ii 2 \pi l_\alpha^\intercal K^{-1}l_\beta) =1$
and the self statistics 
$\exp(\ii  \pi l_\alpha^\intercal K^{-1}l_\alpha) =1$
are trivial for all $\alpha,\beta \in \{1,2,\dots,N\}$.
Hence we derive the correspondence between the
$N$ independent compatible SMG gapping terms and
the $N$ null-braiding statistics $l_\alpha$ vectors
\cite{Haldane1995Stability, KapustinSaulina1008.0654KS, Wang2015Boundary, Levin1301.7355}.
This completes the proof \cite{Wang2013Non-Perturbative}.

\end{enumerate}

\subsection{(2+1)D SMG: Honeycomb Lattice Model}
\label{sec:honeycomb}

\begin{figure}[htbp]
\begin{center}
\includegraphics[width=0.92\columnwidth]{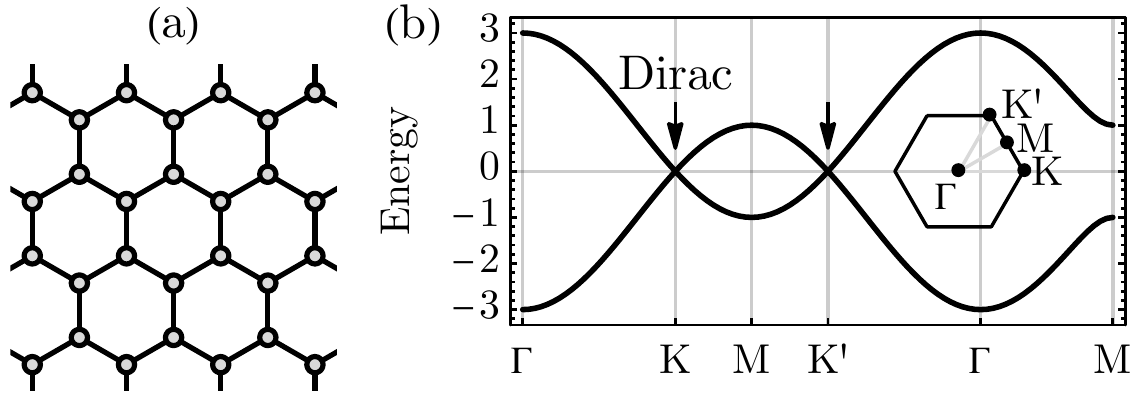}
\caption{(a) Honeycomb lattice. (b) Graphene band structure (inset defines high symmetry points in the Brillouin zone).}
\label{fig:honeycomb}
\end{center}
\end{figure}

The honeycomb lattice model is a simple lattice model for SMG in (2+1)D. The model is defined on a honeycomb lattice, with each lattice site $i$ hosting four complex fermion modes, denoted as $c_{ia}$ (for $a=1,2,3,4$). The model is described by the Hamiltonian
\eq{\label{eq:H_honeycomb}H=-\sum_{a=1}^{4}\sum_{\langle ij\rangle}c_{ia}^\dagger c_{ja}-g\sum_{i}c_{i1}c_{i2}c_{i3}c_{i4}+\text{h.c.},}
where $\langle ij\rangle$ stands for the bond between nearest neighboring sites $i$ and $j$ on the honeycomb lattice, see \figref{fig:honeycomb}(a). The interaction strength $g$ is the only tuning parameter of this model. 

When $g=0$, the free-fermion hopping on the honeycomb lattice gives rise to the graphene band structure \cite{Wallace:1947ei}, as shown in \figref{fig:honeycomb}(b), that produces $4\times2=8$ gapless Dirac fermions $\psi_{Qa}$ at low energy, where $4$ stands for the four internal flavors ($a=1,2,3,4$) and $2$ comes of the fermion doubling ($Q=K,K'$) in the Brillouin zone. They can be described by a low-energy effective field theory Lagrangian
\eq{\label{eq:L_honeycomb}\scL=\sum_{Q=K,K'}\sum_{a=1}^{4}\bar{\psi}_{Qa}\gamma^\mu\partial_\mu \psi_{Qa},}
where $Q=K,K'$ labels the valley (fermion doubling) freedom, $\gamma^\mu=(\sigma^2,\sigma^1,\sigma^3)$ and $\psi_{Qa}$ are complex Grassmann spinors in (2+1)D (with $\bar{\psi}_{Qa}=\psi_{Qa}^\dagger \gamma^0$). They can also be viewed as 16 gapless Majorana fermions equivalently, by decomposing each complex Grassmann field into two real Grassmann fields.

The key physical symmetries that protect these gapless fermions in the non-interacting limit are the lattice translation symmetry and the \emph{anti-unitary} sublattice symmetry  $\dsZ_2^S:c_{i}\to(-)^i c_{i}^\dagger$, $\ii\to-\ii$ (where $(-)^i$ stands for a sign factor that takes $\pm$ on A/B sublattice of the honeycomb lattice) \cite{Schnyder:2008os,Ryu:2010fe,Ludwig:2016pt}. These two symmetries can be combined to create an emergent anti-unitary symmetry at low-energy $\dsZ_4^{TF}:\psi_{Qa}\to\ii\gamma^0 \psi_{Qa}^\dagger, \ii\to-\ii$ \footnote{Because $\scT^2=-1$ on fermions, $\scT^2$ should correspond to the fermion parity operator which need a further square to become identity, therefore $\scT$ generates a four-fold cyclic group $\dsZ_4^{TF}$.}, whose generator $\scT=T_{\vect{R}}^{3/4}\scS$ consists of a sublattice symmetry $\dsZ_2^S$ generator $\scS$ followed by a $3/4$ fraction of the unit-cell translation $T_{\vect{R}}$. Although the 2D lattice translation symmetry $\dsZ^2$ (generated by $T_\vect{R}$ along two linearly independent Bravais lattice vectors $\vect{R}$) cannot be fractionalized on the lattice, yet for the low-energy effective theory \eqnref{eq:L_honeycomb}, {the translation symmetry acting on the low-energy fermions $\psi_Q$ becomes an emergent valley $\U(1)$ symmetry $T_{\vect{R}}:\psi_{K/K'}\to\e^{\pm\ii 2\pi/3}\psi_{K/K'}$} ($\pm$ signs are  associated with $K$ and $K'$ valleys respectively), which can be fractionalized to $T_{\vect{R}}^{3/4}:\psi_{K/K'}\to\pm\ii\psi_{K/K'}$. The anomalous nature of the combined symmetry $\dsZ_4^{TF}$ is manifested by the fact that $\dsZ_4^{TF}$ is only an emergent symmetry at low-energy and becomes ill-defined on the lattice level. 

The (2+1)D Majorana fermion with a $\dsZ_4^{TF}$ internal symmetry (or the $\text{Pin}^{+}$ spacetime-internal symmetry \cite{Kapustin2015Fermionic,Prakash2021Unwinding}, or the DIII symmetry class \cite{Altland1997Nonstandard,Zirnbauer2010Symmetry}) has a $\Omega_4^{\text{Pin}^{+}}(\mathsf{pt})=\dsZ_{16}$ non-perturbative global anomaly based on the cobordism, whose anomaly index corresponds to the number of gapless Majorana fermions. The honeycomb lattice model precisely has 16 Majorana fermions at low-energy, which is free of the $\dsZ_{16}$ non-perturbative global anomaly. However, the $\dsZ_4^{TF}$ symmetry is still restrictive enough to rule out all possible fermion bilinear gapping terms from appearing in \eqnref{eq:L_honeycomb}, which again calls for the SMG mechanism. 

The four-fermion interaction $g$ in \eqnref{eq:H_honeycomb} is one choice of the interaction that drives the desired SMG. On every site, this interaction is the same as the $\SU(4)$ symmetric interaction in \eqnref{eq:H_SU4} (up to unimportant density-density interactions, which do not affect the ground state but only to renormalize the gap size). The interaction $g$ explicitly drives a four-fermion condensation, also known as the charge-4e superconducting order  \cite{Kivelson:1990wy,Talukdar2007Quartet,Berg:2009uq,Radzihovsky:2009bk,Berg:2009nm,Moon:2012ng,Jiang:2016km}. In the large $g$ limit, the many-body ground state of the system is simply the product $\bigotimes_{i}\ket{\mathsf{4e}}_i$ of on-site ground states $\ket{\mathsf{4e}}_i=(\ket{0000}_i+\ket{1111}_i)/\sqrt{2}$, which is symmetric, non-degenerated and gapped, realizing the SMG phase. Therefore, by tuning the strength $g$, one expects to drive an SMG transition at some intermediate $g\sim 1$ (that is comparable with the bandwidth of the lattice fermion). Various numerical simulations of this model (and its variants) \cite{Slagle:2015lo,Ayyar2015Massive,Ayyar:2016fi,Catterall:2016sw,He:2016qy} have suggested the existence of such a direct and continuous SMG transition between the gapless and the gapped phases in (2+1)D. A field theory description of the SMG quantum critical point was proposed in \refcite{You2018Symmetric}, which will be further reviewed in \secref{sec:SU4_parton}.

The honeycomb lattice model also has an explicit $\SU(4)$ symmetry $c_{ia}\to U_{ab}c_{ib}$ that rotates fermions among the four on-site flavors. Although the $\SU(4)$ symmetry does not affect the anomaly analysis in any essential way, it could help to remove all the $\SU(4)$-breaking relevant perturbations (if there were any) at the critical point, which might help to promote a continuous SMG transition.    
Thus the $\SU(4)$ symmetry can also be included in the discussion, and the total internal symmetry becomes $\SU(4)\times_{\dsZ_2^F}\dsZ_4^{TF}$ (where $\SU(4)$ and $\dsZ_4^{TF}$ share the fermion parity $\dsZ_2^F$ subgroup). 

The decomposition of the on-site $\SU(4)$-symmetric interaction into a product of fermion bilinear operators $\Phi_i^{\alpha}\Phi_i^{\alpha}$ as in \eqnref{eq:H_SU4} indicates that one can again introduce a Yukawa field $\phi_\alpha$ to mediate the fermion interaction, \cite{You2018Symmetric}
\eq{\label{eq:honeycomb_YH}\scL_\text{YH}=-\phi_\alpha(\psi_{Ka}^\intercal\ii\gamma^0\tilde{\Gamma}^{\alpha}_{ab}\psi_{K'b}+\text{h.c.})+\frac{1}{2\tilde{g}}\phi^\ast_\alpha\phi_\alpha,}
where $\tilde{\Gamma}^{\alpha}=(\sigma^{12},\sigma^{20},\sigma^{32},\ii\sigma^{21},\ii\sigma^{02},\ii\sigma^{23})$ are the flavor-space matrices for the flavor-sextet pairing that transform as representation $\mathbf{6}$ in $\SU(4)$. Importantly, the Yukawa field must transform as $\dsZ_4^{TF}:\phi_\alpha\to-\phi_\alpha$ to keep $\scL_\text{YH}$ invariant under $\dsZ_4^{TF}$, meaning that directly condensing the Yukawa field $\langle\phi_\alpha\rangle\neq0$ will necessarily break the protecting symmetry $\dsZ_4^{TF}$ (as well as breaking the standing-by $\SU(4)$ symmetry). The SMG interaction thus provides a mechanism to allow the Yukawa field to condense locally without establishing long-range order, so as to maintain the fermion gap without breaking the symmetry.

\subsection{(3+1)D SMG: Chiral Fermion Model}

Lattice regularization of chiral fermions in (3+1)D has always been an important motivation to study SMG in lattice gauge theories. A rich class of SMG transitions has been proposed and analyzed in different 4d chiral fermion models recently \cite{Razamat2021Gapped,Tong2021Comments}. Here we will review one simplest example from \cite{Tong2021Comments} to illustrate the essential features of these models. The example considers a collection of right-handed massless Weyl fermions $\psi$ transforming under an internal $\SU(N)$ symmetry in the ${\ybox(2)}\oplus(N+4)\bar{\ybox(1)}$ representation, described by the Lagrangian
\eq{\scL=\psi^\dagger\ii\bar{\sigma}^\mu\partial_\mu\psi,}
where $\psi$ is the collection of complex Grassmann spinor fields. More explicitly, the fermion field can be split into $\psi=\psi_\lambda\oplus\psi_\psi$, such that $\psi_\lambda$ denotes the single $\ybox(2)$ Weyl fermion and $\psi_\psi$ denotes the $(N+4)$ multiples of $\bar{\ybox(1)}$ Weyl fermions. The goal is to find a path of gapping these fermions without breaking the $\SU(N)$ symmetry.

First of all, the symmetric gapping is possible because all fermions together cancel the $\SU(N)$ 't Hooft anomaly, as can be verified by the following anomaly index calculation
\eqs{&A_{\SU(N)}({\ybox(2)})+(N+4)A_{\SU(N)}(\bar{\ybox(1)})\\
=&(N+4)+(N+4)\times(-1)=0.}
Secondly, the $\SU(N)$ symmetry is restrictive enough to forbid all possible fermion bilinear masses. For chiral fermions, the only available fermion bilinear mass is the Majorana mass that pairs up fermions like $(\psi^\intercal \ii\sigma^2\psi+\text{h.c.})$. Since the Majorana mass is already antisymmetric in the spinor subspace, the flavor subspace must be symmetric. However, the fermion representation of the $\SU(N)$ symmetry guarantees that its symmetric product does not contain a trivial representation $\mathbf{1}$, as
$({\ybox(2)}\oplus(N+4)\bar{\ybox(1)})\times_\scS({\ybox(2)}\oplus(N+4)\bar{\ybox(1)})
\to\,{\ybox(4)}\oplus(N+4){\ybox(1)}\oplus(N+4){\bar{\ybox(1)}\ybox(2)}\oplus(N+4)^2{\overline{\ybox(2)}}\,$.
So any bilinear mass term will transform non-trivially under $\SU(N)$ and cannot be condensed in the presence of the $\SU(N)$ symmetry. Therefore, one must rely on fermion interactions to achieve symmetric gapping, i.e.~the SMG mechanism.

To design the appropriate interaction to realize the SMG transition, one might search for candidate four-fermion interactions by looking for trivial representations in $({\ybox(2)}\oplus(N+4)\bar{\ybox(1)})^{\times 4}$. However, $({\ybox(2)}\oplus(N+4)\bar{\ybox(1)})^{\times 4}\not\to\mathbf{1}$ in general (unless for $N=2,4$), meaning that one might need to look into higher-order fermion interactions, which are even more irrelevant and less interesting to explore. Nevertheless, the three-fermion combination $({\ybox(2)}\oplus(N+4)\bar{\ybox(1)})^{\times 3}$ contains trivial representations of the form $(\psi_\lambda^\intercal\ii\sigma^2\psi_\psi)\psi_\psi$, but this term is fermionic and cannot appear directly in the Lagrangian (otherwise the fermion parity symmetry would be broken). An interesting idea from \cite{Razamat2021Gapped,Tong2021Comments} is to bring down $(N+4)(N+3)/2$ additional fermions $\psi_\chi$ from the high-energy spectrum which transforms trivially under the $\SU(N)$ symmetry, such that a four-fermion interaction 
\eq{\label{eq:L_int_CF}\scL_\text{int}=g((\psi_\lambda^\intercal\ii\sigma^2\psi_\psi)(\psi_\chi^\intercal\ii\sigma^2\psi_\psi)+\text{h.c.})}
can be constructed. In this interaction, the fermion flavors are contracted in the way such that $\psi_\lambda^\intercal\ii\sigma^2\psi_\psi$ and $\psi_\chi^\intercal\ii\sigma^2\psi_\psi$ transform as $\ybox(1)$ and $\bar{\ybox(1)}$ under $\SU(N)$ respectively. 

\begin{table}[htp]
\caption{Representations of the fermion $\psi$ and Yukawa $\phi$ fields in the chiral fermion model under $\SU(N)\times \SU(N+4)$.}
\begin{center}
\setlength{\extrarowheight}{2pt}
\begin{tabular}{ccccc}
& & $\SU(N)$ & $\SU(N+4)$ & \\
\hline
\multirow{3}{*}{$\psi\Bigg\{$} & $\psi_\lambda$ & $\ybox(2)$ & $\mathbf{1}$ & \multirow{3}{*}{physical fermion}\\
& $\psi_\psi$ & $\bar{\ybox(1)}$ & $\bar{\ybox(1)}$ & \\
& $\psi_\chi$ & $\mathbf{1}$ & $\ybox(1,1)$ & \\[2pt]
\hline
$\phi$ & & $\bar{\ybox(1)}$ & $\ybox(1)$ & Yukawa boson
\end{tabular}
\end{center}
\label{tab:repCF}
\end{table}

It is instructive to note that the chiral fermion model has an additional internal symmetry $\SU(N+4)$, under which $\psi_\lambda$, $\psi_\psi$ and $\psi_\chi$ transform as $\mathbf{1}$, $\bar{\ybox(1)}$ and $\ybox(1,1)$ respectively, as summarized in \tabref{tab:repCF}. The representations are so assigned that the fermions are also free of the $\SU(N+4)$ anomaly, as
\eqs{&N A_{\SU(N+4)}(\bar{\ybox(1)})+A_{\SU(N+4)}({\ybox(1,1)})\\
=&N\times(-1)+(N+4-4)=0.}
Moreover, there is no mixed anomaly between $\SU(N)$ and $\SU(N+4)$. The interaction $\scL_\text{int}$ in \eqnref{eq:L_int_CF} also respects the larger $\SU(N)\times\SU(N+4)$ symmetry. \refcite{Tong2021Comments} suggests that $\scL_\text{int}$ is a plausible SMG interaction that drives all fermions $\psi=\psi_\lambda\oplus\psi_\psi\oplus\psi_\chi$ to the gapped phase without breaking the $\SU(N)\times\SU(N+4)$ symmetry. The mechanism will be further reviewed in \secref{sec:CF_SMG_theory}.

The interaction $\scL_\text{int}$ in \eqnref{eq:L_int_CF} can be decomposed into the following Yukawa couplings
\eqs{\label{eq:CF_YH}\scL_\text{YH}=&-(\phi_1^\dagger (\psi_\chi^\intercal\ii\sigma^2\psi_\psi)+\phi_2(\psi_\lambda^\intercal\ii\sigma^2\psi_\psi)+\text{h.c.}) \\
&+\frac{1}{\tilde{g}}(\phi_1^\dagger\phi_2+\text{h.c.}),}
where the Yukawa fields $\phi_1$ and $\phi_2$ are both in the $(\bar{\ybox(1)},\ybox(1))$ representation of $\SU(N)\times\SU(N+4)$. The above Yukawa decomposition again provides an alternative way to understand the SMG transition by driving the Yukawa field into a strong-coupling symmetric phase.

\subsection{SMG in General Dimensions}
\label{sec:generalD}

For SMG to happen in a fermion system with a choice of internal symmetry $G$ in {$d$-dimensional} spacetime, the following two \emph{necessary} conditions must be satisfied:
\begin{itemize}
\setlength\itemsep{0pt}
\item The anomaly index $\nu\in \text{TP}_{d+1}(G)$ of the system must vanish ($\nu=0$), where $\text{TP}_{d+1}(G)$ denotes the classification of invertible topological phases (with low-energy invertible topological field theories)
in $(d+1)$-dimensional spacetime \cite{Freed2016Reflection}. (We may also denote $d+1 \equiv D$ as the bulk dimension.)

\item The single fermion must be in a representation $\rep_\psi^G$ of $G$, such that the \emph{antisymmetric} product (denoted by $\times_\scA$) representation $\rep_\psi^G\times_\scA\rep_\psi^G$ does not contain the trivial representation $\mathbf{1}^G$ in its direct sum decomposition.
\end{itemize}
If the first condition is violated ($\nu\neq 0$), trivially and symmetrically gapping out the fermions is impossible with or without interaction, due to the anomaly obstruction. However, it may be possible to symmetrically gap out the fermions with a symmetric anomalous topological order described by its underlying low-energy topological quantum field theory (TQFT), see Sec.~\ref{sec:Symmetry-Extension}. If the second condition is violated, a trivial and symmetric gap can already be achieved at the free-fermion level by a fermion bilinear condensation (through the trivial channel given by $\rep_\psi^G\times_\scA\rep_\psi^G\to\mathbf{1}^G$), such that fermion interaction is not necessary. When both conditions are satisfied, an interacting (strong-coupling) mechanism is needed to generate the fermion mass without breaking the $G$ symmetry, i.e.~the SMG mechanism. 

These conditions are also sufficient for the existence of an interacting SMG phase (i.e.~the featureless gapped phase). However, they do not necessarily imply a direct and continuous SMG transition of fermions from the gapless phase to the gapped phase. A general design of the four fermion interaction that leads to a single continuous transition between the gapless fermions and the SMG remains a challenging problem. Much of the progress relies on numerical simulations, as to be reviewed in the next section \secref{sec:numerics}.

\section{Numerical Investigations}
\label{sec:numerics}

\subsection{Existence of SMG Phases}
\label{sec:phase}

Because the SMG is generally a non-perturbative interaction effect (with respect to the gapless free-fermion fixed point), numerical simulations play important roles in the study of SMG. Given an anomaly-free gapless fermion system with a proposed gapping interaction, the numerical study has two major goals:
\begin{itemize}
\setlength\itemsep{0pt}
\item Establish the existence of the SMG phase in the strongly interacting limit.
\item Investigate the nature of the SMG transition at the critical interaction strength.
\end{itemize}

The \emph{SMG phase} refers to the strongly interacting trivial gapped state of fermions without symmetry breaking. It was also called the paramagnetic strong-coupling (PMS) phase \cite{Ayyar2015Massive,Butt2018SO4-invariant}, the strongly coupled symmetric gapped (SCSG) phase \cite{You:2014ow,You:2015lj}, or the featureless gapped phase \cite{You2018Symmetric,You2018From} in the literature. Correspondingly, the gapless Dirac/Weyl/Majorana fermion phase in the weak interaction regime is sometimes also called the paramagnetic weak-coupling (PMW) phase \cite{Ayyar2015Massive,Butt2018SO4-invariant}, or the semi-metal (SM) phase \cite{You2018Symmetric,You2018From}. The phase transition separating the SMG phase and the free-fermion phase is called the \emph{SMG transition}.

\begin{table}[htp]
\caption{Summary of numerical studies of SMG phases. (Dim.~- spacetime dimension, Sym.~- internal symmetry in terms of Lie algebra).}
\begin{center}
\begin{tabular}{lllp{40pt}p{40pt}}
Dim. & Sym. & Model & Method & Reference\\
\hline
(1+1)D & $\u(1)$ & $\psi^6$ & DMRG & \cite{Zeng2022Symmetric}\\
& $\su(2)$ & YH & disorder average& \cite{DeMarco2017A-Novel}\\
& $\so(4)$ & $\psi^4$ & QMC & \cite{Slagle:2015lo}\\
& $\su(4)$ & $\psi^4$ & QMC & \cite{Ayyar2017Generating}\\
& $\so(7)$ & YH & RHMC & \cite{Catterall2021Chiral}\\
\hline
(2+1)D & $\so(4)$ & $\psi^4$ & RHMC &  \cite{Catterall:2016sw} \\
& $\so(5)$ & $\psi^4$ & QMC & \cite{Slagle:2015lo}\\
& $\su(4)$ & $\psi^4$ & FBMC & \cite{Ayyar2015Massive,Ayyar:2016fi}\\
& & $\psi^4$ & QMC & \cite{He:2016qy}\\
\hline
(3+1)D & $\so(4)$ & YH & RHMC & \cite{Butt2018SO4-invariant,Catterall2020Exotic}\\
& & QCD & RHMC & \cite{Butt2021Symmetric}\\
& $\su(4)$ & $\psi^4$ & RHMC & \cite{Catterall:2016nh,Schaich2018Phases}\\
& & $\psi^4$ & FBMC & \cite{Ayyar:2016tg,Ayyar:2016ph}\\

\end{tabular}
\end{center}
\label{tab:numerics}
\end{table}

Numerical study of the SMG phases and phase transitions has been performed in various spacetime dimensions with different symmetries, as summarized in \tabref{tab:numerics}. The model Hamiltonian $H=H_0+H_\text{int}$ takes the following general form
\eqs{H_0=&-\sum_{i,j}\sum_{a}t_{ij}\psi_{ia}\psi_{ja}\\
H_\text{int}=&g\sum_{i}\sum_{a,b,c,d,\cdots}V_{abcd\cdots}\psi_{ia}\psi_{ib}\psi_{ic}\psi_{id}\cdots,\\
\text{or }H_\text{int}=&\sum_{i}\sum_{a,b,\alpha}\phi_{i \alpha}\psi_{ia}\Gamma_{ab}^{\alpha}\psi_{ib}+H[\phi_{i\alpha}],}
where $\psi_{ia}$ denotes the fermion of the flavor $a$ on the site $i$ (or in the unit cell $i$). The free-fermion Hamiltonian $H_0$ generates gapless fermions at low-energy, and the interaction Hamiltonian $H_\text{int}$ applies properly designed interactions to drive the SMG. The interaction strength will be generally denoted as $g$ in the following discussion. Different models are mainly distinct in the following aspects:
\begin{itemize}
\setlength\itemsep{0pt}
\item The low-energy fermions can be realized in the lattice model $H_0$ either as the gapless boundary modes of a fermionic SPT state in one higher spacetime dimension \cite{Zeng2022Symmetric}, or as the gapless bulk mode of a semi-metal state in the designated spacetime dimension (such as the honeycomb lattice fermion in (2+1)D \cite{Slagle:2015lo,He:2016qy} and the stagger fermion in general dimensions \cite{Butt2018SO4-invariant,Catterall2020Exotic}). 

\item The interaction $H_\text{int}$ can either be explicit given by multi-fermion local interaction terms (denoted as $\psi^4$ for four-fermion interactions or $\psi^6$ for six-fermion interactions in \tabref{tab:numerics}), or mediated by intermediate Yukawa-Higgs fields (denoted as YH) or non-Abelian gauge fields (denoted as QCD). 

\item The $\so(4)$ (or more precisely $\Spin(3)\times_{\dsZ_2^F}\Spin(4)$), $\so(5)$ (or $\U(1)\times_{\dsZ_2^F}\Spin(5)$), $\su(4)$ (or $\SU(4)\cong\Spin(6)$) interactions can all be viewed as lower-symmetry descendants of the $\so(7)$ (or $\Spin(7)$) Fidkowski-Kitaev interaction, whose relations are discussed in Appendix B of \refcite{You:2015lj}.

\item In even spacetime dimensions, the interaction can be restricted to part of the fermions (forming the mirror sector) to study the SMG in chiral fermion systems \cite{Zeng2022Symmetric,Catterall2021Chiral}.
\end{itemize}

Many different numerical methods have been used to study these models, which include: the density matrix renormalization group (DMRG) \cite{Zeng2022Symmetric}, auxiliary-field quantum Monte Carlo (QMC) \cite{Slagle:2015lo}, rational hybrid Monte Carlo (RHMC) \cite{Clark2005Exact}, and
fermion bag Monte Carlo (FBMC) \cite{Huffman:2017rg}.

The SMG phase has been successfully achieved in studies listed in \tabref{tab:numerics}, which demonstrate the generality of the SMG mechanism in different spacetime dimensions with various symmetry assignments and under different forms of interaction. The SMG phase has the following defining features that can be checked in numerics: 
\begin{itemize}
\setlength\itemsep{0pt}
\item Unique ground state with a gap to all excitations (including both fermionic and bosonic excitations),
\item Absence of spontaneous symmetry breaking (no fermion bilinear condensation),
\item Formation of the four-fermion (or higher multi-fermion) condensate that preserves the symmetry. 
\end{itemize}

\begin{figure}[htbp]
\begin{center}
\includegraphics[width=0.82\columnwidth]{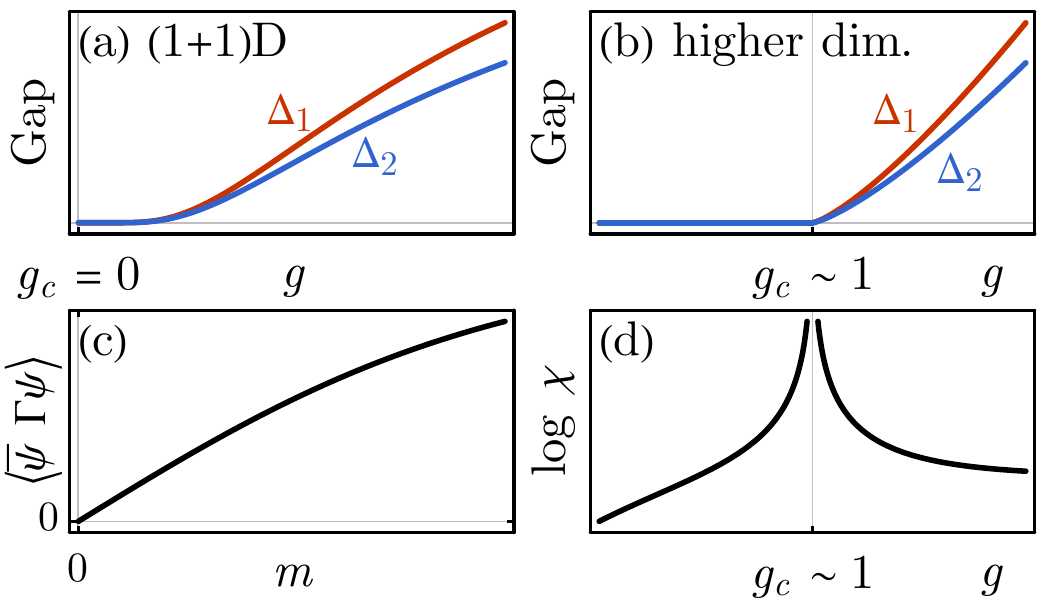}
\caption{Typical features of SMG observed in numerics. Continuous gap opening in (a) (1+1)D SMG driven by four-fermion interactions and (b) higher dimensional SMG in general. (c) Vanishing fermion bilinear expectation $\langle\bar{\psi} \Gamma \psi\rangle$ with its source field $m$ in the SMG phase. (d) Diverging Yukawa field susceptibility $\chi$ at the SMG transition.}
\label{fig:numerics}
\end{center}
\end{figure}

Directly computing the many-body excitation gap is challenging for most numerical approaches. One way to probe the excitation gap is to measure the correlation function in different channels
\eqs{\label{eq:correlation}C_{ij}^{(1)}&=\langle\psi_{ia}\psi_{ja}\rangle\sim \e^{-|\vect{x}_i-\vect{x}_j|/\xi_{1}},\\
C_{ij}^{(2)}&=\langle\psi_{ia}\psi_{ib} \psi_{ja}\psi_{jb}\rangle\sim \e^{-|\vect{x}_i-\vect{x}_j|/\xi_{2}}.\\ }
An exponential decaying correlation function implies a finite excitation gap $\Delta_n\sim 1/\xi_n$ inversely proportional to the correlation length $\xi_n$. In particular, the single-fermion correlation $C^{(1)}$ probes the single-particle gap $\Delta_1$ for fermionic excitations, and the fermion-bilinear correlation $C^{(2)}$ probes the gap $\Delta_2$ for collective bosonic excitations. Various studies \cite{Slagle:2015lo,Catterall:2016sw,Catterall:2016nh,Zeng2022Symmetric} have used the correlation function approach to demonstrate the gap opening (mass generation) in the SMG phase, as illustrated in \figref{fig:numerics}(a,b). \refcite{Slagle:2015lo} observed that the fermion single-particle gap $\Delta_1$ is a bit larger than the bosonic fluctuation gap $\Delta_2$ in the SMG phase, which seems to be consistent with the fluctuating bilinear mass picture (where the bosonic excitations are typically softer than the fermionic excitations), as to be discussed in \secref{sec:FBM}.

More explicitly, \refcite{DeMarco2017A-Novel} investigated the fermion single-particle spectrum of (1+1)D chiral gauge theory coupled to a fluctuating Higgs field with finite spacetime correlation. By choosing the Higgs coupling strength and the Higgs field correlation length appropriately, \refcite{DeMarco2017A-Novel} was able to show that the fermion excitation gap remains open in the SMG phase, despite of the absence of Higgs condensation.

Many numerical works have also confirmed that there is no fermion bilinear condensation in the SMG phase, such that the mass generation is not due to spontaneous symmetry breaking. To check this statement, one can introduce a small source field $m$ that couples to the fermion bilinear term $\bar{\psi} \Gamma \psi$ of interest. By showing that in the large system size limit ($L\to \infty$) the induced fermion bilinear expectation value vanishes as the source field $m$ is turned off 
\eq{\mathop{\mathrm{lim}}_{m\to 0}\mathop{\mathrm{lim}}_{L\to \infty}\langle\bar{\psi} \Gamma \psi\rangle=0,}
one can rule out the fermion bilinear condensation, as illustrated in \figref{fig:numerics}(c). Such check should be performed for all fermion bilinear terms that transform as different representations under the symmetry. This approach was adopted in various studies \cite{Catterall:2016nh,He:2016qy,Schaich2018Phases,Butt2018SO4-invariant,Catterall2020Exotic,Catterall2021Chiral}. Another approach is to measure the correlation function of fermion bilinear operators. If the correlation function decays exponentially, then the bilinear operator is not long-ranged ordered (not condensed). This approach is used in  \refcite{Slagle:2015lo,Zeng2022Symmetric}. 

\subsection{Nature of the SMG Transitions}
\label{sec:transition}

Much effort of the numerical study has been focused on understanding the nature of the SMG transition. Two questions can be asked:
\begin{itemize}
\setlength\itemsep{0pt}
\item Is the SMG transition a \emph{direct} transition (i.e.~without any intermediate phases setting in)?
\item Is the SMG transition \emph{continuous} (i.e.~not first-order)? 
\end{itemize}
If the answers to both questions were yes, the SMG transition would be a quantum critical point which admits a field-theory description in the continuum limit. Such case will be highly interesting from the theoretical perspective, because the phases on both sides cannot be distinguished by a symmetry-breaking order parameter, hence a direct and continuous SMG transition is necessarily an exotic quantum phase transition beyond the Landau-Ginzburg-Wilson paradigm.

\begin{table}[htp]
\caption{Summary of numerical studies of SMG transitions. (Dir.~- direct transition or not, Con.~- continuous transition or not).}
\begin{center}
\begin{tabular}{llllp{75pt}p{40pt}}
Dim. & Sym. & Dir. & Con. & Remarks & Reference\\
\hline
(1+1)D & $\u(1)$ & yes & yes & BKT, $g_c\sim1$ & \cite{Zeng2022Symmetric}\\
& $\so(4)$ & yes & yes & $g_c=0$ & \cite{Slagle:2015lo}\\
& $\su(4)$ & yes & yes & $g_c=0$ & \cite{Ayyar2017Generating}\\
\hline
(2+1)D & $\so(4)$ & yes & yes & & \cite{Catterall:2016sw} \\
& $\so(5)$ & yes & yes & & \cite{Slagle:2015lo}\\
& $\su(4)$ & yes & yes & $\eta=1.05$, $\nu=1.30$ & \cite{Ayyar2015Massive,Ayyar:2016fi}\\
&&&& $\eta=0.7\pm0.1$ & \cite{He:2016qy}\\
\hline
(3+1)D & $\so(4)$ & yes & yes & by frustrating the Yukawa field & \cite{Butt2018SO4-invariant,Catterall2020Exotic,Butt2021Symmetric}\\
& $\su(4)$ & no & - & small intermediate SSB phase & \cite{Schaich2018Phases,Ayyar:2016tg,Ayyar:2016ph}\\

\end{tabular}
\end{center}
\label{tab:transition}
\end{table}

The nature of the SMG transition depends on the spacetime dimension and the symmetry. \tabref{tab:transition} summarizes numerical results in the literature. SMG transitions in (1+1)D are most well understood. As four-fermion interactions are marginal perturbations for (1+1)D massless fermions, the RG equation for the interaction strength $g$ with respect to the RG scale $\ell=\ln\Lambda$ takes the general form of $\dd g/\dd\ell=\alpha g^2$ to the leading order of $g$. If the coefficient $\alpha>0$, $g>0$ is marginally relevant, which flows towards infinity for any finite $g$. This happens to any SMG transition driven by four-fermion interactions in (1+1)D: the system immediately enters the symmetric gapped phase as long as the interaction is turned on (with the correct sign), such that the SMG critical point is at $g_c=0$. The RG equation predicts that the excitation gap should open as (see \figref{fig:numerics}(a))
\eq{\Delta\sim\exp(-\tfrac{1}{\alpha g}),}
which has been verified in \refcite{Slagle:2015lo,Ayyar2017Generating} with different models. 

However, SMG can also be driven by higher-order fermion interactions. For example, the SMG in (1+1)D $\U(1)$-symmetric chiral fermion 3-4-5-0 model is driven by six-fermion interactions, which are irrelevant at the massless free-fermion fixed point. As shown in \refcite{Zeng2022Symmetric}, the SMG transition in this case happens at a finite interaction strength, and the transition is shown to be in the BKT universality class.

Numerical studies have also found evidence for direct and continuous SMG transitions in higher dimensions, as listed in \tabref{tab:transition}. A direct transition can be established by ruling out any intermediate phase between the gapless fermion phase and the SMG phase. The most probable intermediate phase is the spontaneous symmetry breaking (SSB) phase in which the fermion bilinear mass (the Yukawa field) condenses. Such intermediate SSB phase is often observed in (3+1)D systems.\cite{Schaich2018Phases,Ayyar:2016tg,Ayyar:2016ph} However, there are also examples showing that it is possible to shrink the intermediate SSB phase by frustrating the Yukawa field (i.e.~introducing local couplings of the Yukawa field in conflict with its natural ordering tendency), which could foster a direct SMG transition.\cite{Butt2018SO4-invariant,Catterall2020Exotic,Butt2021Symmetric}

Given a direct SMG transition, one can further check whether the transition is continuous or first-order (discontinuous). The numerical evidence for a continuous transition include:
\begin{itemize}
\setlength\itemsep{0pt}
\item Continuous gap opening across the SMG transition. 
\item Universal scaling behavior of physical quantities near the transition. 
\end{itemize}
The excitation gap (or the inverse correlation length) can be extracted from correlation functions (as previously explained around \eqnref{eq:correlation}). \refcite{Slagle:2015lo,Catterall:2016sw} have demonstrated that the gap opens smoothly across the SMG transition, as depicted in \figref{fig:numerics}(b), in support of a continuous transition (i.e.~a quantum critical point). As the correlation length diverges, physical quantities should exhibit universal scaling behaviors near the quantum critical point. The quantity that is often studied in numerics is the uniform static susceptibility of the Yukawa field, which is defined as
\eq{\chi=\frac{1}{L^d}\int\dd^d x\langle\phi_\alpha(x)\phi_\alpha(0)\rangle,}
where the integration is over the spacetime and $L^d$ stands for the spacetime volume. The following universal behavior is observed in \refcite{Ayyar2015Massive,Ayyar:2016fi}
\eq{\chi=L^{2-\eta}f((g-g_c)L^{1/\nu}),}
where $f$ is a universal function, and $\eta$ and $\nu$ are critical exponents. Such that the susceptibility generally diverges as $\chi\sim|g-g_c|^{-\nu(2-\eta)}$ near the SMG critical point, as illustrated in \figref{fig:numerics}(d). The exponent $\eta$ can also be determined from the power-law fitting of the correlation function at the critical point,
\eq{\langle\phi_{i\alpha}\phi_{j\alpha}\rangle\sim|\vect{x}_i-\vect{x}_j|^{-(d-2+\eta)}.}
This approach is used in \refcite{He:2016qy}. Currently, the scaling analysis has only been performed for (2+1)D systems, where the simulation can achieve a linear system size up to $L=60$ \cite{Ayyar:2016fi}, which enables a rather reliable estimate of critical exponents. As a comparison, for (3+1)D systems, the simulation can only reach $L=16$ 
\cite{Schaich2018Phases}. Numerical studies \cite{Ayyar2015Massive,Ayyar:2016fi,He:2016qy} of the (2+1)D $\su(4)$ symmetric SMG found $\eta\sim 1.05$ and $\nu\sim 1.30$, which is close to the (large-$N_f$ limit) exponents ($\eta=\nu=1$) of the (2+1)D Gross-Neveu-Yukawa universality class \cite{Gross:1974fc,Hands1993Four-Fermi}. However, the SMG mechanism is physically distinct from the symmetry-breaking mass generation described by the Gross-Neveu-Yukawa model. Their similar critical exponents motivate the idea to view the SMG transition as a hidden Gross-Neveu-Yukawa transition of fermionic partons, as to be discussed in \secref{sec:fermion_frac}.

\section{Theoretical Understandings}
\label{sec:theory}

\subsection{Fluctuating Bilinear Mass Picture}\label{sec:FBM}

The \emph{fluctuating bilinear mass} (Yukawa field) provides an intuitive physical picture for SMG. It suggests that the SMG can be generally understood in two steps: starting with gapless Weyl/Majorana fermions, first condense a Yukawa field that couples to fermion bilinear mass terms to gap out the fermions at the price of breaking the protecting symmetry, then fluctuate the phase (or direction) of the Yukawa field in the spacetime to restore the symmetry while maintaining the local amplitude of the Yukawa field finite to keep the fermion gap open. The picture can be described by the Yukawa-Higgs model with a symmetry $G$ \cite{Catterall:2016nh,DeMarco2017A-Novel}
\eqs{\label{eq:YH}Z&=\int\DD[\psi,\phi]\e^{-S_\text{Y}[\psi,\phi]-S_\text{H}[\phi]},\\
S_\text{Y}[\psi,\phi]&=\int\dd^d x\;(\bar{\psi}_a
{\ii} \gamma^\mu\partial_\mu\psi_a+\phi_\alpha\bar{\psi}_a\Gamma^{\alpha}_{ab}\psi_b),\\
S_\text{H}[\phi]&=\int\dd^d x\;((\partial_\mu\phi)^2+V_\text{H}(\phi)+\cdots),}
where $\psi_a$ are the Weyl/Majorana fermions (written as real spinor Grassmann fields with $\bar{\psi}_a = \psi_a^\intercal\gamma^0$) and $\phi_\alpha$ are the Yukawa bosons (written as real scalar fields) \footnote{For generality, we assume that all (no matter real or complex) fields are automatically translated into their minimal real embeddings to avoid unnecessary complication of complex conjugations in the discussion.}. Both the fermion $\psi$ and the boson $\phi$ fields are in non-trivial representations of the protecting symmetry $G$, denoted as $\rep_\psi^G$ and $\rep_\phi^G$ respectively. The Yukawa coupling will be loosely denoted as $\phi\,\bar{\psi}\Gamma\psi$ for generality, where the vertex tensor $\Gamma^{\alpha}_{ab}$ is set by the Clebsch-Gordan coefficients of the fusion channel $\rep_\psi^G\times_\scA\rep_\psi^G\to\rep_\phi^G$. Precise forms of the Yukawa coupling can be found in Eqs.\,(\ref{eq:H_YH}, \ref{eq:LLint_YH}, \ref{eq:honeycomb_YH}, \ref{eq:CF_YH}), which vary from model to model. $S_\text{Y}[\psi,\phi]$ describes the gapless fermion $\psi$ coupled to the Yukawa field $\phi$, and $S_\text{H}[\phi]$ describes the dynamics of the Yukawa field with $V_\text{H}(\phi)$ being some Higgs potential. 

Integrating out the Yukawa field $\phi$ leads to a pure fermion model of $\psi$. Assuming $S_\text{H}[\phi]=\frac{1}{2g}\phi^2$ takes the Gaussian form at the ultraviolet (UV) level, the fermion model will be
\eqs{\label{eq:pure}Z&=\int\DD[\psi]\e^{-S[\psi]},\\
S[\psi]&=\int\dd^d x(\bar{\psi}{\ii} \gamma^\mu\partial_\mu\psi+g(\bar{\psi}\Gamma^\alpha\psi)(\bar{\psi}\Gamma^\alpha\psi)).}
It is assumed that the model is free of $G$-anomaly, but the symmetry $G$ is still restrictive enough to forbid all possible bilinear mass terms, paving ways for SMG. The fluctuating Yukawa field effectively mediates the fermion interaction $g$ that is responsible to drive the SMG.

The appearance of the SMG in the Yukawa-Higgs model \eqnref{eq:YH} requires some delicate design of $S_\text{H}[\phi]$ to achieve the appropriate infrared (IR) dynamics. Assuming the Higgs potential $V_\text{H}(\phi)$ pins the Yukawa field $\phi$ to a finite amplitude (e.g.~$|\phi|=1$) throughout the spacetime (as in a non-linear $\sigma$-model), while allows its orientation to fluctuate in its internal flavor space, numerical evidences~\cite{DeMarco2017A-Novel}
show that the fluctuating Yukawa field is possible to gap out the fermions $\psi$ without spontaneously breaking the $G$ symmetry, if the Yukawa field fluctuation has a finite but large correlation length $\xi\gg 1$ (s.t.~$\langle\phi(x)\phi(0)\rangle\sim\e^{-|x|/\xi}$). The fluctuation must be smooth enough so as not to create sharp domain walls that trap gapless domain-wall fermions and close the fermion gap. But the fluctuation must also not be too smooth to establish long-range order of the Yukawa field and break the symmetry. The SMG should be achieved by balancing these two factors.

An alternative way to argue for the SMG is the \emph{topological defect condensation} approach \cite{Metlitski2014Interaction,Wang2014Interacting,You:2014ow,You:2015lj,Catterall2018Topology}. This approach also starts by condensing the Yukawa field with a finite amplitude, but then disordering the Yukawa field orientation by condensing topological point defects (e.g.~vortices in 2D space or monopoles in 3D space) of the Yukawa field. Although the point defects could trap fermion zero modes and leads to gapless fermion excitations in the spectrum, these fermion zero modes can be gapped out by local interactions (such as the FK interaction) uniformly applied throughout the system. A non-trivial check in this approach is to show that for anomaly-free fermion systems (that admits SMG), the point defect always trap $8n$ Majorana zero modes that can be trivialized by interaction. 
The advantage of the topological defect condensation approach is that it does not rely on a delicate tuning of the smoothness of the Yukawa field fluctuation.

Further justification of the fluctuating bilinear mass picture comes from a more explicit trial wave function construction for the SMG state in (0+1)D \cite{You2018Symmetric}. Let $\ket{\Psi[\phi]}$ be the quantum many-body ground state of fermions $\psi$ on a background configuration of the Yukawa field $\phi$. The fluctuating bilinear mass picture \eqnref{eq:YH} suggest that the SMG state (the featureless gapped state in the strong interacting limit) should be described by
\eq{\label{eq:PsiSMG}\ket{\Psi_\text{SMG}}\propto\int\DD[\phi]\e^{-S_\text{H}[\phi]}\ket{\Psi[\phi]}.}
In the (0+1)D example, the ground state of the Yukawa Hamiltonian $H_\text{Y}=-\phi_\alpha \chi_a\Gamma_{ab}^\alpha\chi_b$ is explicitly given by $\ket{\Psi[\phi]}=(1+\frac{1}{|\phi|}\phi_\alpha\chi_a\Gamma_{ab}^\alpha\chi_b+c_1^\dagger c_2^\dagger c_3^\dagger c_4^\dagger)\ket{0000}$ using notations introduced in \secref{sec:FK}. Assuming $S_\text{H}[\phi]$ restricts the $\phi$ vector uniformly on a sphere, the spherical average results in $\ket{\Psi_\text{SMG}}\propto (1+c_1^\dagger c_2^\dagger c_3^\dagger c_4^\dagger)\ket{0000}$ which indeed matches the ground state $\ket{\mathsf{4e}}$ of the FK interaction. The construction also applies to $\su(4)$ symmetric SMGs in all higher dimensions, where the on-site interaction is the same as the (0+1)D case. In the limit that the Yukawa coupling is strong and the Yukawa field $\phi$ fluctuates independently on each site, the trial wave function $\ket{\Psi_\text{SMG}}$ in \eqnref{eq:PsiSMG} approaches to product state $\bigotimes_i\ket{\mathsf{4e}}_i$ --- the exact ground state in the strong coupling ($g\to\infty$) limit. Away from the strong coupling limit, $\ket{\Psi_\text{SMG}}$ might still serve as a variational approximation of the SMG state, but this variational ansatz is less well-controlled beyond $(0+1)$D.

Although the fluctuating bilinear mass picture provides a simple understanding for the SMG phase/state, it breaks down near the SMG transition, where the Yukawa field amplitude is no longer large and stable. A major theoretical challenge is to develop field theory descriptions for the continuous SMG transition, which will be the topic of the following subsection \secref{sec:fermion_frac}.

\subsection{Fermion Fractionalization Field Theory}\label{sec:fermion_frac}

The \emph{fermion fractionalization} \cite{You2018Symmetric,You2018From} provides an overarching theoretical framework to understand various continuous SMG transitions in different dimensions. It unifies different theoretical pictures of SMG in literature. Its key proposal is that the physical fermion $\psi\sim \upeta^n \uppsi $ fractionalizes into bosonic $\upeta$ and fermionic $\uppsi$ partons that deconfine \emph{at and only at} the transition, where the bosonic parton $\upeta$ may appear $n$ multiple times in the fractionalization scheme. Note that upright greek letters will be used to denote parton fields in the parton theory. 

The fermionic parton $\uppsi$ is generally put in the same gapless phase as the physical fermion $\psi$. The physical free fermion phase corresponds to the condensed phase of the bosonic parton $\upeta$, and the SMG phase corresponds to the disordered (gapped) phase of the bosonic parton $\upeta$. Accompanied with the fractionalization, an additional gauge field $a$ must be introduced to bind the partons together and to remove the redundant degrees of freedom introduced by fractionalization. Let $K$ be the emergent gauge group (typically non-Abelian), then the partons are generally charged under an enlarged symmetry-gauge group $G\times K$ (where $G$ is the symmetry group of the physical fermion), as summarized by \tabref{tab:rep}. 

\begin{table}[htp]
\caption{Representations of different fields under $G\times K$, also under spacetime symmetry $\Spin(d)$. Representation assignments must be consistent with $\psi\sim\upeta^n\uppsi$, $\phi\sim\bar{\psi}\Gamma\psi$, and $\upphi\sim\bar{\uppsi}\Gamma\uppsi$.}
\begin{center}
\setlength{\extrarowheight}{2pt}
\begin{tabular}{ccccc}
& $G$ & $K$ & $\Spin$ & meaning\\
\hline
$\psi$ & $\rep_\psi^G$ & $\mathbf{1}^K$ & spinor & physical fermion\\
$\phi$ & $\rep_\phi^G$ & $\mathbf{1}^K$ & scalar & Yukawa boson\\
\hline
$\upeta$ & $\rep_\upeta^G$ & $\bar{\ybox(1)}^K$ & scalar & bosonic parton \\
$\uppsi$ & $\rep_\uppsi^G$ & $\rep_\uppsi^K$ & spinor & fermionic parton (like $\psi$)\\
$a$ & $\mathbf{1}^G$ & $\bar{\ybox(1)}\ybox(1)^K$ & vector & gauge boson\\
\hline
$\upphi$ & $\rep_\upphi^G$ & $\rep_\upphi^K$ & scalar & parton-Higgs boson (like $\phi$)\\
\end{tabular}
\end{center}
\label{tab:rep}
\end{table}

The choice of representations should satisfy the following general rules:
\begin{itemize}
\setlength\itemsep{0pt}
\item The bosonic parton $\upeta$ is always in the (anti)-fundamental representation $\bar{\ybox(1)}^K$ of $K$, such that its condensation can fully Higgs the gauge group $K$.
\item The representation must be assigned in consistent with the fermion fractionalization $\psi\sim\upeta^n\uppsi$, such that the following fusion channel must exist
\eq{\label{eq:rep_condition}(\rep_\upeta^G,\bar{\ybox(1)}^K)^{\times n}\times(\rep_\uppsi^G,\rep_\uppsi^K)\to(\rep_\psi^G,\mathbf{1}^K).}
\item The $K$-gauge field $a$ is always in the trivial representation $\mathbf{1}^G$ of $G$ and the adjoint representation $\bar{\ybox(1)}\ybox(1)^K$ of $K$.
\end{itemize}
Given the representation $(\rep_\psi^G,\mathbf{1}^K)$ of the physical fermion $\psi$, there is still the freedom to choose the gauge group $K$ and the representation $(\rep_\uppsi^G,\rep_\uppsi^K)$ of the fermionic parton $\uppsi$. Once the choice is made, the multiplicity $n$ and the representation $(\rep_\upeta^G,\bar{\ybox(1)}^K)$ of the bosonic parton $\upeta$ are fixed by \eqnref{eq:rep_condition}. There could be multiple parton theories that describe the same SMG transition. Within the present framework, different parton theories are specified by different choices of $K$ and $(\rep_\uppsi^G,\rep_\uppsi^K)$.

The parton theory for the SMG transition generally takes the form of \cite{You2018Symmetric,You2018From}
\eqs{\label{eq:parton}
Z&=\int\DD[\upeta,\uppsi,a]\e^{-S_\text{B}[\upeta,a]-S_\text{F}[\uppsi,a]},\\
S_\text{B}[\upeta,a]&=\int\dd^d x\;(|D_\mu\upeta|^2+r|\upeta|^2+u|\upeta|^4),\\
S_\text{F}[\uppsi,a]&=\int\dd^d x\;\bar{\uppsi} {\ii} \gamma^\mu D_\mu\uppsi,}
where $\uppsi$ and $\upeta$ are treated as real fields (with $\bar{\uppsi}=\uppsi^\intercal\gamma^0$) to avoid unnecessary complications of complex conjugation, and $D_\mu=\partial_\mu-\ii a_\mu$ denotes the covariant derivative (assuming the gauge connection $a$ is automatically represented in the representation of the matter field that it couples to). The parton theory describes the bosonic $\upeta$ and the fermionic $\uppsi$ partons coupled together by the gauge field $a$, and is expected to provide an effective description of the low-energy physics around the SMG critical point. The bosonic parton mass $r$ is treated as the only driving parameter in the parton theory, and is responsible for tuning the SMG transition. $r<0$ corresponds to the gapless phase (the free-fermion fixed point), $r>0$ corresponds to the trivially gapped phase (the strong-coupling fixed point), and $r=0$ corresponds to the critical point where the SMG transition happens. The states of different fields in different cases are summarized in \tabref{tab:states}.

\begin{table}[htp]
\caption{States of different fields across the SMG transition.}
\begin{center}
\begin{tabular}{cccc}
& gapless phase & SMG & gapped phase\\
& $(r<0)$ & $(r=0)$ & $(r>0)$\\
\hline
$\psi$ & gapless & fractionalized & gapped\\
$\phi$ & gapped ($\xi\sim1$) & gapped & gapped\\
\hline
$\upeta$ & condensed & critical & gapped\\
$\uppsi$ & gapless & gapless & gapped\\
$a$ & Higgs & deconfined & Higgs/confined\\
\hline
$\upphi$ & gapped & critical & condensed
\end{tabular}
\end{center}
\label{tab:states}
\end{table}

When $r<0$, the bosonic parton $\upeta$ condenses (i.e.~$\langle\upeta\rangle\neq 0$). As $\upeta$ carries the anti-fundamental representation of the gauge group $K$, the gauge structure is fully Higgs down by the bosonic parton condensation. At the same time, the fermionic parton $\uppsi$ becomes the physical fermion $\psi\sim\langle\upeta\rangle^n\uppsi$ effectively (as the fields $\uppsi$ and $\psi$ only differ by a constant $\langle\upeta\rangle^n$ which simply serves as a quasi-particle weight). The parton theory then reduces to the physical fermion theory $S[\psi]=\int\dd^d x\;\bar{\psi}{\ii} \gamma^\mu\partial_\mu\psi$ in the gapless free-fermion phase.

When $r>0$, the bosonic parton $\upeta$ is gapped and decouples from the gauge field at low-energy. Below the gap of $\upeta$ field, the parton theory is effectively a quantum chromodynamics (QCD) theory $S_\text{F}[\uppsi,a]$ involving the fermionic parton $\uppsi$ coupled to the gauge field $a$. Then it relies on an assumption that the QCD dynamics leads to a featureless gapped ground state (a.k.a.~a trivial ground state), such that both the fermionic partons and the gauge bosons are gapped without breaking the $G$ symmetry or generating topological orders. Of course, one necessary condition for the QCD theory to trivialize is that the theory must have vanishing $G\times K$ anomaly. However, even if the QCD theory is anomaly free, it still depends on the dynamical details to achieve a trivially gapped ground state. The major theoretical effort to understand SMG lies in designing appropriate mechanism to trivialize the QCD theory. Within the fermion fractionalization framework, two mechanisms has been proposed in the literature: (i) parton-Higgs (weak coupling) \cite{You2018Symmetric,You2018From,Tong2021Comments} and (ii) s-confinement (strong coupling) \cite{Razamat2021Gapped,Tong2021Comments}. In some cases, the QCD trivialization may be consistently achieved by both mechanisms.

\subsubsection{Parton-Higgs mechanism}
\label{sec:Parton-Higgs-mechanism}

The {parton-Higgs mechanism} introduces a collection of scalar fields $\upphi$ (parton-Higgs fields) that couple to the fermionic parton $\uppsi$ via Yukawa interactions, such that the QCD theory $S_\text{F}[\uppsi,a]$ is extended to a QCD-Yukawa-Higgs theory
\eqs{\label{eq:QCDYH}S_\text{F}[\uppsi,\upphi,a]=\int\dd^d x\;(&\bar{\uppsi} {\ii} \gamma^\mu D_\mu\uppsi+\upphi_\alpha\bar{\uppsi}\Gamma^\alpha\uppsi\\
+&(D_\mu\upphi)^2+V_\text{H}(\upphi)),}
where the parton-Higgs field $\upphi$ (as fermionic parton bilinear mass) is in the representation $(\rep_\upphi^G,\rep_\upphi^K)$ of $G\times K$. The representation $(\rep_\upphi^G,\rep_\upphi^K)$ must be obtained from an antisymmetric product of fermionic parton representations for consistency
\eq{\label{eq:pY_rule}(\rep_\uppsi^G,\rep_\uppsi^K)\times_\scA(\rep_\uppsi^G,\rep_\uppsi^K)\to(\rep_\upphi^G,\rep_\upphi^K).}
The Yukawa vertex tensor $\Gamma^\alpha$ is fixed by this fusion channel. The QCD-Yukawa-Higgs theory in \eqnref{eq:QCDYH} is reminiscent of the Yukawa-Higgs theory in \eqnref{eq:YH} by promoting the physical fermion $\psi$ and physical Yukawa field $\phi$ to their parton counterparts $\uppsi$ and $\upphi$. However, the key difference lies in the different $G\times K$ representations of $\psi,\phi$ comparing to $\uppsi,\upphi$, as listed in \tabref{tab:rep}. 

Unlike the physical Yukawa field $\phi$, which is solely charged under $G$ in a non-trivial representation $\rep_\phi^G$, such that the condensation of $\phi$ inevitably breaks the $G$ symmetry; the parton-Higgs field $\upphi$ is in a joint representation $(\rep_\upphi^G,\rep_\upphi^K)$ of $G\times K$, which could admit a $G$-symmetric condensation of $\upphi$. The sufficient and necessary condition for the existence of a condensed configuration $\langle\upphi\rangle$ that preserves the $G$ symmetry is that there exists a subgroup $G\times K'$ of $G\times K$, such that when $G\times K$ is broken to $G\times K'$, the representation of $\upphi$ admits a trivial branching channel:
\eq{\label{eq:Higgs_condition}(\rep_\upphi^G,\rep_\upphi^K)\to(\mathbf{1}^G,\mathbf{1}^{K'}).} 
If the condition is met, a Higgs potential $V_\text{H}(\upphi)$ can be constructed to drive the condensation of $\upphi$ in the above trivial branching channel. Such a condensation of the parton-Higgs field $\upphi$ will (i) gap out all fermionic partons $\uppsi$ (that couple to it), (ii) Higgs down the gauge group from $H$ to its subgroup $K'\subseteq K$, and (iii) preserving the symmetry group $G$. Then the only freedom remaining in the theory is the pure gauge fluctuation of  $K'$. If $K'$ remains a non-Abelian group, its gauge coupling can flow strong into the confine phase and gap out the remaining gauge bosons. In this way, all freedoms in the QCD-Yukawa-Higgs theory are symmetrically gapped and the system ends up in the SMG phase. The physical fermion $\psi\sim\upeta\uppsi$ will also be gapped, as its partons are gapped (both $\upeta$ and $\uppsi$ are gapped).

\subsubsection{s-confinement mechanism}
\label{sec:s-confinement-mechanism}

The {s-confinement mechanism} (``s'' for ``smooth'') refers to the strong coupling dynamics of the QCD theory that confines fermionic partons and gauge bosons all together without breaking the $G$ symmetry. This is possible if the 't Hooft anomaly of $G$ is matched up in the theory \cite{tHooft1980}. One motivation for the s-confinement theory comes from the observation that the four-fermion SMG interaction $\psi^4$ (for physical fermions $\psi$) can be broken up into ``3+1'' as $(\psi^3)\psi$. If there is some strong-coupling mechanism that binds the first three fermions into a composite fermion $\psi_\text{comp}\sim\psi^3$ at low-energy, the interaction can then be viewed as a fermion bilinear mass like $\bar{\psi}_\text{free}\psi_\text{comp}$ (where $\bar{\psi}_\text{free}\sim\psi$ stands for the last dangling fermion), which provides a free-fermion understanding for the SMG. This picture is alternative to the ``2+2'' fluctuation bilinear mass picture, where the interaction is broken up into $(\psi^2)(\psi^2)$ with each $\psi^2$ being a composite boson (i.e.~the Yukawa boson $\phi$). However, the Yukawa boson is still interacting, which complicates the analysis of its low-energy dynamics. In contrast, the composite fermion flows to a free fermion critical point, which enables simple argument for gapping. Nevertheless, this still relies on a strong-coupling mechanism for the composite fermion to form in the first place. 
 
The fermion fractionalization plays an important role in understanding the formation of composite fermions in the s-confinement picture. After fractionalization, physical fermions become fermionic partons that couple to an emergent gauge field, then the emergent gauge force can be employed to confine the fermions into the desired composite. Not all fermions are fractionalized in the s-confinement theory. The physical fermions $\psi=\psi_\text{frac}\oplus\psi_\text{free}$ is first divided into two sectors, where $\psi_\text{frac}\sim \upeta^n\uppsi$ are part of the physical fermions that will be fractionalized to bosonic $\upeta$ and fermionic $\uppsi$ partons, and $\psi_\text{free}$ are the remaining physical fermions that do not fractionalize. To drive the SMG transition, the first step is still to gap the bosonic parton $\upeta$, leaving the low-energy freedom in a QCD theory with the fermionic parton $\uppsi$ coupled to an emergent (non-Abelian) gauge field $a$. Under gauge confinement, the fermionic partons $\uppsi$ are bound together in the IR to form fermion composites, denoted as $\psi_\text{comp}\sim (\uppsi)^3$, just like fundamental quarks forming baryons. The necessary condition for the confinement to happen is the existence of a trivializing fusion channel in the gauge sector
\eq{\label{eq:confine_condition}(\rep_\uppsi^K)^{\times 3}\to\mathbf{1}^K,}
such that the gauge charges of $\uppsi^3$ can be neutralized to produce the physical composite fermion $\psi_\text{comp}$.  Then the fermion bilinear mass $\bar{\psi}_\text{free}\psi_\text{comp}$ can be introduced to gap out all fermions together, realizing the SMG. Moreover, the s-confinement mechanism admits the supersymmetric extension \cite{Tong2021Comments}, where addition of supersymmetry provides extra control for the strong-coupling confinement dynamics at low-energy \cite{Intriligator1995Duality}.

\subsection{Examples of Fermion Fractionalization}\label{sec:fermion_frac_example}

In the following, we will briefly exemplify the fermion fractionalization theory in two SMG models introduced in \secref{sec:examples}.

\subsubsection{(2+1)D Honeycomb Lattice Model}\label{sec:SU4_parton}

The SMG in the (2+1)D honeycomb lattice model can be understood using the parton-Higgs mechanism within the framework of fermion fractionalization \cite{You2018Symmetric}. The internal symmetry in consideration is \eq{G=\SU(4)\times_{\dsZ_2^F}\dsZ_4^{TF},}
where $\dsZ_4^{TF}$ is the essential protecting symmetry. The physical fermion $\psi$ is in the $\mathbf{4}_1$ representation of $G$. The proposed fermion fractionalization scheme $\psi\sim\upeta \uppsi$ introduces an emergent gauge group
\eq{K=\SU(4),}
and put the fermionic parton $\uppsi$ in the $(\mathbf{1}_1,\mathbf{4})$ representation of $G\times K$. This choice fully specifies the parton theory. The physical and fractionalized fields in the parton construction are summarized in \tabref{tab:repHL}, such that all consistency conditions are met among their symmetry representations.

\begin{table}[htp]
\caption{Representations of different fields under $(\SU(4)\times_{\dsZ_2^F}\dsZ_4^{TF})_G\times \SU(4)_K$, as well as the Euclidean spacetime symmetry $\Spin(3)$ 
, which are relevant to the SMG in the (2+1)D honeycomb lattice model.}
\begin{center}
\setlength{\extrarowheight}{2pt}
\begin{tabular}{cccc}
& $(\SU(4)\times_{\dsZ_2^F}\dsZ_4^{TF})_G$ & $\SU(4)_K$ & $\Spin$\\
\hline
$\psi$ & $\ybox(1)_1=\mathbf{4}_1$ & $\mathbf{1}$ & spinor (fermion)\\
$\phi$ & $\ybox(1,1)_2=\mathbf{6}_2$ & $\mathbf{1}$ & scalar (boson)\\[2pt]
\hline
$\upeta$ & $\ybox(1)_{0}=\mathbf{4}_{0}$ & $\bar{\ybox(1)}=\bar{\mathbf{4}}$ & scalar (boson) \\
$\uppsi$ & $\mathbf{1}_{1}$ & $\ybox(1)=\mathbf{4}$ & spinor (fermion)\\
$a$ & $\mathbf{1}_{0}$ & $\bar{\ybox(1)}\ybox(1)=\mathbf{15}$ & vector (boson)\\[2pt]
\hline
$\upphi$ & $\mathbf{1}_{2}$ & $\ybox(1,1)=\mathbf{6}$ & scalar (boson)\\
\end{tabular}
\end{center}
\label{tab:repHL}
\end{table}

The SMG can be achieved through the parton-Higgs mechanism by gapping out the bosonic parton $\upeta$ and condensing the parton Yukawa field $\upphi\sim \bar{\uppsi}\Gamma\uppsi$. This breaks $G\times K$ to $G\times K'$, where the new gauge group is
\eq{K'=\Sp(2)\subseteq \SU(4)=K.}
In particular, because the parton Yukawa field $\upphi$ charges 2 under both the $\dsZ_4^{TF}$ symmetry group and the $\dsZ_4^K$ center of the gauge group $K=\SU(4)$, condensing $\upphi$ necessarily locks the $\dsZ_4^{TF}$ and $\dsZ_4^K$ generators together, breaking $\dsZ_4^{TF}\times\dsZ_4^K$ down to $\dsZ_4^{TF}\times\dsZ_2^{K'}$, which is consistent with the $\dsZ_2^{K'}$ center of residual gauge group $K'=\Sp(2)$. As $G\times K$ is broken to $G\times K'$, the representation of $\upphi$ splits into 
\eq{
(\mathbf{1}_2,\mathbf{6}) \to  (\mathbf{1}_0,\mathbf{1})\oplus (\mathbf{1}_0,\mathbf{5}),
}
which admits a trivial representation $(\mathbf{1}_0,\mathbf{1})$ in the decomposition, satisfying the trivializing condition in \eqnref{eq:Higgs_condition} for the parton-Higgs mechanism to work. 

\subsubsection{(3+1)D Chiral Fermion Model}\label{sec:CF_SMG_theory}

The (3+1)D chiral fermion model contains three sectors of physical fermions $\psi= \psi_\lambda\oplus\psi_\psi\oplus\psi_\chi$ transformed under the symmetry group
\eq{G=\SU(N)\times\SU(N+4).}
The SMG in this model can be understood by fermion fractionalization, where only part of the fermions $\psi_\text{frac}=\psi_\psi\oplus\psi_\chi$ are fractionalized: $\psi_\psi\sim\upeta^\dagger\uppsi_\psi$ and $\psi_\chi\sim\upeta^2\uppsi_\chi$, while $\psi_\text{free}=\psi_\lambda$ remains untouched. The fractionalization introduces an emergent gauge group
\eq{K=\SU(N+4).}
\tabref{tab:repCF_full} summarizes the representation of fields satisfying all consistency conditions.

\begin{table}[htp]
\caption{Representations of different fields under $(\SU(N)\times\SU(N+4))_G\times \SU(N+4)_K$, as well as the Euclidean spacetime symmetry $\Spin(4)$, which are relevant to the SMG in the (3+1)D chiral fermion model.}
\begin{center}
\setlength{\extrarowheight}{2pt}
\begin{tabular}{cccc}
& $(\SU(N)\times\SU(N+4))_G$ & $\SU(N+4)_K$ & $\Spin$\\
\hline
$\psi_\lambda$ & $(\ybox(2),\mathbf{1})$ & $\mathbf{1}$ & spinor (fermion)\\
$\psi_\psi$ & $(\bar{\ybox(1)},\bar{\ybox(1)})$ & $\mathbf{1}$ & spinor (fermion)\\
$\psi_\chi$ & $(\mathbf{1},\ybox(1,1))$ & $\mathbf{1}$ & spinor (fermion)\\
$\phi$ & $(\bar{\ybox(1)},\ybox(1))$ & $\mathbf{1}$ & scalar (boson)\\[2pt]
\hline
$\upeta$ & $(\mathbf{1},\ybox(1))$ & $\bar{\ybox(1)}$ & scalar (boson) \\
$\uppsi_\psi$ & $(\bar{\ybox(1)},\mathbf{1})$ & $\bar{\ybox(1)}$ & spinor (fermion)\\
$\uppsi_\chi$ & $(\mathbf{1},\mathbf{1})$ & $\ybox(1,1)$ & spinor (fermion)\\
$a$ & $(\mathbf{1},\mathbf{1})$ & $\bar{\ybox(1)}\ybox(1)$ & vector (boson)\\[2pt]
\hline
$\upphi$ & $(\bar{\ybox(1)},\mathbf{1})$ & $\ybox(1)$ & scalar (boson)\\
\end{tabular}
\end{center}
\label{tab:repCF_full}
\end{table}

To drive the chiral fermions $\psi$ to the SMG phase, the first step is to gap out the bosonic parton $\upeta$. Then the theory contains a free fermion theory of $\psi_\lambda$ and a QCD theory of the fermionic partons $\uppsi_\psi\oplus\uppsi_\chi$ coupled to the $K$ gauge field $a$. The trivialization of the QCD theory can be achieved by either the parton-Higgs mechanism or the s-confinement mechanism (they are equivalent).

In the parton-Higgs mechanism, two parton-Higgs fields $\upphi=\upphi_1\oplus\upphi_2$ are introduced and coupled to the fermions as
\eq{\label{eq:CF_parton_mass}-(\upphi_1^\dagger (\uppsi_\chi^\intercal\ii\sigma^2\uppsi_\psi)+\upphi_2(\psi_\lambda^\intercal\ii\sigma^2\uppsi_\psi)+\text{h.c.}),}
which is closely reminiscent of the Yukawa-Higgs decomposition \eqnref{eq:CF_YH} of the SMG interaction in the chiral fermion model. Both parton-Higgs fields $\upphi_{1,2}$ are in the representation $((\bar{\ybox(1)},\mathbf{1}),{\ybox(1)})$ of $G\times K$. Condensing the parton-Higgs field $\upphi$ in a ``color-flavor'' locked form, the $\SU(N)$ (flavor) subgroup in $G$ will be identified with the $\SU(N)$ (color) subgroup in $K=\SU(N+4)$, such that the color-flavor combined  $\SU(N)$ rotation will form the diagonal $\SU(N)$ group that leaves $\upphi$ invariant. The color-flavor locking breaks the $G\times K$ group to its subgroup $G\times K'$ with \footnote{Some additional $\U(1)$ subgroups will be ignored to simplify the discussion, without hurting the main idea. See \cite{Tong2021Comments} for a more rigorous treatment.}
\eq{K'=\SU(4)\subseteq\SU(N+4)=K.}
Under $G\times K\to G\times K'$, the representation of $\upphi$ decomposes as 
\eq{\upphi:((\bar{\ybox(1)},\mathbf{1}),{\ybox(1)})\to ((\mathbf{1},\mathbf{1}),\mathbf{1})\oplus((\bar{\ybox(1)},\mathbf{1}),\mathbf{4})\oplus((\bar{\ybox(1)}{\ybox(1)},\mathbf{1}),\mathbf{1})}
which contains the trivial representation $((\mathbf{1},\mathbf{1}),\mathbf{1})$ that corresponds to the condensed configuration $\langle\upphi\rangle$, fulfilling the requirement of \eqnref{eq:Higgs_condition}. Moreover, representations of fermions branch as
\eqs{\psi_\lambda&:(({\ybox(2)},\mathbf{1}),\mathbf{1})\to (({\ybox(2)},\mathbf{1}),\mathbf{1}),\\
\uppsi_\psi&:((\bar{\ybox(1)},\mathbf{1}),\bar{\ybox(1)})\to ((\overline{\ybox(2)},\mathbf{1}),\mathbf{1})\oplus((\bar{\ybox(1,1)},\mathbf{1}),\mathbf{1})\oplus((\bar{\ybox(1)},\mathbf{1}),\bar{\mathbf{4}}),\\
\uppsi_\chi&:((\mathbf{1},\mathbf{1}),\ybox(1,1))\to (({\ybox(1,1)},\mathbf{1}),\mathbf{1})\oplus(({\ybox(1)},\mathbf{1}),{\mathbf{4}})\oplus((\mathbf{1},\mathbf{1}),\mathbf{6}).\\}
The fermion representation has transmuted, morphing from a chiral representation of $G\times K$ into a vector-like representation of the surviving $G\times K'$. The parton-Higgs field condensation $\langle\upphi\rangle\neq 0$ will provide mass for almost all fermions via the Yukawa coupling in \eqnref{eq:CF_parton_mass} (by pairing up conjugate fermion representations), with the only exception of the $((\mathbf{1},\mathbf{1}),\mathbf{6})$ fermions in $\uppsi_\chi$. But these fermions are fully neural under the symmetry group $G$ and only charged under the remaining gauge group $K'$. The remaining $K'$ gauge bosons and the $((\mathbf{1},\mathbf{1}),\mathbf{6})$ fermions can be simultaneously removed from the low-energy spectrum once the $K'$ gauge fluctuation flows strong and confines. Hence no low-energy freedom is left over, and the system ends up in the SMG phase.

In the s-confinement mechanism, a gauge-neutral combination of fermionic partons must first exist, as required by \eqnref{eq:confine_condition}. As the fermionic partons $\uppsi_\psi\oplus\uppsi_\chi$ carry the gauge charges $\bar{\ybox(1)}\oplus{\ybox(1,1)}$ respectively (see \tabref{tab:repCF_full}), the gauge neural three-fermion combination can be found in the fusion channel
\eq{{\ybox(1,1)}\times\bar{\ybox(1)}\times\bar{\ybox(1)}\to \mathbf{1}}
So the fermionic partons can be confined into a composite fermion through this confinement channel as
\eq{\psi_\text{comp}\sim (\uppsi_\chi^\intercal\ii\sigma^2\uppsi_\psi)\uppsi_\psi,}
which transforms as $((\overline{\ybox(2)},\mathbf{1}),\mathbf{1})$ under $G\times K$, precisely conjugate to the representation $(({\ybox(2)},\mathbf{1}),\mathbf{1})$ of the remaining physical fermion $\psi_\lambda$. Therefore, the composite fermion $\psi_\text{comp}$ can pair up with the remaining fermion $\psi_\lambda$  to produce a $G$-symmetric mass term $\psi_\lambda^\intercal \ii \sigma^2\psi_\text{comp}$, that gaps out all fermions from the low-energy spectrum, leading to the SMG state.

\subsection{Symmetry Extension Construction}
\label{sec:Symmetry-Extension}

\subsubsection{Symmetry Breaking vs Symmetry Extension vs SMG}
\label{sec:sym-breaking-extension-smg}

\begin{figure}[htbp]
\begin{center}
\includegraphics[width=0.95\columnwidth]{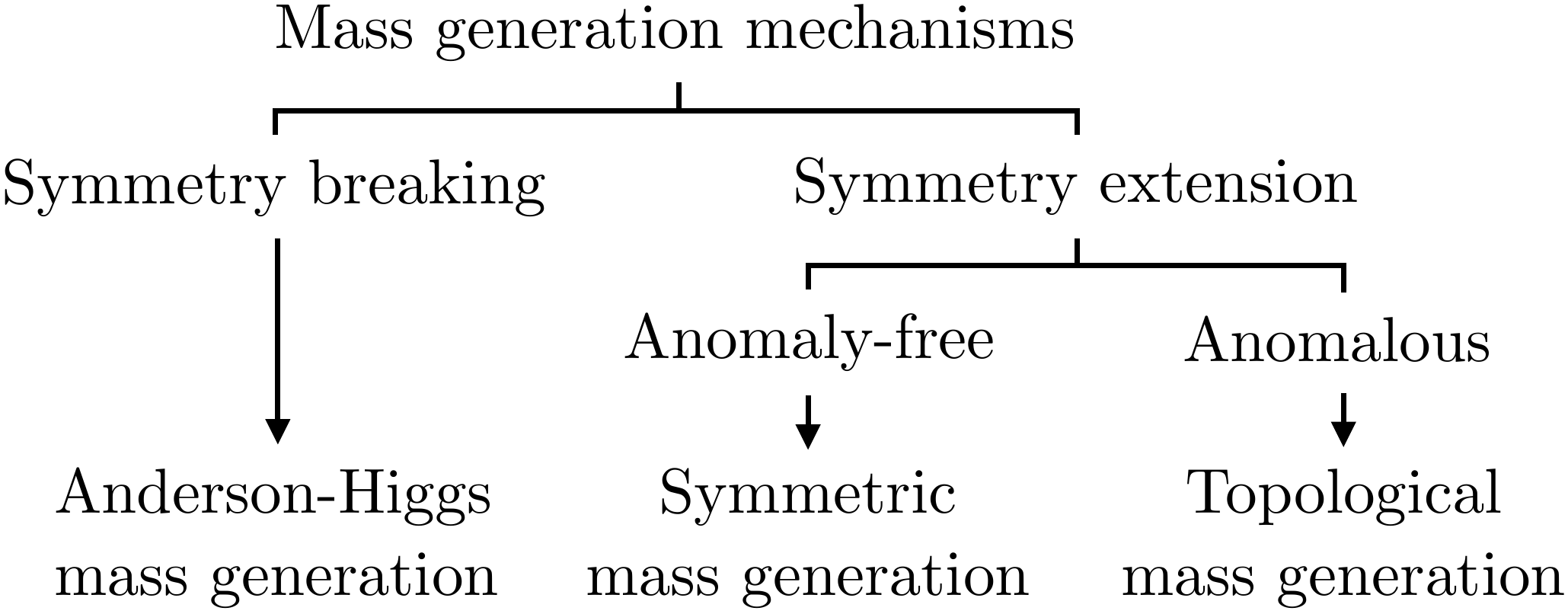}
\caption{Classification of mass generation mechanisms.}
\label{fig:MG}
\end{center}
\end{figure}

From the symmetry perspective, the fermion mass generation generally falls into two categories: (i) the \emph{symmetry breaking} mechanism and (ii) the \emph{symmetry extension} mechanism \cite{Witten:2016yb, Wang2018Symmetric}, see \figref{fig:MG}. At a high level, the SMG is essentially an \emph{symmetry extension} mechanism 
in the \emph{absence} of the 't Hooft anomaly. (Strictly speaking, the SMG mechanism does not enlarge the physical symmetry, but rather extends it to a symmetry-gauge structure. Nevertheless, we will still call it by symmetry extension in a general sense.)

\begin{enumerate}[leftmargin=2.mm]

\item 
{\bf Symmetry/Gauge breaking}: Anderson-Higgs mechanism, chiral symmetry breaking, 
Dirac mass and Majorana mass are induced by the {symmetry breaking} --- breaking either 
global symmetries or gauge structures, by condensing a Yukawa-Higgs field that couples to a fermion bilinear mass term.
More precisely, starting from a symmetry group $G$ (specifically here an internal symmetry, global or gauged), $G$ is broken down to an appropriate subgroup $G' \subseteq G$ to induce quadratic mass term for fermions. Mathematically, it can be described by an \emph{injective} homomorphism ${\iota}$:
\eq{G' \xrightarrow{\iota} G.}

Here are some explicit examples: \\
$\bullet$ Bardeen-Cooper-Schrieffer (BCS) type $\Z_2$-gauged superconductor with a low energy $\Z_2$ TQFT, we have $G'=\Z_2$ and $G=\U(1)$ electromagnetic gauge group. \\
$\bullet$ The Standard Model electroweak Higgs mechanism breaks 
$G = G_{\text{SM}_q} \equiv (\SU(3) \times   \SU(2) \times \U(1)_{\text{Y}})/\Z_q$ with $q=1,2,3,6$ and the appropriate greatest common divisor (gcd), down to $G'=(\SU(3) \times \U(1)_\text{EM})/\Z_{\gcd(q,3)}$.\\
$\bullet$ (3+1)D Dirac mass pairs two Weyl fermions ($\psi_L$ and $\psi_R$) via 
the Dirac mass term $m_{\rm D}(\psi_L^\dagger \psi_R+\psi_R^\dagger \psi_L)$ 
which breaks the unitary internal $G=\U(2)$ symmetry of two Weyl fermions
down to a vector $G' = \U(1)$ symmetry.\\ 
$\bullet$ (1+1)D Dirac mass pairs two Weyl fermions ($\psi_L$ and $\psi_R$) via 
the Dirac mass term $m_{\rm D}(\psi_L^\dagger \psi_R+\psi_R^\dagger \psi_L)$ 
which breaks the unitary internal $G=\U(1)_L \times \U(1)_R$ symmetry of two Weyl fermions
down to a vector $\U(1)$ symmetry, so $G' = \U(1)$.\\
$\bullet$ (3+1)D Majorana mass pairs a single Weyl fermion $\psi_L$ to itself,
 $\psi_R=\ii \sigma^2 \psi_L^{*}$, 
so the Majorana mass term $m_{\rm M} \bar \psi \psi = 
m_{\rm M} (\psi_L^\dagger (\ii \sigma^2) \psi^{*}_L
+{\psi^{\rm T}_L} (-\ii \sigma^2) \psi_L)$,
which breaks the unitary internal $G=\U(1)$ symmetry of a Weyl fermion
down to a fermion parity $G' = \Z_2^F$ symmetry.

\item 
{\bf Symmetry/Gauge extension}: In contrast, the {symmetry extension} \cite{Wang2018Symmetric} or gauge enhancement \cite{Wang2019Gauge} provides a fermion mass generation mechanism that preserves the symmetry. 
Symmetry extension construction of gapped phases 
first appears in \cite{Witten:2016yb} based on the gauge bundle descriptions,
then \refcite{Wang2018Symmetric} refines the idea to the lattice models, 
group-cohomology cocycle or continuum field theory descriptions.
It extends the symmetry group $G$ (that can include the spacetime-internal symmetry, global or gauged) to a larger group $\tilde{G}$ by enlarging the Hilbert space with additional/redundant degrees of freedom, in order to trivialize the 't Hooft anomaly or to lift any other symmetry obstruction towards a gapped phase with a mass generation. Mathematically, it can be described by a \emph{surjective} homomorphism ${r}$:
\bea \label{eq:pullback}
\tilde G \xrightarrow{r} G,
\eea
which can be understood as part of the group extension in an exact sequence (see \eqnref{eq:Gext}) \cite{Wang2018Symmetric}, or can be further generalized as the fibrations of their classifying spaces and higher classifying spaces 
\cite{Tachikawa2017gyf1712.09542, Wang1801.05416, Wan2018djlW2.1812.11955}. The symmetry extension can induce the \emph{topological mass} in the presence of anomaly, and the \emph{symmetric mass} in the absence of anomaly \cite{Prakash2021Unwinding}.

$\bullet$ {\bf Topological Mass Generation (TMG)} --- If the fermion system has a non-trivial 't Hooft anomaly in $G$, the anomaly will post an obstruction toward \emph{trivial} gapping, which already rules out SMG, leaving the possibility to TMG. A non-vanishing \emph{perturbative local anomaly} disallows any symmetric gapped phase (even with topological order), also it can never be trivialized by a symmetry extension. So, in order to implement the symmetry extension construction, 
the non-vanishing $G$-anomaly must be a \emph{nonperturbative global anomaly} in $G$. 

For simplicity, the discussion below focuses on a limited special case of \eqref{eq:pullback}.
If the global anomaly in $G$ can become anomaly-free 
in $\tilde{G}$, by pulling the $G$ group back to the extended $\tilde{G}$ group via a short exact sequence
\eq{\label{eq:Gext} 1\to K\to\tilde{G} \xrightarrow{r} G \to 1,}
and if the normal subgroup $K$ is a discrete finite group, then the fermion can acquire a topological mass upon gauging $K$,
which gives rise to a discrete $K$ gauge TQFT 
in appropriate spacetime dimensions and under appropriate criteria detailed in \secref{sec:more-on-sym-extension}. 
This is also called the group extension of the original quotient group $G$ extended by a normal subgroup $K$ to the total group $\tilde{G}$.
The topological mass refers to the energy gap of a TQFT matching the 't Hooft anomaly of $G$. Finding the group extension $\tilde{G}$ often requires the essential use of algebraic topology criteria, such as the Lydon-Hochschild-Serre spectral sequence method \cite{Wang2018Symmetric}. 
\refcite{Wang2018Symmetric, Wang1801.05416, Prakash2018Unwinding, Guo2020Fermionic, Prakash2021Unwinding, 
KobayashiOhmoriTachikawa1905.05391} provide several explicit lattice Hamiltonian
or lattice path integral constructions.

$\bullet$ {\bf Symmetric Mass Generation (SMG)} --- In the case of SMG, the fermion system is already anomaly free, but the physical symmetry $G$ is too restrictive to allow any symmetric fermion-bilinear mass. However, with the symmetry extension described by the following short exact sequence
\eq{1\to K \to\tilde{G}=G\times K \xrightarrow{r} G \to 1,}
the fermion-bilinear mass (i.e.~the parton-Higgs field $\upphi$) can be charged under both $G$ and $K$, which can possibly be condensed, breaking only $K$ to its subgroup $K'$ without breaking $G$, as long as the condensed fermion-bilinear mass transform trivially under $G\times K'\subseteq G\times K$. Moreover, the extended normal subgroup $K$ (and $K'$) can be (and will always be) gauged, such that the actual physical symmetry will not be enlarged by the symmetry extension. This is simply a rephrasing of the parton-Higgs mechanism in the fermion fractionalization framework discussed in \secref{sec:fermion_frac}. 

\begin{figure}[htbp]
\begin{center}
\includegraphics[width=\columnwidth]{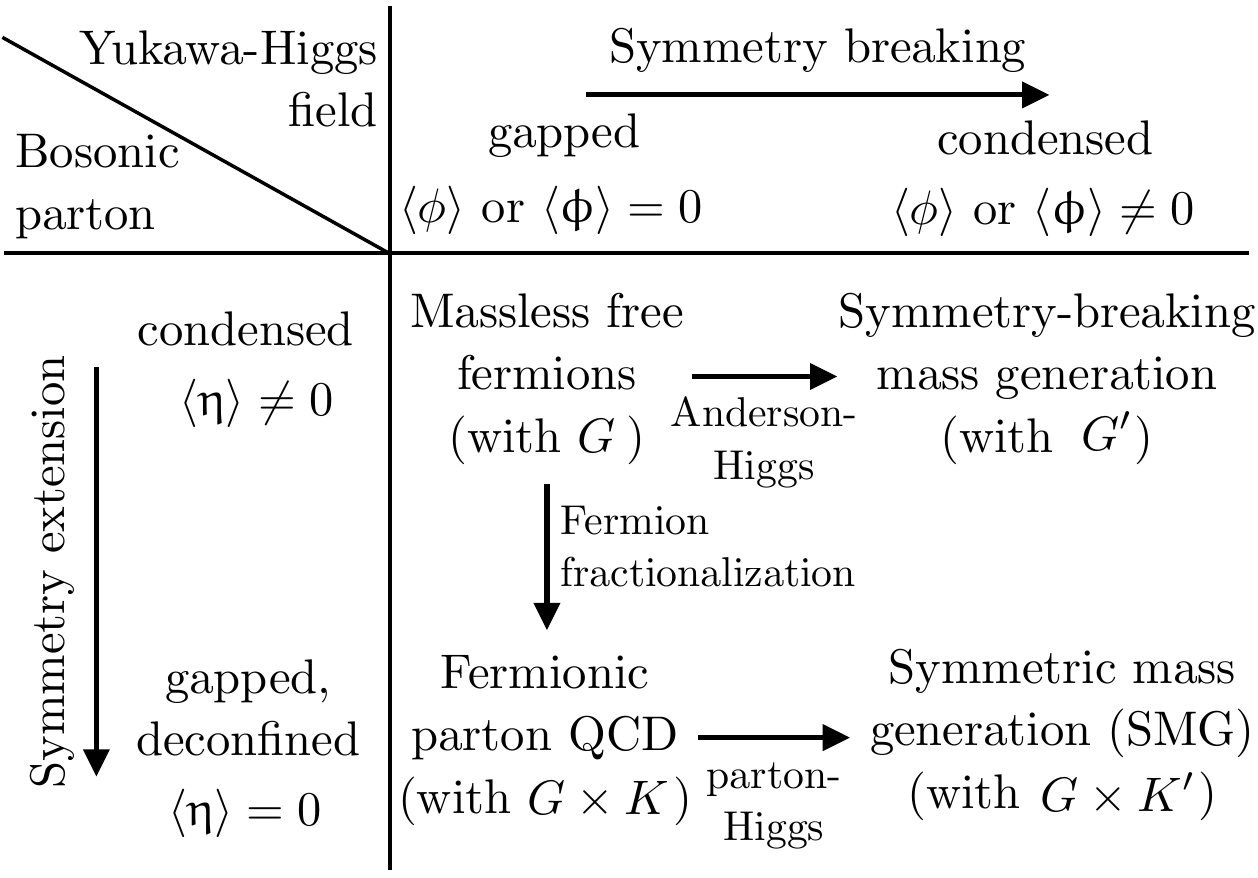}
\caption{Symmetry-gauge groups in different phases of the Yukawa-Higgs field and the bosonic parton field.}
\label{fig:sym}
\end{center}
\end{figure}

\figref{fig:sym} concludes how the symmetry-gauge group is extended/broken in different phases. A few general requirements for the SMG to happen are summarized as follows:
\begin{itemize}
\setlength\itemsep{0pt}
\item The full gauge group $K$ must be large enough to counteract any non-trivial action of the symmetry group $G$ on the parton-Higgs field $\upphi$, i.e.~$G$ must acts \emph{projectively} on $\upphi$, such that the $G$ symmetry can remain unbroken under the condensation of $\upphi$.
\item To achieve SMG in this framework, the deformation path must pass through the strong coupling regime where neither the bosonic parton $\upeta$ nor the parton-Higgs field $\upphi$ has an expectation value, and the full gauge group $K$ is unbroken.
\item After the parton-Higgs field $\upphi$ condenses, the remaining unbroken gauge group $K'$ would better be either trivial or non-Abelian, such that either there is no residual $K'$ gauge fluctuation or the residual $K'$ gauge fluctuation can be confined automatically. Otherwise, if $K'$ is Abelian, it becomes possible that the SMG critical point will expand into a critical phase, described by an Abelian $K'$-gauge theory.
\end{itemize}

\end{enumerate}

\subsubsection{More on Symmetry Extension Construction}
\label{sec:more-on-sym-extension}

The {\bf symmetry extension construction} 
$\tilde G \xrightarrow{r} G$ in \eqref{eq:pullback}
based on the pullback $G$-symmetry to the extended $\tilde G$-symmetry
can be interpreted 
in different languages for different communities \cite{Wang2018Symmetric}:

\noindent
$\bullet$ For condensed matter, a nontrivial SPT state in the $G$ symmetry cannot 
be deformed to a trivial tensor product state via a finite-depth of local unitary transformations without breaking the $G$-symmetry. But the successful symmetry extension means that we can find an appropriate $\tilde{G}$
such that the SPT state in the extended $\tilde{G}$ symmetric Hilbert space 
can  be deformed to a trivial tensor product state via a finite-depth of local unitary transformations still 
preserving the $\tilde{G}$-symmetry.

\noindent
$\bullet$ For quantum field theory or high-energy physics,
the successful symmetry extension means that the 't Hooft anomaly in $G$-symmetry
becomes anomaly-free in $\tilde{G}$-symmetry.

\noindent
$\bullet$ For mathematics,
the successful symmetry extension means that a nontrivial class of cocycle, cohomology, or cobordism 
of the $G$-symmetry
becomes a trivial class in the $\tilde{G}$-symmetry.
Suppose the nontrivial class of $G$-symmetry cocycle, cohomology, or cobordism in the $D$ dimensions 
is labeled by $\omega_{D}^G$, then the trivialization means that its pullback (namely $r^* \omega_{D}^G$)
becomes a $\tilde{G}$-symmetry coboundary in the $D$ dimensions 
(namely $\omega_{D}^{\tilde{G}}=\delta \beta_{D-1}^{\tilde{G}}$) 
which splits to the $\tilde{G}$ cochain (namely $\beta_{D-1}^{\tilde{G}}$) 
in the $D-1$ dimensions \cite{Wang2018Symmetric}. 
In summary, given the $\omega_{D}^G$ of the $G$-symmetry,
the successful symmetry extension requires
to find a solution of both the extended $\tilde{G}$ and $\beta_{D-1}^{\tilde{G}}$ to satisfy
\bea
r^*\omega_{D}^G = \omega_{D}^{\tilde{G}}=\delta \beta_{D-1}^{\tilde{G}}.
\eea
Here the bulk dimension is $D=d+1$, while the theory (on the boundary) with 't Hooft anomaly has its dimension $D-1=d$. 
Below some examples of the symmetry extension construction of gapped phases in various dimensions
based on the simplest short exact sequence in \eqref{eq:Gext} 
$$
1\to K\to\tilde{G} \xrightarrow{r} G \to 1
$$
with a finite group $K$, are considered. 
In particular, two issues should be addressed:\\
(1) Given a theory with 't Hooft anomaly in $G$ symmetry 
(which we shall also call it a $d$D anomalous boundary theory of a $(d+1)$D bulk SPT), can $\tilde{G}$ be found to trivialize the anomaly? If so, what is the minimal $\tilde{G}$? \\
(2) If the $\tilde{G}$ is found, there are two implications of the construction based on \eqref{eq:Gext} \cite{Wang2018Symmetric}: \\
$\bullet$ {\bf $\tilde{G}$-symmetric extended gapped phase} 
(as a $\tilde{G}$-symmetric gapped boundary of the bulk $G$-SPT). 
In this case, all $K, \tilde{G}$ and $G$ are not dynamically gauged.\\
$\bullet$ {\bf ${G}$-symmetric gapped dynamical $K$-gauge theory
with a 't Hooft anomaly in $G$}. 
When $K$ is dynamically gauged, 
in some cases, the $G$ is preserved at IR; in other cases, the $G$ becomes spontaneous symmetry breaking (SSB) at IR. One should also be careful to distinguish the two different kinds of dynamics.\\

\refcite{Wang2018Symmetric} proves that ``\emph{given any unitary or anti-unitary finite group $G$,
for $d \geq 1$,
there always exists a finite $K$-extension to 
a $\tilde{G}$-symmetric extended gapped phase to trivialize 
any nonperturbative global anomaly in $G$}
(see also the proof given later in \cite{Tachikawa2017gyf1712.09542}).''
It is physically meaningful to consider gauging $K$ for $d \geq 2$. 
Once $K$ is dynamically gauged, \refcite{Wang2018Symmetric} 
finds that \emph{${G}$-symmetric gapped dynamical $K$-gauge theory
with a 't Hooft anomaly in $G$} can only be obtained for $d \geq 3$,
but also finds that the $G$-symmetry is spontaneously broken for $d=2$.

The above result is established when $G$ is an ordinary 0-symmetry that acts on the 0D point operator. The symmetry extension construction can also be generalized to higher global symmetries \cite{Gaiotto2014kfa1412.5148}
such as the 1-symmetry that acts on the 1D line operator (see a recent review \cite{McGreevy2204.03045}).
Here are some examples (Table \ref{Table:G} for a summary) including ordinary and higher symmetries:


\begin{table*}[!thb]
		\begin{center}
			\hspace{-12.5mm}
			\begin{tabular}{|c|c|c| c| c| c|}
				\hline
				$\begin{array}{c} 
				\text{Dim}\\ 
				\text{$d$}
				\end{array}$
				& 
				\multicolumn{2}{|c|}
				{Original System. $G$} 
				& $K \to \tilde{G} \to G$   & 
					\setlength{\extrarowheight}{2pt}
				$\begin{array}{c} \text{Reduced class in $\tilde{G}$}\\ 
				\text{of $d+1$D $G$-iTQFT}\\
				\text{or $d$D $G$-anomaly} \end{array}$
				&
				$\begin{array}{c} \text{Gauge $K$}\\ 
				\text{Dynamics} 
				\end{array}$\\
				\hhline{======}
				\multirow{5}{*}[-16pt]{0+1} & 
				\multirow{2}{*}{
				$
				\begin{array}{c} 
				\text{bdry of Haldane chain:}\\
				\text{a doublet or qubit}
				\end{array}
				$}
				& {$\SO(3)$} & 
				$\Z_2 \to \SU(2)   \to G$ 
				& \multirow{2}{*}{$\mathbb{Z}_2 \Rightarrow 0$}  & 
				\multirow{5}{*}[-16pt]{No}\\
				\cline{3-4}
				 & & 
				 {$\Z_2^T$} & 
				$\Z_2 \to \mathbb{Z}_4^T   \to G$ 
				& &\\ 
				\cline{2-5}
				 & 
				{$
				\begin{array}{c} 
				\text{bdry of 4 Kitaev chains:}\\
				\text{4 Majorana modes}
				\end{array}
				$}
				 & \multirow{2}{*}{$\Z_2^T \times \Z_2^F$} & 
				$\Z_2 \to \mathbb{Z}_4^T \times \Z_2^F  \to G$ 
				& $\mathbb{Z}_8 \Rightarrow \mathbb{Z}_4$ &\\ 
				\cline{2-2} \cline{4-5}
			     &  
			     {$
				\begin{array}{c} 
				\text{bdry of 2 Kitaev chains:}\\
				\text{2 Majorana modes}
				\end{array}
			     $}
			     &  & $\Z_2 \to \mathbb{D}_8^{F,T} \to G$  & $\mathbb{Z}_8 \Rightarrow \mathbb{Z}_2$ & 
			              \\ 
			    \cline{2-5}	
				 & 
				$\begin{array}{c} 
				\text{bdry of 2 Kitaev chains:}\\
				\text{2 Majorana modes}
				\end{array}$
				 & $\Z_4^{TF}$ & $\Z_4 \to  \mathbb{M}_{16}^{F,T} \to G$  & $\mathbb{Z}_2 \Rightarrow 0$ & \\
				\hline
				1+1 &
				$\begin{array}{c} 
				\text{bdry of (2+1)D CZX model:}\\
				\text{(1+1)D edge modes}
				\end{array}$
				& 
				$\Z_2$
				& 
				$\begin{array}{c}
				\text{$\Z_2 \to  \Z_4 \to G$}\\
				\text{Symm-extension, but SSB.}
				\end{array}
				$
				& $\Z_2 \Rightarrow 0$ & $G$ SSB\\
				\hline
				\multirow{3}{*}[-10pt]{2+1} &
				$\begin{array}{c} 
				\text{bdry of (3+1)D $w_1(TM)^4$ SPT:}\\
				\text{(2+1)D surface state}
				\end{array}$
				& 
				$\Z_2^T$
				& 
				$\Z_2 \to  \Z_4^T \to G$
				& $\Z_2 \Rightarrow 0$ & 
				\multirow{1}{*}[+4.8pt]{
				$
				\begin{array}{c} 
				\text{$G$-symmetric}\\
				\text{$K$-gauge TQFT}
				\end{array}
				$
				}
				\\
				\cline{2-6}
				& 
				\multirow{2}{*}[-6pt]
		    	        {$
				\begin{array}{c} 
				\text{bdry of (3+1)D $k\mathcal{P}(B_{2})$ higher-}\\
				\text{SPT: (2+1)D surface state}
				\end{array}$
				}
				& 
				\multirow{2}{*}[-6pt]{$\SO \times \Z_{2,[1]}$}
				& 
				$\begin{array}{c}
				\text{even $k$:}\\
				\Z_2 \to \Spin \times \Z_{2,[1]} \to G\\
				\text{but SSB}
				\end{array}
				$
				& 
				\multirow{2}{*}[-6pt]{$k \in \Z_4 \Rightarrow k \in \Z_2$}
				& $G$ SSB 
				\\
				\cline{4-4} \cline{6-6}
				& & & 
				$\begin{array}{c}
				\text{odd $k$:}\\
				\text{No symm-extension.}
				\end{array}
				$
				&  & No
				\\
				\hline
				\multirow{7}{*}[-14pt]{3+1} &
				$\begin{array}{c} 
				\text{bdry of (4+1)D $A^5$ SPT:}\\
				\text{(3+1)D bdry state}
				\end{array}$
				& 
				$\Z_2^T$
				& 
				$\Z_2 \to  \Z_4 \to G$
				& $\Z_2 \Rightarrow 0$ & 
				\multirow{3}{*}[-8pt]{
				$\begin{array}{c} 
				\text{$G$-symmetric}\\
				\text{$K$-gauge TQFT}
				\end{array}$
				}
				\\
				\cline{2-5}
				&
				$\begin{array}{c} 
				\text{bdry of (4+1)D $w_3(TM) B_2^e$}\\
				\text{higher-SPT}
				\end{array}$
				& 
				$\SO \times \Z_{2,[1]}^e$
				& 
				$\Z_2 \to  \Spin \times \Z_{2,[1]}^e \to G$
				& $\Z_2 \Rightarrow 0$ & \\
				\cline{2-5}
				& 
				{\multirow{2}{*}[-6pt]
		    	        {$
				\begin{array}{c} 
				\text{bdry of (4+1)D $k A \mathcal{P}(B_{2}^e)$}\\
				\text{higher-SPT}
				\end{array} 
				$}
				}
				&
				{\multirow{2}{*}[-6pt]{
				$\begin{array}{c} 
				\text{$\Spin \times_{\Z_2^F} \Z_8$}\\ 
				\text{$\times \Z_{2,[1]}^e$}
				\end{array} 
				$}}
				& 
				$\begin{array}{c}
				\text{even $k$:}\\
				\Z_2 \to \Spin \times \Z_8 \times \Z_{2,[1]}^e \to G
				\end{array}
				$
				& 
				\multirow{2}{*}[-6pt]{$k \in \Z_4 \Rightarrow k \in \Z_2$}
				&  
				\\
				\cline{4-4} \cline{6-6}
				& & & 
				$\begin{array}{c}
				\text{odd $k$:}\\
				\text{No symm-extension.}\\
				\text{No-go obstruction.}
				\end{array}
				$
				&  & No
				\\
				\cline{2-6}
				& {$
				\begin{array}{c} 
				\text{bdry of (4+1)D $\frac{1}{2} \tilde{w}_1(TM) \mathcal{P}(B_{2}^e)$}\\
				\text{higher-SPT:} \\
				\text{(3+1)D $\SU(2)_{\theta=\pi}$ YM}
				\end{array} 
				$}  &
				$\Z_2^T \times \Z_{2,[1]}^e$
				 &
				 $\begin{array}{c}
				\text{$\Z_2 \to \Z_2^T \times \Z_{4,[1]}^e \to G$}\\
				\text{Symm-extension, but SSB.}
				\end{array}
				$
				 &
				 {$\Z_2 \Rightarrow 0$} & $G$ SSB
				\\
				\cline{2-6}				
				& 
				\multirow{2}{*}[0pt]{$
				\begin{array}{c} 
				\text{bdry of (4+1)D}\\ 
				\text{$(-N_f)\eta_{\rm 4d}(\PD(A))$-SPT:}\\
				\text{(3+1)D $15N_f$-fermion SM}
				\end{array} 
				$}
				 & 
				 \multirow{2}{*}[0pt]{
				 $
			         \Spin \times_{\Z_2^F} \Z_{4,X}
				$}
				& 
				$\begin{array}{c}
				\text{even $N_f$:}\\
				\Z_2 \to \Spin \times \Z_4 \to G,\\
				\Z_2 \to \Spin \times \Z_8 \to \Spin \times \Z_4
				\end{array}
				$
				& 
				\multirow{2}{*}[-8pt]{ 
			         {$\Z_{16} \Rightarrow \Z_2$}
				}
				&
				{
				$\begin{array}{c} 
				\text{$G$-symmetric}\\
				\text{$K$-gauge TQFT}
				\end{array}$
				}\\
				\cline{4-4}	 \cline{6-6}	
				&  & 
				& 
				$\begin{array}{c}
				\text{odd $N_f$:}\\
				\text{No symm-extension.}\\
				\text{No-go obstruction.}
				\end{array}
				$ 
				& & No\\
				\hline
			\end{tabular}
		\end{center}
		\caption{Summary of the symmetry-extension construction based on 
	 $1\to K\to\tilde{G} \xrightarrow{r} G \to 1$ in \eqref{eq:Gext}. 
	 The first and second column show the spacetime dimension $d$ and the $G$-anomalous theory (as a boundary [bdry] of a bulk $d+1$D SPT). 
	 The third column shows whether the $\tilde G$-symmetry-extended gapped phase exists.
	 The fourth column shows how the boundary $G$-anomaly or bulk $G$-SPT classification is reduced in $\tilde G$.
	 The last column shows the dynamics after gauging $K$: Even if $\tilde G$-symmetry-extended gapped phase exists,
	 the $K$-gauged dynamics can induce either a $G$ spontaneous symmetry-breaking (SSB)
	 or a $G$-symmetric $K$ gauge theory.
}
		\label{Table:G}
\end{table*}


%
\begin{enumerate}[leftmargin=2.mm]
\item $d=1$, a (0+1)D anomalous theory and a (1+1)D bulk:

$\bullet$ Two Kitaev's chains:
{In \secref{sec:FK}, the discussion was limited to gapping 
(0+1)D eight Majorana modes or (1+1)D eight Kitaev's chains by the SMG preserving the $\Z_2^{T}\times\Z_2^{F}$,
which is free from the $\dsZ_8$ global anomaly. Beyond SMG, even for 
(0+1)D two Majorana modes or (1+1)D two Kitaev's chains with $2 \mod 8$ class of $\Z_8$ global anomaly,
it is still possible to gap the whole system, leaving a single ground state \emph{without breaking} $G=\dsZ_2^{T}\times\dsZ_2^{F}$, but instead by
\emph{extending} the symmetry to a dihedral group of order 8 as $\tilde{G}=\mathbb{D}_8^{F,T} \equiv\dsZ_4^{T}\rtimes\dsZ_2^{F}$ \cite{Prakash2021Unwinding}. Namely, when $T^2=+1$ is extended to a time-reversal symmetry fictionalization 
$T^2=-1$ and $T^4=+1$, the $\tilde{G}=\mathbb{D}_8^{F,T}$ symmetric interactions can lift up the degenerate Majorana zero modes.
Crucially the time-reversal $\dsZ_2^{T}$ generator $T$ does not commute with the fermion parity $\Z_2^{F}$ generator
$(-1)^F$, but $T(-1)^F T^{-1}= - (-1)^F$. This means that $T$ switches a bosonic sector $| B \rangle$ and a fermionic sector $| F \rangle$ in the Hilbert space.
This is called \emph{supersymmetry extension} that trivializes this (0+1)D fermionic anomaly and also trivializes the (1+1)D fermionic SPT state
\cite{Prakash2021Unwinding}. Another related property is that 
the preserved symmetry demands the anomalous boundary theory of $\pm 2 \mod 8$ Majorana zero modes must be $\scN=2$ supersymmetric quantum mechanics with two supercharges \cite{PrakashWang2011.12320}.
}

Two Kitaev's chains can also allow $G=\Z_4^{TF}$ symmetry with $T^2=(-1)^F$ \cite{Gu1308.2488}.
When $T^2=(-1)^F$ is extended to a time-reversal symmetry fictionalization 
$T^4=-1$ and $T^8=+1$, a non-abelian order-16 finite $\tilde{G}=\mathbb{M}_{16}^{F,T}$ symmetric interactions can lift up the degenerate Majorana zero modes \cite{PrakashWang2011.12320}.

$\bullet$ Four Kitaev's chains and a Haldane chain:
{For (0+1)D four Majorana modes or (1+1)D four Kitaev's chains, its $4 \mod 8$ class of $\Z_8$ global anomaly is actually equivalent to
a single (1+1)D Haldane's chain \cite{You:2014ho} (tensor product with a trivial gapped fermionic product state) 
with $1 \mod 2$ class of $\Z_2$ global anomaly. 
The $G=\Z_2^T$-symmetric Haldane chain can be trivialized in a bosonic $\tilde{G}=\Z_4^T$ \cite{Prakash2018Unwinding}.
The $G=\SO(3)$-symmetric Haldane chain can be trivialized in a $\tilde{G}=\SU(2)$ \cite{Prakash2018Unwinding}.
}

$\bullet$ Related studies on the fractionalized symmetries on the boundary of the layers of Kitaev chains
can also be found in \cite{Gu1308.2488, Dijkgraaf1804.03275, Montero2008.11729, Turzillo2012.04621, Delmastro2101.02218}.
In particular, \refcite{Dijkgraaf1804.03275, Turzillo2012.04621} studies the pure $\Z_2$ class global gravitational anomaly 
on the boundary of a single Kitaev chain (which is an invertible fermionic topological order beyond the SPT,
known as the mathematical Arf invariant). 
The pure gravitational anomaly cannot be trivialized by any symmetry extension, 
but may be ``trivialized'' by coupling to a gravitational theory \cite{Dijkgraaf1804.03275}.

Here a (0+1)D theory has no parity $P$ but only at most time-reversal $T$, and Majorana fermion has no charge conjugation $C$ symmetry;
so only the $T$ fractionalization is found. In higher dimensions, the common theme along the direction of 
this phenomenon is the $C$-$P$-$T$ fractionalization \cite{Wang2109.15320}.

\item $d=2$, a (1+1)D anomalous theory and a (2+1)D bulk:

$\bullet$ The (1+1)D edge of a (2+1)D CZX model as a $\Z_2$-SPT state
is known to allow a symmetric gapless or a symmetry-breaking gapped boundary \cite{ChenLiuWen1106.4752}.
However, the $G=\Z_2$ can be extended to give a $\tilde{G}=\Z_4$-symmetry-extended gapped boundary.
Unfortunately, gauging the normal subgroup $K=\Z_2$ results in a (1+1)D discrete $K$-gauge theory 
with $G=\Z_2$ spontaneous symmetry breaking,
which is consistent with the standard lore that there is no (1+1)D non-invertible intrinsic topological order, at least in the bosonic systems.

Other applications of the symmetry-extension construction on the (1+1)D gauge theories and orbifolds can be found in a recent survey \cite{Sharpedecomposition2204.09117}.

\item $d=3$, a (2+1)D anomalous theory and a (3+1)D bulk:

$\bullet$ The (2+1)D surface of a (3+1)D $\Z_2^T$-SPT state (topological superconductor)
allows a symmetric gapless, symmetry-breaking, or symmetric gapped surface topological order boundary \cite{Vishwanath2013Physics,WangSenthil1302.6234, Kapustin1404.6659}.
The $G=\Z_2^T$ can be extended to give a $\tilde{G}=\Z_4^T$-symmetry-extended gapped boundary \cite{Wang2018Symmetric}.
Gauging the normal subgroup $K=\Z_2$ results in a (2+1)D discrete $K$-gauge theory with both electric and magnetic gauge charges
are Kramers doublet with $T^2=-1$.

$\bullet$ The (2+1)D surface state of (3+1)D $k\mathcal{P}(B_{2})$ higher-SPT state with $k \in \Z_4$:
This higher-SPT state is protected by a 1-form symmetry (denoted as $G=\SO \times \Z_{2,[1]}$) which couples to a 2-form background field $B_{2}$.
The $\mathcal{P}(B_{2})$ means the Pontryagin square of $B_{2}$.
The symmetry-extension construction can obtain a gapped phase for an even $k$ via extending to $\tilde G=\Spin \times \Z_{2,[1]}$
(although gauging $K=\Z_2$ results in $G$ SSB), but
the symmetry-extension trivialization is proven to not exist for an odd $k$ \cite{Wan2018djlW2.1812.11955}.
Later \refcite{Cordova2019bsd1910.04962} proves a no-go theorem that the symmetry-preserving TQFT also does not exist for an odd $k$.
This means the (2+1)D surface state must be either symmetric gapless or symmetry-breaking  for an odd $k$.

\item $d=4$, a (3+1)D anomalous theory and a (4+1)D bulk:

$\bullet$ The (3+1)D boundary of a (4+1)D $\Z_2$-SPT state 
allows a symmetric gapless, symmetry-breaking, or symmetric gapped surface topological order boundary.
The $G=\Z_2$ can be extended to give a $\tilde{G}=\Z_4$-symmetry-extended gapped boundary \cite{Wang2018Symmetric}.
Gauging the normal subgroup $K=\Z_2$ results in a (3+1)D discrete $K$-gauge theory with both electric and magnetic gauge charges
carries a fractional $G$ charge.

$\bullet$ The (3+1)D boundary of a (4+1)D $w_3(TM) B_2^e$ higher-SPT state 
is protected by a 1-form electric symmetry (denoted as $G=\Z_{2,[1]}^e$, coupled to a 2-form $B_{2}^e$ field),
while the $w_j(TM)$ is the $j$th Stiefel-Whitney class of the tangent bundle $TM$ of spacetime manifold $M$.
The corresponding anomaly occurs as a part of the anomaly of (3+1)D SU(2) Yang-Mills theory coupled to
two Weyl fermions in the adjoint representation of SU(2) (below called this theory as the adjoint QCD$_4$, see \cite{Anber2018iof1805.12290, Cordova2018acb1806.09592DumitrescuClay, BiSenthil1808.07465} and \cite{Wan2018djlW2.1812.11955}).
The $G=\Z_{2,[1]}^e$ can be extended to give a $\tilde{G}=\Z_2^F \times \Z_{2,[1]}^e$-symmetry-extended gapped boundary \cite{Wan2018djlW2.1812.11955}.
Gauging the normal subgroup $K=\Z_2$ results in a (3+1)D discrete $K$-gauge theory 
such that its electric gauge charge has 
fermionic statistics.

$\bullet$ The (3+1)D boundary of a (4+1)D $k A \mathcal{P}(B_{2}^e)$
higher-SPT state: Again this is part of the anomaly of the adjoint QCD$_4$ \cite{Cordova2018acb1806.09592DumitrescuClay}.
The higher-SPT is protected by a $\Z_8$-axial symmetry 
(coupled to a 1-form $A$ field, 
with its fourth power of the symmetry generator equals to $(-1)^F$)
and
a 1-form electric symmetry $\Z_{2,[1]}^e$ (coupled to a 2-form $B_{2}^e$ field), which can be denoted as a spacetime-internal symmetry
$G=\Spin \times_{\Z_2^F}  \Z_8 \times \Z_{2,[1]}^e$. There is a $k \in \Z_4$ class.
The even $k$ class can be trivialized by a $\Z_2$-extension to 
$\tilde G=\Spin \times  \Z_8 \times \Z_{2,[1]}^e$.
The odd $k$ class cannot be trivialized by any symmetry extension \cite{Wan2018djlW2.1812.11955}.
Later \refcite{Cordova2019bsd1910.04962} proves a no-go theorem that the symmetry-preserving TQFT also does not exist.
The above results \cite{Wan2018djlW2.1812.11955, Cordova2019bsd1910.04962}
turn out to rule out certain UV-IR duality proposal of the adjoint QCD$_4$ hypothesized in \cite{BiSenthil1808.07465}.

$\bullet$ The (3+1)D boundary of a (4+1)D $\frac{1}{2} \tilde{w}_1(TM) \mathcal{P}(B_{2}^e)$
higher-SPT state: The boundary turns out to associate with the anomaly
of the (3+1)D SU(2) Yang-Mills gauge theory with a ${\theta=\pi}$ $F \wedge F$ topological term
(denoted as $\SU(2)_{\theta=\pi}$ YM) \cite{Gaiotto2017yupZoharTTT1703.00501,WanWangZheng1812.11968,Wan2019oyr1904.00994},
while the $\tilde w_1(TM)$ is the twisted first Stiefel-Whitney class of $TM$ such that $\frac{1}{2} \tilde{w}_1(TM) \mathcal{P}(B_{2}^e)$
is a mod 2 class.
The $\SU(2)_{\theta=\pi}$ YM kinematically at UV has time-reversal $\Z_2^T$ and 1-form $\Z_{2,[1]}^e$ symmetries. The 
$G=\Z_2^T \times \Z_{2,[1]}^e$ can be extended to $\tilde G=\Z_2^T \times \Z_{4,[1]}^e$ to trivialize the anomaly,
thus the $\tilde G$-symmetric extended gapped phase can be constructed \cite{WanWangZheng1812.11968, Wan2019oyr1904.00994}. 
However, upon gauging $K=\Z_{2,[1]}$,
this induces the $G$ spontaneous symmetry breaking (SSB) \cite{Wan2019oyr1904.00994}.
Later \refcite{Cordova2019bsd1910.04962} proves a no-go theorem that the symmetry-preserving TQFT also does not exist.
The above results together demand that the IR fate of $\SU(2)_{\theta=\pi}$ YM must be either symmetric gapless or symmetry-breaking only.

$\bullet$ The (3+1)D boundary of a (4+1)D 
$k\eta_{\rm 4d}(\PD(A))$ SPT state:
This is a $k \in \Z_{16}$ class of (3+1)D anomaly and (4+1)D SPT state,
protected by a unitary $\Z_{4,X}$ symmetry such that $X^2=(-1)^F$;
in terms of a spacetime-internal symmetry $G= \Spin \times_{\Z_2^F} \Z_{4,X}$.
The $\eta_{\rm 4d}(\PD(A))$ is a 4d Atiyah-Patodi-Singer eta invariant $\eta_{\rm 4d}$
evaluated on the 4-manifold Poincar\'e dual (PD) to the $(\Z_{4,X}/\Z_2^F)$-gauge field $A$.
The $\eta_{\rm 4d}(\PD(A))$ is a cobordism invariant of the bordism group $\Omega_5^{\Spin \times_{\Z_2^F} \Z_{4,X}}=\Z_{16}$, 
see \cite{GarciaEtxebarriaMontero2018ajm1808.00009, Hsieh2018ifc1808.02881} and \cite{Guo2020Fermionic}.
It turns out that $k=(-N_f)$ of such SPT state captures a global
anomaly of (3+1)D $15N_f$-Weyl-fermion Standard Model (SM) where $N_f$ is the number of families of quarks and leptons
 \cite{GarciaEtxebarriaMontero2018ajm1808.00009,WW2019fxh1910.14668}.
If $k$ is odd, \refcite{Cordova1912.13069} proves an obstruction, so the symmetry-gapped TQFT is not possible to saturate this odd $k$ anomaly.
If $k$ is even, \refcite{Hsieh2018ifc1808.02881, JW2006.16996, JW2012.15860} show that two layers of symmetry extensions
can construct the
$\tilde G = \Spin \times {\Z_{8}}$-symmetry extended gapped phase:
the first layer $ 1 \to \Z_2 \to \Spin \times {\Z_{4}} \to \Spin \times_{\Z_2^F} {\Z_{4,X}} \to 1$ and the second layer
$ 1 \to \Z_2 \to \Spin \times {\Z_{8}} \to  \Spin \times {\Z_{4}}  \to 1$. These constructions
may have applications to beyond the SM physics \cite{JW2006.16996, JW2008.06499, JW2012.15860}.
				
\end{enumerate}

\section{Features and Applications}
\label{sec:other}

\subsection{Green's Function Zeros}
\label{sec:zeros}

Given different mass generation mechanisms discussed above, a key question is how to diagnose the SMG mechanism: if a mass gap is observed in an interacting fermionic system, how do we know the mass gap is opened up through the SMG mechanism, rather than a more conventional symmetry-breaking mechanism, say by condensing a fermion-bilinear mass. 
Over the years it was gradually realized that one of the characteristic features of the SMG phase is that, the fermion Green's function $G(\ii\omega)=-\langle\psi(\ii\omega)\psi^\dagger(\ii\omega)\rangle$ in the Matsubara frequency space should have a zero at $\omega \rightarrow 0$, i.e. $\det G(\ii \omega \rightarrow 0) = 0$. This applies to the SMG phase either in the bulk or on the boundary of a fermionic system. 

There are various ways to argue the necessary existence of this Green's function zero in a system with SMG. For the Green's function defined in the bulk, one general argument for the existence of zero is based on the topological number associated with the free fermionic SPT state~\cite{Gurarie:2011gl,gurarie2,Kaplan2021Index}. 
For a fermionic system with translation symmetry, the topological number can be defined in the Matsubara frequency and momentum space using the Euclidean spacetime full fermion Green's function~\cite{volovik1,volovikbook,wanggreen1,wanggreen2,wanggreen3}:  
\eq{\label{eq:n}n \sim \int \dd\omega \dd^{{d-1}}k \ \mathrm{tr} [ B (G^{-1} \partial G) \wedge (G^{-1} \partial G) \cdots ],} where $G$ is the matrix of the full fermion Green's function, and $B$ is certain matrix in the flavor space. For a Chern insulator at even spatial dimension, $B$ is an identity matrix, and $n$ is equivalent to the Chern number. One can prove mathematically that the number $n$ must be a quantized integer, and it can only change discontinuously when the Green's function has singularity.

\begin{figure}[htbp]
\begin{center}
\includegraphics[width=0.72\columnwidth]{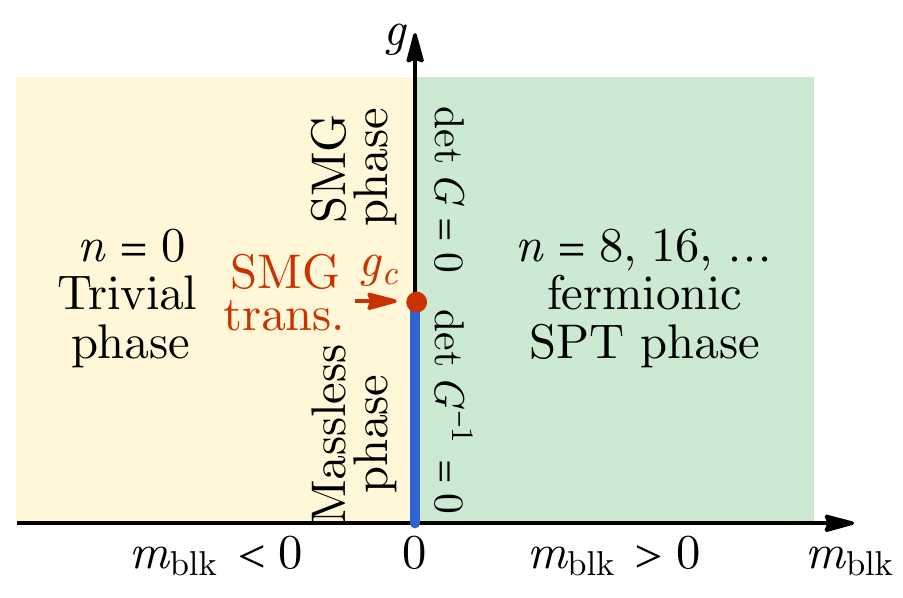}
\caption{Schematic phase diagram between a trivial phase and a (fake) fermionic SPT phase that can be trivialized by interaction. Massless bulk fermions along the critical line in the weak-coupling limit undergoes the SMG as the interaction $g$ exceeds a critical value $g_c$. Although the two phases (technically one phase) are smoothly connected through the strong-coupling regime, the topological number $n$ must still jump across the missing ``phase boundary'', along which the Green's function must have a zero.}
\label{fig:Gzero}
\end{center}
\end{figure}

The singularity of $G$ happens at two types of ``transitions". The first type of transition is a physical transition where $\det G^{-1}(\ii \omega = 0)=0$ vanishing to zero, i.e. the fermions become gapless. This corresponds to the quantum critical line between the fermionic SPT and trivial phases in the weak-coupling regime, as shown in \figref{fig:Gzero}.
However, note that in the definition of the topological number $n$ in \eqnref{eq:n}, $G^{-1}$ and $G$ appear on an equal footing, hence $n$ can also change when $\det G(\ii \omega = 0) = 0$, i.e.~when the Green's function has a zero. Hence when the free fermion SPT phase is trivialized by interactions, although there is no unavoidable phase transition between the SPT and the trivial phase, the topological number $n$ still has to change discontinuously somewhere in the phase diagram, as shown in \figref{fig:Gzero}. Since there is no real physical transition, the number $n$ has to change through the zero of the full interacting fermion's Green's function. This must hold throughout the SMG phase (line) regardless of the interaction strength $g$ (as long as $g>g_c$).

Computing the full fermion's Green's function for a given interacting Hamiltonian is generally a challenging task. To argue that the Green's function zero is a general feature of the SMG state independent of microscopic details, one relies on the \emph{topological defect condensation} construction for the SMG state, introduced in \secref{sec:FBM}. Starting with the SMG of eight Majorana zero modes in (0+1)D, the fermion Green's function can be explicitly computed~\cite{You:2014br,Slagle:2015lo}:  
\eq{
G_{ab}(\ii\omega) \sim \frac{\ii\omega
\delta_{ab}}{(\ii\omega)^2-m^2}, }
where $m$ is an effective mass proportional to the strength of the fermion interaction. Obviously, in (0+1)D, $G_{ab}(\ii\omega)$ approaches zero as $\ii\omega \rightarrow 0$. Now one can construct a higher dimensional SMG state by decorating every topological point defect of the Yukawa field with eight interacting Majorana fermions in its SMG state, and then proliferating the point defects to put the Yukawa field in a disordered phase, following the strategy of the decorated domain wall construction \cite{Chen2014Symmetry-protected}. The fermion Green's function can be evaluated in the spacetime by patching the (0+1)D Green's functions along the world line of the topological defect, and then path integrating all possible world line configurations. Following this approach, \refcite{Xu2021Greens} was able to show that the fermion Green's function takes the general form of
\eq{\label{eq:G}G_{ab}(k)\sim\frac{k_\mu\gamma^\mu\delta_{ab}}{k_\mu k^\mu-m^2},}
which universally exhibits a zero $G(k_\mu=0)=0$ in zero momentum-energy limit. Before the non-perturbative prove by \refcite{Xu2021Greens}, \eqnref{eq:G} was first obtained in \refcite{You:2014br} by a perturbative calculation, and later argued in \refcite{BenTov:2015lh,You2018Symmetric,Catterall2018Topology} using the fluctuating bilinear mass picture. 

%

\subsection{Deconfined Quantum Criticality}
\label{sec:DQCP}

While the Green's function zero provides a key diagnosis of the SMG phase, what about the diagnosis for the SMG transition? The theoretical framework of fermion fractionalization indicates that if the SMG transition is direct and continuous, it should be a deconfined quantum critical point (DQCP) \cite{You2018Symmetric}, where the physical fermions $\psi$ fractionalize into \emph{deconfined} bosonic $\upeta$ and fermionic $\uppsi$ partons \emph{at and only at} the critical point. 

The concept of DQCP \cite{Senthil:2004wj,Motrunich:2004hh,Senthil:2004qm}  was originally introduced to describe the direct continuous transition between the antiferromagnetic phase and the valence bond solid phase in (2+1)D quantum spin models. The two phases break distinct symmetries (spin-rotation and lattice-rotation symmetries respectively) and cannot be connected by a single quantum critical point without fine-tuning in the Landau-Ginzburg-Wilson paradigm. The DQCP provides an explanation for this exotic quantum critical point by fractionalizing the physical spin into deconfined spinons (partons) at and only at the critical point. The continuous SMG transition is also an exotic quantum critical point that requires the DQCP description. However, Unlike the conventional (bosonic) DQCP that fractionalizes bosonic degrees of freedoms, the continuous SMG transition is a \emph{fermionic} version of DQCP, as it fractionalizes fermionic degrees of freedoms.

\begin{figure}[htbp]
\begin{center}
\includegraphics[width=0.9\columnwidth]{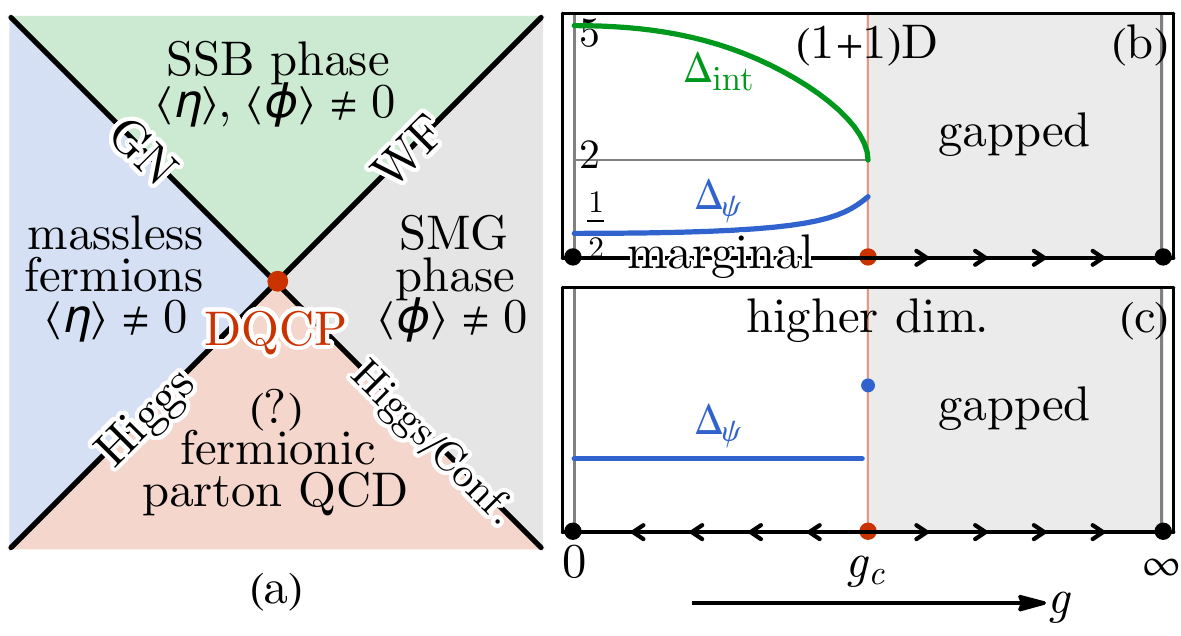}
\caption{(a) Schematic phase diagram determined by whether $\upeta$ and $\upphi$ are condensed or not. The phase transitions belongs to the following universality classes: Gross-Neveu (GN), Wilson-Fisher (WF), Higgs, or confinement (Conf.). The DQCP emerges at the intersection all phases. (b,c) Flow of coupling constant $g$ and the scaling dimensions for physical fermions $\Delta_\psi$ and/or the interaction term $\Delta_\text{int}$ in (b) the (1+1)D 3-4-5-0 model, (c) higher dimensional models.}
\label{fig:DQCP}
\end{center}
\end{figure}

The SMG critical point can be viewed as the intersection among four phases:
\begin{enumerate}[label=(\roman*),itemsep=0pt]
\item the massless fermion phase ($\langle\upeta\rangle\neq 0, \langle\upphi\rangle= 0$),
\item the SMG phase ($\langle\upeta\rangle= 0, \langle\upphi\rangle\neq 0$),
\item the spontaneous symmetry breaking (SSB) massive phase ($\langle\upeta\rangle\neq 0, \langle\upphi\rangle\neq 0$),
\item the fermionic parton QCD phase (if stable) ($\langle\upeta\rangle= 0, \langle\upphi\rangle= 0$).
\end{enumerate}
The four phases corresponds to the four different states of the bosonic parton $\upeta$ and the parton Higgs $\upphi$ fields, as summarized in \figref{fig:sym}. \figref{fig:DQCP}(a) shows a schematic phase diagram. The DQCP emerges at the origin of the phase diagram, where both $\upeta$ and $\upphi$ are critical. It seems that a direct transition between the massless fermion phase to the SMG phase requires fine tuning through the DQCP. However, if the fermionic parton QCD theory is dynamically unstable, the QCD phase will shrink to a single transition line between the massless fermion and the SMG phases, such that a direct continuous SMG transition can persist without fine tuning.

Strictly speaking, the DQCP picture for the SMG transition  is only valid and necessary in (2+1)D or higher dimensions. The SMG in lower dimensions (i.e.~(0+1)D and (1+1)D) are exceptional and can be understood without involving fermion fractionalization or DQCP. 
\begin{itemize}[itemsep=0pt]
\item In the (0+1)D Fidkowski-Kitaev model, the four-fermion interaction will immediately opens the gap, which can be understood by solving the quantum mechanical problem in \eqnref{eq:H_FK} exactly. There is no notion of phase transition and quantum criticality in (0+1)D, not to mention DQCP. 
\item In (1+1)D, take the 3-4-5-0 model for example, the SMG can be understood within the Luttinger liquid framework as a BKT transition. The six-fermion SMG interaction in \eqnref{eq:LLint_fermion} has a bare scaling dimension $\Delta_\text{int}|_{g=0}=5$ in the free fermion limit, which is  perturbatively irrelevant. However, with a non-perturbative (finite) interaction strength, under the RG flow, the Luttinger parameters (as exact marginal parameters) will be modified by the interaction, leading to the decrease of the scaling dimension of the interaction term. When the scaling dimension drops below 2 (which is the spacetime dimension), the SMG interaction will become relevant, driving the system into the SMG phase.\cite{Tong2021Comments,Zeng2022Symmetric} So the SMG transition is triggered right at $\Delta_\text{int}|_{g=g_c}=2$, see \figref{fig:DQCP}(b). 
\item In higher dimensions ((2+1)D and above), interactions are always perturbatively irrelevant at the free fermion fixed point, such that an infinitesimal interaction $g$ will not immediately drive the SMG transition. Therefore, the transition generally requires a finite critical interaction strength $g_c$. The critical point ($g=g_c$) is expected to be an unstable fixed point under RG, which either flows to the free fermion fixed point ($g=0$) or the SMG (gapped phase) fixed point ($g=\infty$), as illustrated in \figref{fig:DQCP}(c). 
\end{itemize}

One implication of fermion fractionalization at the DQCP is that the physical fermion scaling dimension will generally be larger at the SMG transition compared to the free fermion fixed point \cite{You2018Symmetric}. Massless free fermion (Dirac/Weyl/Majorana) in $d$-dimensional spacetime has the scaling dimension $\Delta_{\psi}|_{g=0}=(d-1)/2$. At the SMG transition point ($g=g_c$), the fermion fractionalizes to the critical bosonic parton $\upeta$ and the gapless fermionic parton $\uppsi$. In the large-$N$ limit, the scaling dimensions of partons are $\Delta_\upeta=d/2-1$ and $\Delta_\uppsi=(d-1)/2$, such that the leading order estimation of the physical fermion scaling dimension becomes $\Delta_{\psi}|_{g=g_c}=\Delta_\upeta+\Delta_\uppsi=d-3/2$. For $d>2$, the large-$N$ estimation implies
\eq{\Delta_{\psi}|_{g=g_c}>\Delta_{\psi}|_{g=0},}
i.e.~the fermion scaling dimension will jump to a higher value right at the SMG transition, as shown in \figref{fig:DQCP}(c). For $d=2$, the  above naive dimension counting seems to indicate $\Delta_{\psi}|_{g=g_c}=\Delta_{\psi}|_{g=0}$ (which is not correct, unless $g_c=0$). More careful Luttinger liquid RG analysis \cite{Zeng2022Symmetric} shows that the fermion scaling dimension increases continuously with the interaction until the transition happens, as shown in \figref{fig:DQCP}(b). Thus, the statement $\Delta_{\psi}|_{g=g_c}>\Delta_{\psi}|_{g=0}$ still holds for the (1+1)D model.

\subsection{Symmetric Mass Generation of the Standard Model}\label{sec:SM}

One important motivation driving the study of SMG is to seek for the lattice regularization of chiral gauge theories, in particular the Standard Model of particle physics. The goal is to gap the fermion doublers (mirror fermions) without affecting the original chiral fermions
\cite{Eichten1986Chiral, Wenanomalies1303.1803, Wen:2013kr, You:2015lj, Wang2018A-Non-Perturbative}. Possible routes to gap the Standard Model with either 15$N_f$ or 16$N_f$ Weyl fermions via the SMG are reviewed in the following, where $N_f$ stands for the family (or generation) number and can be taken to be $N_f=3$.


\subsubsection{Symmetry Extension of the 15$N_f$- or 16$N_f$-Weyl-Fermion Standard Model}

\label{sec:15vs16NfSM}

The (3+1)D Standard Model (SM) has a chiral internal symmetry group of 
the Lie algebra $\su(3) \times \su(2) \times \u(1)_{Y}$, which could correspond to either of the four versions of Lie groups for $q=1,2,3,6$:
\eq{G_{\SM_q} \equiv \frac{\SU(3) \times   \SU(2) \times \U(1)_{Y}}{\Z_q},}
which are all 
compatible with all known particle representation data to date.
The following discussion is applicable to any of the four versions of SM for $q=1,2,3,6$. 
For any specific version, the SM contains $N_f=3$ families (generations) of matter fermions. In each family, there can be either 15 or 16 left-handed Weyl fermions. The corresponding SM phases will be denoted as 15$N_f$-SM and 16$N_f$-SM respectively. The first 15 Weyl fermions transform in the following representations of $\su(3) \times \su(2) \times \u(1)_{Y}$
\bea \label{eq:SMrep}
&&\bar{d}_R \oplus {l}_L  \oplus q_L  \oplus \bar{u}_R \oplus   \bar{e}_R  \cr
&&=(\overline{\bf 3},{\bf 1})_{2} \oplus ({\bf 1},{\bf 2})_{-3}  
\oplus
({\bf 3},{\bf 2})_{1} \oplus (\overline{\bf 3},{\bf 1})_{-4} \oplus ({\bf 1},{\bf 1})_{6}\quad
\eea 
in each family. Here $u_R$ and $d_R$ are up and down types of right-handed quarks.
The $q_L$ is the $\su(2)$ doublet of up and down types of left-handed quarks.
The $e_R$ is the right-handed electron.
The $l_L$ is the $\su(2)$ doublet of neutrino and electron types of left-handed leptons. The 16th Weyl fermion corresponds to the sterile neutrino $\bar{\nu}_R=({\bf 1},{\bf 1})_{0}$, which can be appended to \eqnref{eq:SMrep} in any of the families.

For simplicity, all SM fermions are assumed to be massless, without including the electroweak 
Higgs and its symmetry breaking.
To apply SMG to 15$N_f$-SM or 16$N_f$-SM, 
one shall confirm that the two necessary conditions in \secref{sec:generalD} are satisfied.
For the second condition in \secref{sec:generalD}, it is true in the SM.
For the first condition, several recent works had checked the cobordism group classification of anomalies in the SM \cite{GarciaEtxebarriaMontero2018ajm1808.00009,
2019arXiv191011277D, WW2019fxh1910.14668, JW2006.16996}.
Given the spacetime-internal symmetry $G=\Spin \times G_{\SM_q}$, the anomaly index $\nu\in \text{TP}_{5}(G)$ belongs to the cobordism group
\bea
 \TP_5(\Spin \times G_{\SM_q})
=\left\{\begin{array}{ll} \Z^5  \times\Z_2,&q=1,3,\\
\Z^5 ,&q=2,6.
\end{array}
\right.
\eea
As checked in \refcite{2019arXiv191011277D, WW2019fxh1910.14668, JW2006.16996},
The perturbative local anomalies ($\Z$ classes) and nonperturbative global anomalies ($\Z_n$ classes) all vanish ($\nu=0$) for both the 15$N_f$-SM and 16$N_f$-SM, such that the SMG is possible in either cases. 
However, if an additional continuous baryon minus lepton symmetry $\U(1)_{\bf B - L}$ is to be preserved, the spacetime-internal symmetry is enlarged to $G=\Spin \times_{\Z_2^F} \U(1)_{\bf B - L} \times G_{\SM_q}=\Spin^c \times G_{\SM_q}$, then the anomaly index $\nu\in \text{TP}_{5}(G)$ belongs to a different cobordism group
\eq{ \TP_5(\Spin^c \times G_{\SM_q})
=\Z^{11}
, \quad q=1,2,3,6,}
which only vanishes ($\nu=0$) for $16N_f$ Weyl fermions \cite{2019arXiv191011277D, JW2006.16996}. Therefore, the SMG preserving an additional $\U(1)_{\bf B - L}$ only works for the $16N_f$-SM.

%
Razamat and Tong (\refcite{Razamat2021Gapped} 
and \refcite{Tong2021Comments}) showed that both the 
15$N_f$-SM and 16$N_f$-SM can be embedded into a left-right (LR) model with 27 Weyl fermions (denoted as the 27$N_f$-LR) 
preserving the $G_{\SM_q}$ symmetry. The key idea is to bring additional fermions down from high-energy  that are vector-like under $G_{\SM_q}$
to mix with the low-energy SM chiral fermions in each family:
\bea\label{eq:SM-SUSY}
\hspace{-6mm}
\setlength{\extrarowheight}{2pt}
\begin{array}{| l | c | ccc | c}
\cline{1-5}
(\overline{\bf 3},{\bf 1})_{2} & ({\bf 1},{\bf 2})_{-3}  &\multicolumn{1}{c|}{({\bf 3},{\bf 2})_{1}}   & \multicolumn{1}{c|}{ (\overline{\bf 3},{\bf 1})_{-4}} & ({\bf 1},{\bf 1})_{6} & \\
\cline{3-6}
(\overline{\bf 3},{\bf 1})_{2} & ({\bf 1},{\bf 2})_{-3}  & &  & & \multicolumn{1}{c|}{ {({\bf 1},{\bf 1})_{0}}}\\
\cline{1-2}
({\bf 3},{\bf 1})_{-2} & ({\bf 1},{\bf 2})_{+3}  & &  & & \multicolumn{1}{c|}{{({\bf 1},{\bf 1})_{0}}}\\
\cline{1-2} \cline{6-6}
\end{array}.
\eea
The first row of \eqnref{eq:SM-SUSY} corresponds to the original 15 left-handed Weyl fermions in \eqnref{eq:SMrep}, and the second and third rows of \eqnref{eq:SM-SUSY} correspond to 6 left-handed and 6 right-handed
additional Weyl fermions in total in a vector-like theory, which add up to $27$ Weyl fermions per family.  
In \eqref{eq:SM-SUSY}, all right-handed fermions are complex conjugated to be written as their anti-particles,
so all fermions are written in the left-handed versions. 
The sterile neutrino $\bar{\nu}_R$ (the 16th Weyl fermion) corresponds to one of the $(\mathbf{1},\mathbf{1})_0$ representation in \eqnref{eq:SM-SUSY}.

Hereafter for the left-handed and right-handed notations,
we always use the italic font $L$ and $R$ to denote that of spacetime symmetry (Spin group),
while use the text font L and R for that of internal symmetry.

The 27$N_f$-LR model has an enlarged symmetry $G_{\LR_{q,p}}$ with $q,p \in \{ 1,2,3,6 \}$, totally sixteen versions,
\eqs{&G_{\LR_{q,p}}\equiv \frac{G_{\SM_q}\times\SU(2)_\mathrm{R}\times\U(1)_\mathrm{R}}{\dsZ_p}\\
&=\frac{\SU(3)\times\SU(2)_\mathrm{L}\times\SU(2)_\mathrm{R}\times\U(1)_\mathrm{L}\times\U(1)_\mathrm{R}}{\dsZ_q\times\dsZ_p},}
where $\SU(2)\times\U(1)_Y$ in $G_{\SM_q}$ is renamed to $\SU(2)_\mathrm{L}\times\U(1)_\mathrm{L}$ in $G_{\LR_{q,p}}$. The 27 Weyl fermions can be organized by the $\SU(2)_\mathrm{R}$ irreducible representations, as indicated by the framed boxes in \eqnref{eq:SM-SUSY}, or 
written as the representation of 
${\su(3) \times  \su(2)_{\rm L} \times  \su(2)_{\rm R} 
\times \u(1)_{\rm L}  \times \u(1)_{\rm R}}$:
\eqs{\label{eq:SMrepLR}
&\big(\bar{d}_R \oplus {l}_L  \oplus q_L  \oplus \bar{u}_R \oplus   \bar{e}_R \oplus \bar{\nu}_R \big) \oplus {d}'_L \oplus \bar{l}'_R =\\
&\big((\overline{\bf 3},{\bf 1},{\bf 2})_{2,-1} 
\oplus ({\bf 1},{\bf 2},{\bf 2})_{-3,3}  
\oplus ({\bf 3},{\bf 2},{\bf 1})_{1,-2} 
\oplus (\overline{\bf 3},{\bf 1},{\bf 1})_{-4,2}\\ &
\oplus ({\bf 1},{\bf 1},{\bf 1})_{6,-6} 
\oplus {({\bf 1},{\bf 1},{\bf 2})_{0,-3}}\big) 
\oplus ({\bf 3},{\bf 1},{\bf 1})_{-2,4}  
\oplus ({\bf 1},{\bf 2},{\bf 1})_{3,0}.}
This essentially promotes $\bar{d}_R,l_L,\bar{\nu}_R$ to $\SU(2)_\mathrm{R}$ doublets and introduces new $\SU(2)_\mathrm{R}$ singlet fermions ${d}'_L,\bar{l}'_R$.

\begin{figure}[htbp]
\begin{center}
\includegraphics[width=0.5\columnwidth]{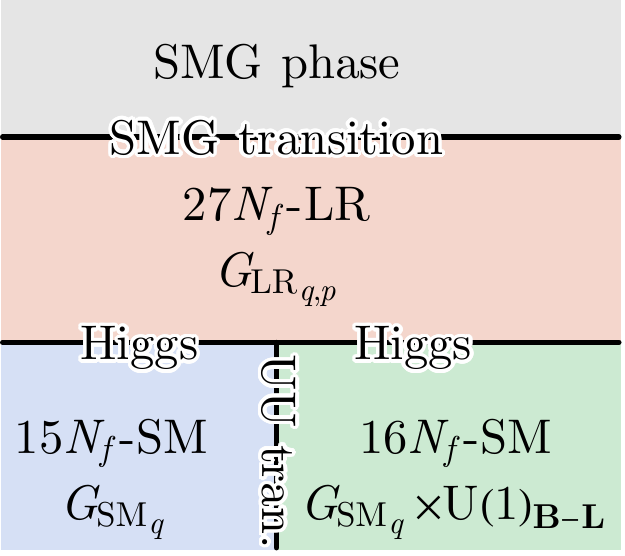}
\caption{A schematic phase diagram among the SMG, the 27$N_f$-LR, the 15$N_f$-SM or the 16$N_f$-SM phases. The SMG phase is separated from the 27$N_f$-LR phase by the SMG transition. The 15$N_f$-SM and 16$N_f$-SM phases are separated by the Ultra Unification (UU) transition \cite{JW2006.16996, JW2012.15860, JW2008.06499}, 
and they can be both obtain from the 27$N_f$-LR model phase by different Higgs transitions.}
\label{fig:SMLR}
\end{center}
\end{figure}

The 27$N_f$-LR phase can be viewed as an intermediate quantum phase between the 15$N_f$-SM or 16$N_f$-SM and the SMG phases, as shown in \figref{fig:SMLR}. The transition from the 27$N_f$-LR phase to the 16$N_f$-SM phase can be driven by condensing a scalar Higgs field $h_\mathrm{R}=(\mathbf{1},\mathbf{1},\mathbf{2})_{0,-3}$ that couples to the Weyl fermions by the following Higgs term
\eq{\label{eq:27-16Higgs}
\big(\epsilon_R h_\mathrm{R}(\bar{d}_R d'_L+
\bar{l}'_R l_L)+(h_\mathrm{R}^\dagger \bar{\nu}_R)(h_\mathrm{R}^\dagger \bar{\nu}_R)\big)+\text{h.c.},}
where $\epsilon_R$ is an anti-symmetric tensor in the $\SU(2)_\mathrm{R}$ doublet subspace. 
The $\epsilon_L$ and $\epsilon$ tensor for the $\SU(2)_\mathrm{L}$ and Lorentz $\su(2)$ subspace are omitted.
All Lagrangian terms become a scalar in a trivial singlet representation of both spacetime and internal symmetries.
For each family, the Higgs condensation $\langle h_\mathrm{R}\rangle\neq 0$ 
(only its \emph{upper} $\SU(2)_\mathrm{R}$ doublet component is nonzero and condenses) 
lifts 11 Weyl fermions with a mass gap 
(the \emph{lower} half of both $\SU(2)_\mathrm{R}$ doublets of $\bar{d}_R$ and $l_L$ get mass,
while the dimension-5 term $(h_\mathrm{R}^\dagger\bar{\nu}_R)(h_\mathrm{R}^\dagger\bar{\nu}_R)$ 
gives Majorana mass to only 
the \emph{upper} half of the $\SU(2)_\mathrm{R}$ doublet $\bar{\nu}_R$), leaves 16 Weyl fermions at low energy, and breaks the symmetry from $G_{\LR_{q,p}}$ down to $G_{\SM_q}\times\U(1)_{\bf B - L}$. 

To further lift the sterile neutrino, an additional scalar Higgs field $h'_\mathrm{R}=(\mathbf{1},\mathbf{1},\mathbf{1})_{0,6}$ should be introduced ({with $\U(1)_{\bf B - L}$ charge $-1$}), such that the Higgs term becomes
\eq{\label{eq:27-15Higgs}
\big(\epsilon_R h_\mathrm{R}(\bar{d}_R d'_L+\bar{l}'_R l_L)+
h'_\mathrm{R}(\epsilon_R \bar{\nu}_R {\nu}_R)\big)+\text{h.c.}}
In each family, the Higgs condensation $\langle h_\mathrm{R}\rangle,\langle h'_\mathrm{R}\rangle\neq 0$ 
leaves 15 Weyl fermions at low energy
(the $h'_\mathrm{R}(\epsilon_R \bar{\nu}_R {\nu}_R)$ gives a
Dirac mass to both upper and lower components of the $\SU(2)_\mathrm{R}$ doublet $\bar{\nu}_R$), 
breaks the symmetry from $G_{\LR_{q,p}}$ down to $G_{\SM_q}$
with no $\U(1)_{\bf B-L}$, 
and drives the transition from the 27$N_f$-LR phase to the 15$N_f$-SM phase.

Embedding the 15$N_f$-SM into the 27$N_f$-LR amounts to the following symmetry extension
\eq{\frac{\SU(2)_\mathrm{R}\times \U(1)_\mathrm{R}}{{\Z_{\gcd(p,2)}}}\to\Spin\times G_{\LR_{q,p}}\to\frac{\Spin\times_{}G_{\SM_q}}{{\Z_{\gcd(p,3)}}},}
and embedding the 16$N_f$-SM into the 27$N_f$-LR amounts to the following symmetry extension
\eq{\SU(2)'_\mathrm{R}\to\Spin\times G_{\LR_{q,p}}\to\frac{\Spin^c\times_{}G_{\SM_q}}{\Z_{\gcd(p,3)}},}
where $\SU(2)'_\mathrm{R}=\frac{\SU(2)_\mathrm{R}\times\U(1)_\mathrm{R}}{\Z_{\gcd(p,2)} \times \U(1)_{\bf B-L}}$.

%

\subsubsection{Symmetric Mass Generation in a 27$N_f$-Weyl-Fermion Left-Right Model}  

\label{sec:27NfSM}

The SMG in the 27$N_f$-LR model
can be achieved by both the parton-Higgs mechanism in \secref{sec:Parton-Higgs-mechanism}
and the s-confinement mechanism in \secref{sec:s-confinement-mechanism}.

\begin{enumerate}[leftmargin=2.mm]

\item The s-confinement mechanism: According to Razamat and Tong \cite{Razamat2021Gapped}, one first supersymmetrizes the fermions in \eqnref{eq:SM-SUSY} to their corresponding $\CN=1$ chiral multiplets as
$\bar{d}_R \oplus {l}_L  \oplus q_L  \oplus \bar{u}_R \oplus   \bar{e}_R \oplus \bar{\nu}_R \oplus {d}'_L \oplus \bar{l}'_R \to {\rm D}\oplus{\rm L} \oplus {\rm Q}\oplus {\rm U}\oplus {\rm E}\oplus {\rm N}\oplus {\rm D}'\oplus {\rm L}'$.
%
Then gauge the $\SU(2)_{\rm R}$ symmetry by turning on a dynamical $\SU(2)_{\rm R}$ gauge field that couples 
to the 
$\su(2)_{\rm R}$ doublet: ${\rm D}, {\rm L}, {\rm N}$.
A  \emph{dangerously irrelevant}
superpotential $\CW_{\rm UV}$ at UV 
\bea \label{eq:W-potential-UV}
\CW_{\rm UV}&=&{\rm L}{\rm L}{\rm E}+{\rm D}{\rm D}{\rm U}+{\rm L}{\rm D}{\rm Q}+{\rm L}{\rm N}{\rm L}'+ {\rm D}{\rm N}{\rm D}'+
{\rm h.c.}\quad\;
\eea
becomes a $G_{\SM_q}$-symmetric \emph{relevant} deformation
that pairs the 15 mesons ($\tilde{\rm E},\tilde{\rm U},\tilde{\rm Q},\tilde{\rm L},\tilde{\rm D}$) formed by ${\rm D}, {\rm L}, {\rm N}$ via s-confinement) 
with the remained 15 superfields 
(namely ${\rm Q}, {\rm U}, {\rm E}, {\rm D}', {\rm L}'$) in a quadratic manner $\CW_{\rm IR}$ at IR, consequently gapping out all fields as SMG:
\bea \label{eq:W-potential-IR}
\CW_{\rm IR}&=&\tilde{\rm E}{\rm E}+\tilde{\rm U}{\rm U}+\tilde{\rm Q}{\rm Q}+\tilde{\rm L}{\rm L}'+\tilde{\rm D}{\rm D}' +
{\rm h.c.}
\eea
Here the $\su(3)$ color and $\su(2)_\mathrm{L}$ flavor indices are suppressed, with the understanding that they should be contracted properly to make the Lagrangian a singlet. As $\CW_{\rm IR}$ flows strong, all fermions are gapped from low-energy, resulting in the SMG phase.
When there are multiple families, independent $\SU(2)_\mathrm{R}$ gauge fields are introduced in each family, such that the total gauge group is $\SU(2)_{\rm R_1} \times \SU(2)_{\rm R_2} \times \cdots \times \SU(2)_{\mathrm{R}_{N_f}}$. This guarantees that the s-confinement can induce the fully gapped SMG phase in each family independently.

\item
The parton-Higgs mechanism: Tong \cite{Tong2021Comments} shows
that $N_f$ families of 27 Weyl-fermion model
can be fully gapped by preserving \emph{not only} 
the SM internal symmetry group $G_{\SM_q}$ for $q=1,2,3,6$,
\emph{but also} an additional continuous baryon minus lepton symmetry $\U(1)_{\bf B - L}$.

The parton-Higgs mechanism introduces the scalar Higgs fields
${\phi}={({\bf 1},{\bf 2},{\bf 2})_{-3,3}}$. \refcite{Tong2021Comments} suggests 
to fully gap (the 27 Weyl fermions per family) 
to achieve the SMG by adding
\bea 
\label{eq:Yukawa-full-gap-phi}
\big(\phi (
   {{\bar{d}_R}}  {{q_L}} 
+ {{\bar{\nu}_R}} {{\bar{l}'_R}}
+ {{\bar{e}_R}} {{l_L}})
+\phi^2  {{{d'_L}} {{\bar{u}_R}}}
\big) 
+ {\rm h.c.}
\eea
when the condensation $\langle \phi \rangle \neq 0$ occurs.
The internal symmetry breaking pattern of this SMG deformation is
\bea
&& G_{\SM_q} \times_{\Z_{\gcd(p,3)}} K  
= G_{\SM_q} \times_{\Z_{\gcd(p,3)}}  \frac{\SU(2)_\mathrm{R}\times \U(1)_\mathrm{R}}{{\Z_{\gcd(p,2)}}}  \cr
&&\mapsto \frac{\SU(3) \times \SU(2)_{\text{diagonal}}  \times \U(1)_{\text{diagonal}}}{\Z_q} \times \U(1)_{\bf B-L}, \quad\quad 
\eea
which leaves a continuous baryon minus lepton number symmetry preserved.

\end{enumerate}


\section{Summary and Outlook}
\label{sec:summary}

We have reviewed the topic of symmetric mass generation (SMG) which has attracted considerable ongoing
interests from both the condensed matter and high energy physics communities in recent years. 
The SMG transition is beyond the classic Landau-Ginzburg-Wilson paradigm, which potentially involves fractionalization of physical fermions at and only at the critical point.
We have reviewed various aspects related to SMG, including interacting topological insulators (TI), topological superconductors (TSC)  \cite{2010RMP_HasanKane, 2011_RMP_Qi_Zhang}, SPT states \cite{Senthil1405.4015, WenZoo1610.03911},
anomalies, lattice regularization of chiral gauge theories, and the current status of numerical efforts from both the lattice gauge theory and condensed matter communities. We have also discussed theoretical understanding related to SMG, such as the s-confinement mechanism, the connection between SMG and DQCP, etc.

Various numerical works have suggested that the SMG could be a continuous transition, and the critical exponents have been measured in some of the numerical works. These works pose a challenge to further analytical understanding of SMG. Unlike the standard Higgs-Yukawa types of theories where a large-$N$ or small-$\epsilon$ expansion can be applied, theoretically we do not yet have a controlled theory where we can perform a reliable analytical calculation for the critical exponents for SMG, and compare with the numerical simulation. If the SMG is indeed a continuous transition and corresponds to a certain type of conformal field theories, then the conformal Bootstrap method can also provide very helpful insights into the nature of the transition. All these require further efforts from different disciplines of theoretical physics. 

The problem of the nonperturbative regularization of chiral fermion and chiral gauge theories also potentially bring the high-energy lattice, mathematical physics, quantum information, numerical simulation, and condensed matter communities 
to work closer together. 
What else can stimulate different communities to work together other than tackling a profound mysterious problem?
Looking back at science history, the phenomena of the anomaly-inflow  \cite{1984saCallanHarvey, Witten2019bou1909.08775} 
and domain wall fermion \cite{Kaplan1992A-method, Kaplan2009Chiral} (as the precursors
of the TI/TSC/SPT 
states) had an old tradition 
rooted in the high-energy theory and lattice community in the mid 1980s and mid 1990s.
Closely related phenomena like integer and fractional quantum Hall states are studied in the 
condensed matter community already in the early 1980s. Yet the acceleration came much later until the
concrete materialization discovery of TI/TSC in 2005 
(in both theories and experiments \cite{2010RMP_HasanKane, 2011_RMP_Qi_Zhang})
ignites the serious interest in the classification of the interacting many-body quantum systems of SPT states 
in the early 2010s \cite{Schnyder:2008os,Kitaev2009Periodic,Ryu:2010fe,Senthil1405.4015, WenZoo1610.03911}.
This generates crossing-disincline interests in the classification of anomalies, cobordism classes, 
and the TQFTs in quantum matter,
high-energy string theory, and mathematical physics 
\cite{Wenanomalies1303.1803,Kapustin2014Anomalies, Wang2015Field-Theory, Kapustin2015Fermionic, Witten1508.04715,
Freed2016Reflection}. Look back and ponder this history, 
some curious minds might ask: Why were the discovery of SMG and classification of TI/TSC/SPT states not made in the 
high-energy lattice community even earlier, 
although the lattice gauge theory 
pioneers like Wilson and Kogut already had earlier achievements on the related lattice topics 
in the 1970s \cite{Kogut1979}?
Perhaps the inputs of the condensed matter material experiments are crucial, 
also perhaps the crossing-boundary interdisciplinary ideas are much more welcomed in the 
recent condensed matter developments.
As Preskill summarized: Combing more ideas together is better than each of the isolated ideas \cite{Preskill1811.10085}.
Hopefully combing the ideas of (1) domain wall fermions \cite{Kaplan1992A-method}, 
(2) gapping the mirror fermions \cite{Eichten1986Chiral}, 
and (3) the anomaly-free SMG \cite{Wenanomalies1303.1803, Wen:2013kr, You:2015lj, Wang2018A-Non-Perturbative}, already gives the ample insights to fully 
solve the lattice definition of chiral fermion, chiral gauge theory, and chiral Standard Model problems.

\begin{acknowledgments}

The authors are listed in alphabetical order by a standard convention.
We would like to acknowledge Cenke Xu, who initiated this review project, contributed to the writing of Secs.\,\ref{sec:FK}, \ref{sec:zeros}, \ref{sec:summary}, and suggested numerous improvements. We would also like to thank David Tong for patiently explaining a few points of his works to us, and John McGreevy for reading through our draft and providing many helpful suggestions.  
JW thanks Yuta Hamada for discussions.
JW is supported by Center for Mathematical Sciences and Applications at Harvard University and
NSF Grant DMS-1607871 ``Analysis, Geometry and Mathematical Physics.'' 
YZY is supported by a startup fund at UCSD.

\end{acknowledgments}

\bibliography{ref}
\end{document}